\newcommand{\commentt}[2]{#1}
\newcommand{\comment}[1]{}

\newcommand{\cov}{{\rm Cov}}
\newcommand{\var}{{\rm Var}}

\commentt{
\newcommand{\citet}{\citeasnoun}
\documentclass[a4paper,11pt]{article}
\usepackage{graphicx,slashbox,amsthm}
\usepackage[full]{harvard}

\usepackage[ps,dvips]{xy}
\usepackage{pgfplots}
\title{Randomized Dimension Reduction for Monte Carlo Simulations}
\author{Nabil Kahal\'e
\thanks{\emph{ESCP Europe, Labex ReFi, Big data research center,  75011 Paris,
France; {e-mail: }{nkahale@escpeurope.eu}.}}
}
\date{\today}
\usepackage{amstext}
\usepackage{amssymb}
\usepackage{amsmath}
\usepackage{setspace}
\usepackage[margin=1in]{geometry}
\usepackage[english]{babel}
\bibliographystyle{dcu}
\usepackage{color}
\usepackage{textcomp}
\usepackage{varioref}
\begin{document}

\newtheorem{example}{Example}[section]
\newtheorem{theorem}{Theorem}[section]
\newtheorem{conjecture}{Conjecture}[section]
\newtheorem{lemma}{Lemma}[section]
\newtheorem{proposition}{Proposition}[section]
\newtheorem{remark}{Remark}[section]
\newtheorem{corollary}{Corollary}[section]
\newtheorem{definition}{Definition}[section]
\numberwithin{equation}{section}
\maketitle
\newcommand{\ABSTRACT}[1]{\begin{abstract}#1\end{abstract}}
\newcommand{\citep}{\cite}
}
{
\documentclass[mnsc]{informs3} 
\usepackage[margin=1in]{geometry}
\usepackage{textcomp}
\usepackage{pgfplots}
\DoubleSpacedXII



\usepackage[round]{natbib}
 \bibpunct[, ]{(}{)}{,}{a}{}{,}%
 \def\bibfont{\small}%
 \def\bibsep{\smallskipamount}%
 \def\bibhang{24pt}%
 \def\newblock{\ }%
 \def\BIBand{and}%

\TheoremsNumberedThrough     
\ECRepeatTheorems

\EquationsNumberedThrough    

\MANUSCRIPTNO{} 


\begin{document}
\RUNAUTHOR{Kahal\'e}
\RUNTITLE{Randomized Dimension Reduction}
\TITLE{Randomized Dimension Reduction for Monte Carlo Simulations}
\ARTICLEAUTHORS{%
\AUTHOR{Nabil Kahal\'e}

\AFF{ESCP Europe, Labex R\'efi, Big data research center, 75011 Paris, France, \EMAIL{nkahale@escpeurope.eu} \URL{}}
} 
\bibliographystyle{informs2014} 
}

\ABSTRACT{
We present  a new unbiased algorithm that  estimates the expected value of $f(U)$ via Monte
Carlo  simulation, where $U$ is a vector of $d$ independent random variables, and $f$ is
a function of \(d\) variables. We assume that \(f\) does not depend equally on all its arguments.  Under certain conditions we prove that, for the same computational cost, the variance of  our estimator is lower than the variance of the standard Monte Carlo estimator by a factor of order  \(d\).  Our method can be used to obtain  a low-variance unbiased estimator  for the expectation  of a function of the state of  a Markov chain at a given time-step. We study applications to  volatility forecasting and time-varying queues. Numerical experiments show that our algorithm dramatically improves upon the standard Monte Carlo method for large values of \(d\), and is highly resilient to discontinuities.
     }
\commentt{
}{\maketitle}
\section{Introduction}
Markov chains arise in a variety of fields such as finance, queuing  theory,  and social networks.   While much research has been devoted to the study of steady-states of Markov chains, several practical applications rely on the transient behavior of Markov chains. For example, the volatility of an index can be modelled as a Markov chain using the GARCH model~\citep[Ch. 23]{Hull14}. Financial institutions conducting  stress tests may need to estimate the probability that the volatility  exceeds a given level in a few years from now. Also, due to  the nature of human activity, queuing systems  in  areas such as health-care, manufacturing, telecommunication and transportation networks, have often time-varying features and do not have a steady-state. For instance, empirical data show significant daily variation in  traffic in wide-area networks~\citep{paxson1994empirically,thompson1997wide} and vehicular flow on roads~\citep{TrafficJamOPRE2003}. Estimating the expected delay of packets in    a wide-area network at a specific time of the day (12pm, say) could be used to dimension such  networks. Similarly, estimating the velocity of cars in a region at 6pm could be used to design  transportation       
networks. In the same vein, consider the problem of estimating the queue-length at the end of a business day
in a call center that operates with fixed hours. In such call centers, knowing how many
calls would still need to be answered at 5pm could be an important metric that would
be needed in estimating their staffing requirements.    Methods to determine appropriate
staffing levels in call centers and other
many-server queueing systems with time-varying
arrival rates have been designed in~\citep{whitt2008staffing}. Also, approximation tools have been developed to study  time-varying queues (see~\citep{whitt2017timeVarqueues} and references therein).  However, in many situations, there are no  analytical tools, except Monte Carlo simulation,  to study accurately systems modeled by a  Markov chain.   A drawback of Monte Carlo simulation is its high computation cost. This motivates the need to design efficient simulation tools to study the transient behavior of  Markov chains, with or without time-varying features.

This paper gives a new  unbiased algorithm to estimate
\(E(f(U))\),   where      \(U=(U_{1},\ldots,U_{d})\) is a vector of \(d\) independent  random variables
    \(U_{1},\ldots,U_{d}\)
  taking values in a measurable space \(F\), and  \(f\)  is a real-valued Borel-measurable function on \(F^{d}\) such that 
\(f(U)\) is square-integrable. For instance,  \(F\) can be equal to \(\mathbb{R}\) or to any vector space over \(\mathbb{R}\). Under certain conditions, we show that  our  algorithm yields substantial lower variance than the standard Monte Carlo method for the same computational effort.  Our techniques can be used to efficiently estimate the expected value of a function of the state of a  Markov chain at a given time-step \(d\), for a class of  Markov chains driven by independent random variables. An alternative  algorithm for Markov chains estimation, based on Quasi-Monte Carlo  sequences, that substantially improves upon standard Monte Carlo in certain numerical  examples, is given in~\citep{Ecuyer2008}, with 
bounds on  the variance proven for special situations where the state space of the chain is a subset of the real
numbers. 

In a standard Monte Carlo scheme, \(E(f(U))\)    is estimated by simulating  \(n\)  independent vectors in \(F^{d}\) having the same distribution as \(U\),   and taking the average of \(f\) over the \(n\) vectors.  In the related  Quasi-Monte Carlo method (see~\citep[Ch.~5]{glasserman2004Monte}),  \(f\) is evaluated at a predetermined deterministic sequence of points.
  In several  applications, the efficiency of   Quasi-Monte Carlo algorithms  can be improved by reordering the \(U_{i}\)'s and/or making a change of variables, so that the value of  \(f(U)\) depends  mainly on the first few  \(U_{i}\)'s.
For instance, the Brownian bridge construction and principal components analysis have been
used~\citep{CMO1997,ABG1998,AL2000}  to reduce the error in the valuation of  financial derivatives via Quasi-Monte Carlo methods (see~\citep{caflisch1998}
for related results). The relative importance of the first variables can  formally be measured by calculating the effective dimension in the truncation sense, a concept defined in~\citep{CMO1997}: when the first variables are important, the effective dimension in the truncation sense  is low in comparison to the nominal dimension.  It is proven in~\citep{SW1998}  that Quasi-Monte Carlo methods are  effective for a class of functions where the importance of \(U_{i}\) decreases with \(i\). \citet{EcuyerLemieux2000}  apply Quasi-Monte Carlo methods to   queueing simulation    and option pricing, and examine their connection
to the effective dimension.  The truncation dimension and a related notion, the effective dimension in the superposition sense, are studied in~\citep{sobol2001global,owen2003,LiuOwen2006}. It is shown in~\citep{WangFang2003,WangSloan2005,Wang2006}  that the Brownian bridge and/or principal components analysis algorithms substantially reduce the truncation dimension of certain financial instruùments. Alternative linear transformations have been
proposed in~\citep{ImaiTan2006,wangSloan2011,wangTan2013}  to reduce the effective dimension of financial derivatives and improve the performance of Quasi-Monte Carlo methods. 

Other previously known variance reduction techniques have exploited the  importance of certain variables  or states. 
For instance,     
stratified sampling along    important directions
 is used in pricing  path-dependent options~\cite[Section 4.3.2]{glasserman1999asymptotically,glasserman2004Monte}.  Importance sampling methods aim to increase the number of  samples that hit an important set via a change of measure technique~\cite[Section V.5]{asmussenGlynn2007}.   
When \(d=2\) and  \(f(U_{1},U_{2})\) is more influenced by \(U_{1}\) than by \(U_{2}\),   and the expected time to generate \(U_{1}\) is much lower than the expected time to generate \(U_{2}\), the splitting   technique~\cite[Section V.5]{asmussenGlynn2007} simulates several independent copies of \(U_{1}\) for each copy of \(U_{2}\).  \citet[Section V.5]{asmussenGlynn2007} give
the variance of the splitting estimator and the optimal  number of copies of \(U_{1}\),
and show that the splitting technique is related to the conditional  Monte Carlo method. Multilevel splitting techniques are often  used for variance reduction in  the estimation of rare event probabilities~\cite[VI.9]{asmussenGlynn2007}. The idea is to  split each path that reaches an important region into a number of subpaths in order to produce more paths that hit the rare event set. The rare event probability is then evaluated via a telescoping product. \citet{ermakov1995design} analyse  multilevel splitting techniques that estimate  functionals of Markov chains with a discrete state space and   of  ergodic Markov chains in their steady state.   \citet{glasserman1999multilevel} analyse the performance of multilevel splitting techniques for rare event estimation and give, under certain conditions,  the optimal degree of splitting as the probability of the event goes to \(0\). Multilevel splitting methods have had many applications, such as the estimation of network reliability~\citep{botev2013static} and of  rare events in Jackson networks~\citep{blanchet2011analysis}. Multilevel splitting techniques for rare event simulation with finite time constraints are analysed in~\citep{jiang2017importance}. A comprehensive survey  on multilevel splitting techniques with applications to rare event simulations, sampling from complicated distributions, Monte Carlo counting, and randomized optimization, can be found in~\cite[Ch. 9]{rubinstein2016simulation}. 

     Another technique, the multilevel Monte Carlo 
(MLMC) method introduced in~\citep{Giles2008},  which relies on  low dimensional approximations of the function to be estimated,  dramatically reduces the computational complexity of estimating an expected value arising
from a stochastic differential equation.  Related randomized multilevel methods that produce  unbiased estimators" for equilibrium expectations of functionals defined on homogeneous Markov chains have been provided in~\citep{glynn2014exact}. These methods apply to    the class of positive Harris recurrent Markov chains, and to chains that are contracting on average.   It is shown in~\citep{GlynnRhee2015unbiased} that similar randomized multilevel methods can be used to efficiently compute  unbiased estimators for expectations of functionals of solutions to stochastic differential equations. The MLMC method has had numerous other applications (e.g.,~\citep{Staum2017}).

  The basic idea behind our algorithm is that, if  \(f\)  does not depend  equally on all its arguments, the standard Monte Carlo method can be inefficient because it simulates all \(d\) arguments of \(f\) at each iteration. Assuming that  the expected time needed to simulate \(f(U)\) is  of order  \(d\) and that the variance of  \(f(U)\) is upper and lower-bounded by constants, the expected time needed to achieve  variance  \(\epsilon^{2}\) by standard Monte Carlo simulation is  \(\Theta(d\epsilon^{-2})\). In contrast, our algorithm simulates  at each iteration  a random subset  of  arguments of \(f\), and reuses the remaining arguments from the previous iteration.
Under certain conditions,
we show that  our algorithm estimates \(E(f(U))\) with variance  \(\epsilon^{2}\) in \(O(d+\epsilon^{-2})\) expected time. We also establish central limit theorems
on the statistical error of our algorithm.
When \(d=2\), our method is very similar to the splitting technique in~\cite[Section V.5]{asmussenGlynn2007}. Our approach can thus be viewed as a   multidimensional version of this technique. However, in contrast with
existing multilevel splitting algorithms
where splitting decisions   typically depend on the   current state, in our method, the  arguments of \(f\) to be redrawn  are independent of  previously generated copies of  \(U\).

In order to  optimize  the performance of our estimator, we minimize the asymptotic product of the variance and expected running time, in the same spirit as stratified sampling~\cite[Section 4.3.1]{glasserman2004Monte}, the splitting technique~\cite[Section V.5]{asmussenGlynn2007}),   MLMC \citep{Giles2008}, and related methods~\citep{GlynnRhee2015unbiased}.
   This minimization is performed via a new geometric algorithm that solves in \(O(d)\) time a \(d\)-dimensional optimisation problem. Our geometric algorithm is of independent interest and can be used to solve an optimization problem of the same type that was  solved in~\citep[Section 3]{GlynnRhee2015unbiased} in   \(O(d^{3})\) time. We are not aware of other previous algorithms that solve this problem.
 This work extends the research in~\citep{kahNIPS16}, where the variance properties of the randomized estimator presented in this paper were announced without  proof.

Our method has the following features:

\begin{enumerate}
\item Under certain conditions,
it estimates \(E(f(U))\) with variance  \(\epsilon^{2}\) in \(O(d+\epsilon^{-2})\) expected time. We are not aware of any previous method that achieves, under the same conditions, such a tradeoff between the  expected running time and accuracy. In contrast with stratified sampling, which can be performed in practice  only along a small number of dimensions~\cite[Example 4.3.4] {glasserman2004Monte}, our method 
is targeted at high-dimensional problems. The efficiency of our method typically increases with \(d\), even though it can be used in principle for any  \(d\ge2\). \item 
It is easy to implement, does not make any continuity assumptions on \(f\), nor does it    require  a detailed knowledge of the structure of \(f\) or \(U\).
In contrast,  importance sampling,  multilevel    splitting  and multilevel Monte Carlo methods can achieve substantial variance reduction by exploiting the  structure of the simulated model. The standard Quasi-Monte Carlo method does not    necessitate  a detailed knowledge of the model structure, but makes regularity assumptions on the function to be integrated, and its efficiency does not  increase with \(d\). 

\end{enumerate}        
  
The
rest of the paper is organized as follows.   
\S\ref{se:main} presents our generic randomized dimension reduction  algorithm and analyses its performance. \S\ref{se:timeVarBound}  describes the aforementioned geometric algorithm and gives a numerical  implementation of the randomized dimension reduction algorithm.    \S\ref{se:examples}  provides applications to Markov chains. \S\ref{se:deterministic} presents and analyses a deterministic version of our algorithm. \S\ref{se:ML}   compares our algorithm to a class of MLMC algorithms. 
 \S\ref{se:numerExper} gives numerical simulations. \S\ref{se:conclu} contains concluding remarks. Most proofs are contained in the appendix\commentt{}{ or the on-line supplement}.
The connection between our approach and the ANOVA decomposition and truncation dimension is studied in the appendix.  The \commentt{appendix}{on-line supplement} explores further the relation between our method,   the splitting technique, and the conditional Monte Carlo method, and contains more numerical simulations.

    \section{The  generic randomized dimension reduction algorithm}\label{se:main} 
\subsection{The algorithm description}
\label{sub:generalAlgorithm}We assume that all random variables in this paper are defined on the same probability space \((\Omega,\mathcal{F},\mathbb{P})\). Our algorithm
   estimates \( E(f(U))\) by performing \(n\) iterations, where \(n\) is
   an arbitrary positive integer. The algorithm samples more often the first arguments of \(f\) than the last ones. It implicitly assumes that, roughly speaking, the importance  of the $i$-th argument of  \(f\)  decreases with $i$.   In many Markov chain examples, the last random variables are more important than the first ones, but our algorithm  can still be used efficiently after re-ordering the random variables, as described in detail in \S\ref{se:examples}. A general  approach to rank input variables according to their importance is described in~\citep{sobol2001global}, but we will not   use such an approach in our examples.

Let \begin{displaymath}
A=\{(q_{0},\dots,q_{d-1})\in\mathbb{R}^{d}:1=q_{0}\ge q_{1}\ge\cdots\ge q_{d-1}>0\}.
\end{displaymath}Throughout the paper,      \(q=(q_{0},\dots,q_{d-1})\) denotes
an element of \(A\).  Our   generic algorithm takes such a vector      \(q\) as parameter. Let \((N_{k}\)), 
\(k\geq 1\),  be    a sequence of  independent random integers in \([1,d]\) such that \(\mathbb{P}(N_{k}>i)=q_{i}\) for \(0\leq i\leq d-1\) and
\(k\geq 1\).  The algorithm simulates \(n\) copies \(V^{(1)},\dots,V^{(n)}\) of \(U\) and consists of
the following steps:
\begin{enumerate}
\item 
First iteration.
Simulate a vector \(V^{(1)}\) that has the same distribution as \(U\) and calculate \(f(V^{(1)})\).

\item Loop. In iteration \(k+1\), where \(1\leq k\leq n-1\), let \(V^{(k+1)}\) be the vector obtained from  \(V^{(k)}\) by redrawing the first \(N_{k}\) components of \(V^{(k)}\), and keeping the remaining components unchanged.
Calculate \(f(V^{(k+1)})\).
\item Output the average  of \(f(V^{(1)}),\dots,f(V^{(n)})\). \end{enumerate}

More formally, 
 consider a sequence \((U^{(k)}\)), 
\(k\geq 1\),  of independent  copies of \(U\) such that the two sequences  \((N_{k})\), 
\(k\geq 1\), and  \((U^{(k)}\)), 
\(k\geq 1\), are independent.   Define the sequence \((V^{(k)}\)),  \(k\geq1\),  in \(F^{d}\)    as follows: \(V^{(1)}=U^{(1)}\) and, for \(k\geq1\), the first \(N_{k}\) components of \(V^{(k+1)}\) are the same as the corresponding components of \(U^{(k+1)}\), and the remaining components of \(V^{(k+1)}\)  are the same as the corresponding components of \(V^{(k)}\).   The algorithm then outputs \begin{displaymath}
f_{n}\triangleq\frac{f(V^{(1)})+\dots+f(V^{(n)})}{n}.
\end{displaymath}
Note that \(f_{n}\) is an unbiased estimator of \( E(f(U))\) since \(V^{(k)}\,{\buildrel d \over =}\,U\) for   \(1\leq k\leq n\). 
\subsection{Performance analysis}
For ease of presentation, we  ignore the time needed to generate \(N_{k}\) and the running time of the third step of the algorithm. For \(1\leq i\leq d\), let \(t_{i}\)  be  the expected time needed to generate \(V^{(k+1)}\) and calculate \(f(V^{(k+1)})\) when \(N_{k}=i\).  Equivalently,  \(t_{i}\) is  the expected time needed to perform Step~2 of the algorithm when \(N_{k}=i\). Thus, \(t_{i}\)  is  the expected time needed to re-draw the first \(i\) components of \(U\) and  recalculate  \(f(U)\), and  \(t_{d}\) is  the expected
time needed to simulate \(f(U)\).
By co\textsf{}nvention,  \(t_{0}=0\). We  assume for simplicity that \(t_{i}\) is a strictly increasing function of \(i\). In many examples (see  \S\ref{sub:Lipschitz} and \S\ref{se:examples}), it can be shown that \(t_{i}=O(i)\).         As \(\mathbb{P}(N_{k}=i)=q_{i-1}-q_{i}\) for \(1\leq i\leq d\) and \(k\geq1\), where \(q_{d}=0\), the expected running time of a single iteration of our algorithm, excluding the first one, is equal to \(T\), where
\begin{equation}\label{eq:TimeVarianceUpperBound}
T\triangleq\sum^{d}_{i=1}(q_{i-1}-q_{i})t_{i}=\sum^{d-1}_{i=0}q_{i}(t_{i+1}-t_{i}).
\end{equation}  
For \(0\leq i\leq d\), define  \begin{equation*}
C(i)\triangleq\var(E(f(U)|U_{i+1},\ldots,U_{d})).
\end{equation*}Thus, \(C(0)=\var (f(U))\), while \(C(d)=0\), and we can interpret \(C(i)\) as the variance captured by the last \(d-i\) components
of \(U\). Note that if \(f\) depends only on its first \(i\) arguments, then \(f(U)\) is independent of \((U_{i+1},\ldots,U_{d})\), and
so \begin{displaymath}
E(f(U)|U_{i+1},\ldots,U_{d})=E(f(U)),
\end{displaymath} which implies that \(C(i)=0\). More generally, if the last  \(d-i\) arguments of \(f\) are not important, the conditional expectation \(E(f(U)|U_{i+1},\ldots,U_{d})\) is ``almost'' constant, and its  variance    \(C(i)\)  should be  small. Thus \(C(i)/C(0)\) can be used to measure the importance of
the last \(d-i\) components
of \(U\). As shown in the appendix, when  \(U\) is uniformly distributed on the \(d\)-dimensional unit cube \([0,1]^{d}\), this ratio coincides with a global sensitivity index for the subset \(\{i+1,\dots,d\}\),  defined in~\cite[Definition 3]{sobol2001global}
 in terms  of the ANOVA decomposition of \(f\). Proposition~\ref{pr:covf(U)f(U')} below shows that \((C(i))\), \(0\leq i\leq d\),  is always a decreasing sequence,  and gives an alternative   expression for \(C(i)\),  which can be viewed as a  variant of Theorem 2 of~\citep{sobol2001global}.   
\begin{proposition}\label{pr:covf(U)f(U')}
The sequence \((C(i))\), \(0\leq i\leq d\), is decreasing. If \(U'_{1},\ldots ,U'_{i}\) are  random variables such that  \(U'_{j}\,{\buildrel d \over =}\,U_{j}\)  for \(1\leq j\leq i\), and   \(U'_{1},\ldots ,U'_{i}\),  \(U\) are independent, then   
\begin{equation}\label{eq:ViCov}
C(i)=\cov(f(U),f(U'_{1},\ldots ,U'_{i},U_{i+1},\ldots,U_{d})).
\end{equation}
\end{proposition}
Theorem~\ref{th:variance} below establishes a formal relationship between the variance of \(f_{n}\) and the \(C(i)\)'s. Let  \(\nu^{*}\) be the vector of \(\mathbb{R}^{d+1}\) with  \(\nu^{*}_{0}=C(0)\) and \(\nu^{*}_{i}=2C(i)\) for \(1\leq i\leq d\). 
\begin{theorem}\label{th:variance}For \(n\geq1\),
\begin{equation}\label{eq:lemmaVariance}
n\var(f_{n})\le \sum^{d-1}_{i=0}\frac{\nu^{*}_{i}-\nu^{*}_{i+1}}{q_{i}}.
\end{equation}Furthermore, the LHS of~\eqref{eq:lemmaVariance} converges to its RHS as \(n\) goes to infinity. 
\end{theorem}
As, for \(\nu=(\nu_{0},\dots,\nu_{d})\in\mathbb{R}^{d}\times\{0\}\),   \begin{equation}\label{eq:numonotone}
\sum^{d-1}_{i=0}\frac{\nu _{i}-\nu _{i+1}}{q_{i}}=\nu _{0}+\sum^{d-1}_{i=1}\nu _{i}(\frac{1}{q_{i}}-\frac{1}{q_{i-1}}),
\end{equation}the RHS of~\eqref{eq:lemmaVariance} is a weighted combination of the \(C(i)'s\), with positive weights. Thus, the smaller  the \(C(i)\)'s, the smaller  the RHS of~\eqref{eq:lemmaVariance}. Furthermore, as \(C(i)\) is the variance of   the conditional expectation \(E(f(U)|U_{i+1},\ldots,U_{d})\), which can be considered as  a smoothed version of \(f(U)\), we expect our algorithm to be resilient to discontinuities of \(f\). 

Denote by \(\mathbb{R}_{+}\)  the set of nonnegative real numbers.
For \(q\in A\), and \(\vartheta=(\vartheta_{0},\dots,\vartheta_{d})\in\{0\}\times \mathbb{R}_{+}^{d}\), and \(\nu=(\nu_{0},\dots,\nu_{d})\in\mathbb{R}_{+}^{d}\times\{0\}\),     set  
   \begin{equation}\label{eq:RDef}
R(q;\vartheta,\nu)=(\sum^{d-1}_{i=0}q_{i}(\vartheta_{i+1}-\vartheta_{i}))(\sum^{d-1}_{i=0}\frac{\nu _{i}-\nu _{i+1}}{q_{i}}).
\end{equation}
The expected time needed to perform \(n\) iterations of the algorithm, including  the first one, is  \(T_{n}=(n-1)T+t_{d}\). Theorem~\ref{th:variance} and~\eqref{eq:TimeVarianceUpperBound} imply that    \(T_{n}\var(f_{n})\) converges to \(R(q;t,\nu^{*})\) as \(n\) goes to infinity, where \(t=(t_{0},\dots,t_{d})\).  By~\eqref{eq:numonotone},
 \(R(q;\vartheta,\nu)\) is an increasing function with respect to 
 \(\nu\), i.e. 
 \(R(q;\vartheta,\nu)\leq R(q;\vartheta ,\nu')\) for \(\nu\leq\nu'\), where the symbol \(\leq\) between vectors represents componentwise inequality. Similarly,  it is   easy to see that  \(R(q;\vartheta ,\nu)\) is increasing with respect to \(\vartheta\). Let   \(T^{\text{tot}}(q,\epsilon)\) be   the total expected time  it takes for our algorithm to   guarantee that \(\text{Std}(f_{n})\leq\epsilon\).     Corollary~\ref{cor:Ttotal} below gives an upper bound on   \(T^{\text{tot}}(q,\epsilon)\)  in terms of \(R(q;t,\nu^{*})\). It also  implies that,  if \(R(q;t,\nu^{*})\) is upper bounded by a constant independent of \(d\),  and    \(\var(f(U))=\Theta(1)\),   then our algorithm outperforms the standard Monte Carlo algorithm by a factor of order \(t_{d}\). More precisely, running our algorithm for     \(n=\lceil t_{d}T^{-1}\rceil\) iterations has the same expected cost, up to a constant, as a single iteration of the standard Monte Carlo method, but produces an unbiased estimator of  \(E(f(U))\) with \(O(1/t_{d})\) variance.           
\begin{corollary} 
\label{cor:Ttotal}
For \(\epsilon>0\),
\begin{equation}\label{eq:corTotq}
T^{\text{tot}}(q,\epsilon)\le t_{d}+ R(q;t,\nu^{*})\epsilon^{-2}. 
\end{equation} Furthermore, if   \(n=\lceil t_{d}T^{-1}\rceil\),  the expected running time of \(n\) iterations of the algorithm is at most \(2t_{d}\), and\begin{equation}\label{eq:corvarfn}
\var(f_{n})\leq\frac{R(q;t,\nu^{*})}{t_{d}}.
\end{equation}   
\end{corollary}
\commentt{\begin{proof}}{\proof{Proof.}} Theorem~\ref{th:variance} and~\eqref{eq:TimeVarianceUpperBound} imply that \(n\var(f_{n})T\leq R(q;t,\nu^{*})\). Thus,  \(\text{Std}(f_{n})\leq\epsilon\) for \(n=\lceil R(q;t,\nu^{*})T^{-1}\epsilon^{-2}\rceil\). The expected time needed to calculate \(f_{n}\) is  \(T_{n}\), which is upper-bounded by $t_{d}+ R(q;t,\nu^{*})\epsilon^{-2}\) since \(n-1\leq  R(q;t,\nu^{*})T^{-1}\epsilon^{-2}\). Hence~\eqref{eq:corTotq}.
On the other hand, if    \(n=\lceil t_{d}T^{-1}\rceil\),
then \(T_{n}\leq2t_{d}\) since  \((n-1)T\leq t_{d}\), and~\eqref{eq:corvarfn} holds since \(nT\geq t_{d}\). \commentt{\end{proof}}{\Halmos\endproof}

Theorem~\ref{th:CLT} below establishes a central limit theorem on \(f_{n}\). It also establishes a central limit theorem on the estimate of  \(E(f(U))\) that can be obtained with a  computational budget \(c\), using the framework described by~\citet{glynn1992asymptotic}. Denote by \(\tilde N (c)\)  the number of iterations generated  by our algorithm in    $c$ units
of computation time. In other words,   \(\tilde N (c)\) is the maximum integer \(n\) such that \(f_{n}\) is calculated   within \(c\)  time (with \(f_{0}\triangleq0\)). As the time
to calculate \(f_{n}\) is random,   \(\tilde N (c)\)   is a random integer. Let \(\Rightarrow\)  denote weak convergence (see~\citep{billingsley1999convergence}).
\begin{theorem}\label{th:CLT}
As   \(n\rightarrow\infty\), 
\begin{equation}\label{eq:simpleCLT}
\ \sqrt{n}(f_{ n}-E(f(U))) \Rightarrow N(0,\sigma^{2} ), 
\end{equation}
where\begin{displaymath}
 \sigma^{2}=\sum^{d-1}_{i=0}\frac{\nu^{*}_{i}-\nu^{*}_{i+1}}{q_{i}}.
\end{displaymath}
Furthermore, as \(c\rightarrow\infty\),
\begin{equation}\label{eq:CLTComputingTime}
\sqrt{c}(f_{\tilde N(c)}-E(f(U))) \Rightarrow N(0,R(q;t, \nu^{*})). 
\end{equation}
\end{theorem} In light of above, we will use \(R(q;t,\nu^{*})\)  to  measure the performance of our algorithm. The smaller the \(C(i)\)'s and \(t_{i}\)'s, the smaller  \(R(q;t,\nu^{*})\), and the better the performance
of our algorithm.  Proposition~\ref{pr:upperBoundingCi}  below  shows that  \(C(i)\) is small if \(f\) is well-approximated  by a function of its first \(i\) arguments.      
\begin{proposition}\label{pr:upperBoundingCi}
For  \(1\leq i\leq d\), if   \(f_{i}\) is a  measurable function from \(F^{i}\)  to \(\mathbb{R}\) such that \(f_{i}(U_1,\dots,U_i)\) is square-integrable, then 
\begin{equation*}
C(i)\le \var(f(U)-f_{i}(U_1,\dots,U_i)).
\end{equation*}
\end{proposition}

 \subsection{Explicit and semi-explicit distributions}     
An optimal choice for \(q\) is a one that minimizes \(R(q;t,\nu^{*})\). A numerical algorithm that performs such minimization is presented in \S\ref{se:timeVarBound}.  
This subsection gives explicit or semi-explicit choices for \(q\),  with corresponding
upper-bounds on  \(R(q;t,\nu^{*})\).
  
Proposition~\ref{pr:powerLawCi} below gives upper bounds on \(R(q;t,\nu^{*}) \)    if    \(t_{i}=O(i)\)  and \((C(i))\)  decreases at a sufficiently high rate. It implies in particular that, if    \(t_{i}=O(i)\)   and \(C(i)=O((i+1)^{\gamma})\) with  \(\gamma<-1\), then   \(T^{\text{tot}}(q,\epsilon)=O(d+ \epsilon^{-2})\). 
\begin{proposition}\label{pr:powerLawCi}
Assume that \(d\geq2\) and there are constants \(c\) and \(c'\) and \(\gamma<0\) independent of \(d\) such that \(t_{i}\le ci\) and \(C(i)\leq c'\,(i+1)^{\gamma}\) for \(0\leq i\leq d\). Then, for \(q_{i}=(i+1)^{(\gamma-1)/2}\),    \(0\leq i\leq d-1\),  there     is a  constant \(c_{1}\) independent of \(d\) such that \begin{equation}\label{eq:UpperRPower}
R(q;t,\nu^{*}) \le \begin{cases} c_{1}, & \gamma<-1, 
\\ c_{1}\ln^2(d), &  \gamma=-1, 
\\ c_{1}\, d^{\gamma+1}, &  -1<\gamma<0. 
\end{cases} 
\end{equation}
\comment{ and 
\begin{equation}\label{eq:UpperTPower}
T^{\text{tot}}(q,\epsilon) \le \begin{cases} c_{2}(d+\epsilon^{-2}), &  \gamma<-1, 
\\ c_{2}(d+\ln^2(d)\epsilon^{-2}), &  \gamma=-1, 
\\ c_{2}(d+ d^{\gamma+1}\epsilon^{-2}), & -1<\gamma<0. 
\end{cases} 
\end{equation}
}
\end{proposition}
Below is a simple example where \(C(0)=1\) and the \(C(i)\)'s do not meet the conditions of
Proposition~\ref{pr:powerLawCi}.  
\begin{example}
\label{ex:Sum}Suppose  that \(F=\mathbb{R}\) and that  \(U_{1},\dots,U_{d}\) are square-integrable real-valued  random variables with unit variance. Assume  that  \(f(x_{1},\dots,x_{d})=d^{-1/2}(\sum^{d}_{j=1}x_{j})\) for \((x_{1},\dots,x_{d})\in\mathbb{R}^{d}\).  As  \(f\) depends equally on its arguments,  our algorithm does not improve upon the standard Monte Carlo method.    Since  \begin{displaymath}
E(f(U)|U_{i+1},\ldots,U_{d})=d^{-1/2}(E(U_{1}+\cdots+U_{i})+U_{i+1}+\cdots+U_{d}),
\end{displaymath}\(C(i)=(d-i)/d\). The conditional variance $C(i)$ decreases very slowly \(i\) since \(C(d/2)\) has the same order of magnitude as \(C(0)\). Thus the \(C(i)\)'s do not meet the conditions of
Proposition~\ref{pr:powerLawCi}.  
\end{example}

When  upper-bounds on the \(C(i)\)'s and \(t_{i}\)'s satisfying a convexity condition are known, Proposition~\ref{pr:qiChoice} below gives an explicit vector \(q\) together
with an upper bound on \(R(q;t,\nu^{*})\). 
\begin{proposition}\label{pr:qiChoice}
Assume that \(t_{i}\leq\vartheta_{i}\) for \(0\leq i\leq d\), where \(\vartheta_{0},\dots,\vartheta_{d}\) is a strictly increasing sequence with \(\vartheta_{0}=0\). Assume further that     \(\nu_{0},\dots,\nu_{d-1}\) are positive real numbers such that  \(C(i)\leq\nu_{i}\) for \(0\leq i\leq d-1\), and  that the sequence 
\begin{equation*}
\theta_{i}=\frac{\nu_{i+1}-\nu_{i}}{\vartheta_{i+1}-\vartheta_{i}},
\end{equation*}  \(0\leq i\leq d-1\), is  increasing (by convention, \(\nu_{d}=0\)). Then, for  \(q_{i}=\sqrt{\theta_{i}/\theta_{0}}\),   \(0\leq i\leq d-1\),
\begin{equation*}
R(q;t,\nu^{*})\leq 2\left(\sum_{i=0}^{d-1}\sqrt{(\nu_{i}-\nu_{i+1})(\vartheta_{i+1}-\vartheta_{i})}\right)^{2}.
\end{equation*} 
\end{proposition}
\commentt{\begin{proof}}{\proof{Proof.}} 

 We first observe that \(\theta_{i}\le\theta_{d-1}<0\) for \(0\leq i\leq d-1\). Thus \(q\)  is well-defined and belongs to \(A\).  Let \(\vartheta=(\vartheta_{0},\dots,\vartheta_{d})\). As \(t\leq\vartheta\) and  \(\nu^{*}\leq2\nu\), and since  \(R(q;.,.)\) is increasing with respect to its second and third  arguments, we have  \(R(q;t,\nu^{*})\leq R(q;\vartheta,2\nu)\).    This concludes the proof. 
\commentt{\end{proof}}{\Halmos\endproof}
 Proposition~\ref{pr:varfnSqrCi}  below  yields an upper bound on  \(\sqrt{R(q;t,\nu^{*})}\) in terms of a weighted sum of the square roots of  the \(C(i)\)'s, for a semi-explicit vector \(q\).      
      \begin{proposition}
\label{pr:varfnSqrCi}Assume  that \(C(d-1)>0\). If, for \(0\leq i\le d-1\), \begin{equation*}
q_{i}=\sqrt{\frac{t_{1}C(i)}{t_{i+1}C(0)}},
\end{equation*}then 
\begin{equation}\label{eq:qiforExplicitBoundGen}
R(q;t,\nu^{*})
\leq8\left(\sum^{d-1}_{i=0}(\sqrt{t_{i+1}}-\sqrt{t_i})\sqrt{C(i)}\right)^2.
\end{equation}
\end{proposition}
Proposition~\ref{pr:explicitDistribution} below  gives an explicit  distribution which is   optimal up to a logarithmic factor, without  requiring any prior knowledge on the \(C(i)\)'s.       
\begin{proposition}\label{pr:explicitDistribution}
For  any  \(q\in A\), 
\begin{equation}\label{eq:prWLowerB}
R(q;t,\nu^{*})\geq \sum^{d-1}_{i=0}C(i)(t_{i+1}-t_{i}).
\end{equation}  Furthermore, if \(q_{i}=t_{1}/t_{i+1}\) for \(0\leq i\leq d-1\),  then \begin{equation}\label{eq:prWUpperB}
R(q;t,\nu^{*})\leq 2(1+\ln (\frac{t_{d}}{t_{1}}))\sum^{d-1}_{i=0}C(i)(t_{i+1}-t_{i}).
\end{equation}
\end{proposition}

\subsection{A Lipschitz function example}\label{sub:Lipschitz}
Assume  that \(F=\mathbb{R}\) and that  \(U_{1},\dots,U_{d}\) are square-integrable real-valued  random variables, with \(\sigma_{1}\geq \cdots\geq\sigma_{d}>0\),
where \(\sigma_{i}\) is the standard deviation of \(U_{i}\). Assume also that  \(f(x_{1},\dots,x_{d})=g(\sum^{d}_{j=1}x_{j})\) for \((x_{1},\dots,x_{d})\in\mathbb{R}^{d}\), where \(g\) is a real-valued \(1\)-Lipschitz function on \(\mathbb{R}\) that can be calculated in constant time. For instance,  \(f(x_{1},\dots,x_{d})=\max(\sum^{d}_{j=1}x_{j}-K,0)\), where \(K\) is a constant, satisfies this condition. Assume further that  each \(U_{i}\)   can  be simulated in constant time.  For \(1\leq k\leq n\), let \(S_{k}\) be the sum of all components of \(V^{(k)}\). Thus \(S_{k+1}\) can be calculated recursively in \(O(N_{k})\) time by adding to \(S_{k}\)  the first  \(N_{k}\) components of \(V^{(k+1)}\) and subtracting  the first  \(N_{k}\) components of \(V^{(k)}\). Hence  \(t_{i}\le ci\), for some constant \(c\). 

In order to bound the \(C(i)\)'s, we show that \(f(U)\) can be approximated by \(f_{i}(U_{1},\dots,U_{i})\), where  \(f_{i}(x_{1},\dots,x_{i})=g(\sum^{i}_{j=1}x_{j}+\sum^{d}_{j=i+1}E(U_{j}))\)  for \((x_{1},\dots,x_{i})\in\mathbb{R}^{i}\). Let \(||Z||=\sqrt{E(Z^{2})}\) for a real-valued random variable \(Z\). By Proposition~\ref{pr:upperBoundingCi}, \begin{eqnarray}\label{eq:ciLipschitz}
C(i)&\le&||f(U)-f_{i}(U_{1},\dots,U_{i})||^{2}\nonumber
\\&\leq& ||\sum^{d}_{j=i+1}(U_{j}-E(U_{j}))||^{2}\nonumber
\\&=& \var(\sum^{d}_{j=i+1}U_{j})\nonumber
\\&=&\sum^{d}_{j=i+1}\sigma_{j}^{2}.
\end{eqnarray} 
The second equation follows from the assumption that \(g\) is  \(1\)-Lipschitz. By applying Proposition~\ref{pr:qiChoice}, with \(\vartheta_{i}=ci\) and \(\nu_{i}=\sum^{d}_{j=i+1}\sigma_{j}^{2}\), and setting      \(q_{i}=\sigma_{i+1}/\sigma_{1}\),  \(0\leq i\leq d-1\), we infer that
\begin{equation*}
R(q;t,\nu^{*})\le2c(\sum^{d}_{i=1}\sigma_{i})^{2}.
\end{equation*}  Thus, if  \(\sigma_{i}=O(i^{\gamma}\)), with \(\gamma<-1\), then \(R(q;t,\nu^{*})=O(1)\) and \(T^{\text{tot}}(q,\epsilon)=O(d+ \epsilon^{-2})\).
Also, by~\eqref{eq:ciLipschitz} and a standard calculation, \(C(i)=O((i+1)^{2\gamma+1})\)  for \(0\leq i\leq d\). As \(2\gamma+1<-1\), Proposition~\ref{pr:powerLawCi} is also applicable in this case. \comment{
\subsubsection{A polynomial counter-example}Assume that \begin{displaymath}
f(x_{1},\dots,x_{d})=\sum^{d}_{j=1}{x_{j}x_{1}}^{j}.
\end{displaymath} In general, in a standard implementation,  \(t_{1}=\Omega(d)\),   since calculating \(f(U)\) after redrawing \(U_{1}\)  typically takes  \(\Theta(d)\) steps. Thus, \(T=\Omega(d)\), for any \(q\in A\). On the other hand, by~\eqref{eq:numonotone}, the RHS of~\eqref{eq:lemmaVariance} is lower-bounded by \(\var(f(U))\). Thus, \(R(q;t,\nu^{*})=\Omega(d\var(f(U))\), which corresponds to to the time-variance product of the standard Monte Carlo algorithm. Therefore, a  standard implementation of the Randomized Dimension Reduction algorithm does not outperform the standard Monte Carlo algorithm in this example.  }   
\section{The optimal distribution}\label{se:timeVarBound}
We now seek to calculate a vector \(q\) that minimizes \(R(q;t,\nu^{*})\). Given a vector  \(\nu\)   in \(\mathbb{R}^{d}\times\{0\}\) whose first \(d\) components are positive,  Theorem~\ref{th:OptTimeVariance} below gives a geometric algorithm that finds in \(O(d)\) time a vector \(q^{*}\) that minimizes \(R(q;t,\nu)\) under the constraint that \(q\in A\). Note that the vector \(q^{*}\) depends on \(\nu\).
   In \citep[Section 3]{GlynnRhee2015unbiased}, a dynamic programming algorithm that  calculates such a vector \(q^{*}\) in  \(O(d^{3})\) time   has been described. 

\begin{figure}
\caption{Lower hull}
\begin{center}
\begin{tikzpicture}
\begin{axis}[
    title={},
    xlabel={$i$},
    xmin=0, xmax=6,
    ymin=0, ymax=30,
    xtick={0,1,2,3,4,5,6},
    ytick={0,5,10,15,20,25},
    legend pos=north east,
    ymajorgrids=true,
    grid style=dashed,
]
 
\addplot[
    color=blue,
    mark=square,
    ]
    coordinates {
    (0,20)(1,21)(2,13)(3,8)(4,7)(5,2)(6,0)
    };
 \addplot[
    color=red,
    mark=star,
    ]
    coordinates {
    (0,20)(1,16)(2,12)(3,8)(4,5)(5,2)(6,0)
    };
    \legend{$\nu$,$\nu'$}

\end{axis}
\end{tikzpicture}
\end{center}
\label{fig:hull}
\end{figure}
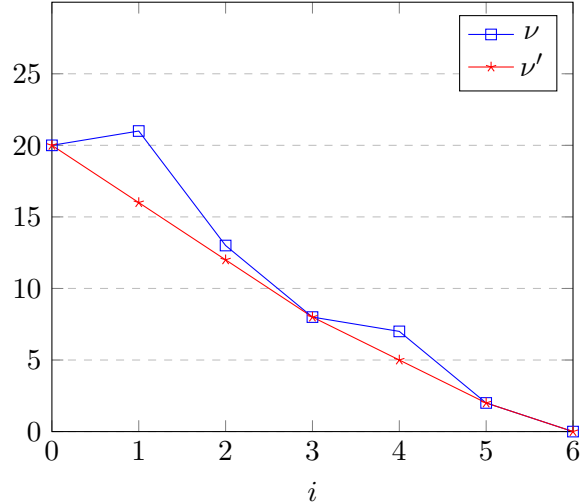

Let \(\nu'=(\nu'_{0},\dots,\nu'_{d})\in\mathbb{R}^{d+1}\) be such that the set  \(\{(t_{i},\nu'_{i}):0\leq i\leq d\}\)
forms the lower hull of the set  \(\{(t_{i},\nu_{i}):0\leq i\leq d\}\). In other words,  \(\nu'\) is the supremum  of all sequences in \(\mathbb{R}^{d+1}\) such that \(\nu'\le\nu\) and  the sequence   \((\theta_{i})\) is  increasing, where  
\begin{equation}\label{eq:deftheta}
\theta_{i}=\frac{\nu'_{i+1}-\nu'_{i}}{t_{i+1}-t_{i}},
\end{equation}  \(0\leq i\leq d-1\).  For instance, if \(d=6\), with \(t_{i}=i\) and \(\nu=(20,21,13,8,7,2,0)\), then  \(\nu'=(20,16,12,8,5,2,0)\), as illustrated in Fig.~\ref{fig:hull}. \S\ref{sub:hull} shows how to calculate \(\nu'\) in \(O(d)\) time. 

\begin{theorem}\label{th:OptTimeVariance}
Let   \(\nu\)    be a vector in \(\mathbb{R}^{d}\times\{0\}\) whose first \(d\) components are positive. For \(0\leq i\leq d-1\),  set  \(q^{*}_{i}=\sqrt{\theta_{i}/\theta_{0}}\), where \(\theta_{i}\) is given by~\eqref{eq:deftheta}, and let \(q^{*}=(q^{*}_{0},\dots,q^{*}_{d-1})\). Then \(q^{*}=\arg\min_{q\in A}R(q;t,\nu)\), and 
\begin{equation}\label{eq:optTimeVariance}
 R(q^{*};t,\nu)=\bigg(\sum_{i=0}^{d-1}\sqrt{(\nu'_{i}-\nu'_{i+1})(t_{i+1}-t_{i})}\bigg
)^2.\end{equation}  
\end{theorem}
\subsection{Lower hull calculation}\label{sub:hull}
   Given \(\nu\),   the following algorithm, due to \citep{Andrew79},  first generates recursively a subset \(B(j)\) of  \(\{1,\dots,d\}\), \(2\leq j\leq d\), then calculates \(\nu'\) via \(B(d)\). The algorithm runs in \(O(d)\) time.\begin{enumerate}
\item 
Set \(B(2)=\{1,2\}\).
\item For \(j=3\) to \(d\), denote by \(i_{1}<\cdots<i_{m}\)  the elements of \(B(j-1)\).
Let \(k\) be the largest element of   \(\{2,\dots,m\}\) such that \((t_{i_k},\nu_{i_{k}})\) lies below the segment
\([(t_{i_{k-1}},\nu_{i_{k-1}}),(t_{j},\nu_{j})]\), if such \(k\) exists, otherwise let \(k=1\). Set  \(B(j)=\{i_{1},\dots,i_{k},j\}\). 
\item
For \(i=1\) to \(d\), let \(i'\) and \(i''\) be two elements of \(B(d)\) with \(i'\leq i\leq i''\). Set \(\nu'_{i}\) so that \((t_{i},\nu'_{i})\) lies on the segment \([(t_{i'},\nu_{i'}),(t_{i''},\nu_{i''})]\). \end{enumerate}
\subsection{Estimating the \(C(i)\)'s}
The calculation of a vector \(q^{*}\in A\)  that minimizes  \(R(q;t,\nu^{*})\) requires the knowledge of the \(C(i)'s\). Proposition~\ref{pr:CIVarReduc} below can be used
to estimate  \(C(i)\)    via Monte Carlo simulation. Assuming that \(f(U)\) can be approximated by a function of its first \(i\) arguments, we expect that both components of the product in the RHS of~\eqref{eq:controlVariate}  to be small, on average. Thus, \eqref{eq:controlVariate} can be considered as a ``control variate'' version of~\eqref{eq:ViCov}, and should yield a more accurate estimate of \(C(i)\) via Monte Carlo simulation for large values of \(i\).
\begin{proposition}\label{pr:CIVarReduc}
Assume that \(U'_{1},\ldots ,U'_{d}\), and  \(U''_{i+1},\ldots,U''_{d}\), are  random variables such that \(U'_{j}\,{\buildrel d \over =}\,U_{j}\)  for \(1\leq j\leq d\), and  \(U^{''}_{j}\,{\buildrel d \over =}\,U_{j}\)  for \(i+1\leq j\leq d\), and   \(U'_{1},\ldots ,U'_{d}\),  \(U \),  \(U^{''}_{i+1},\ldots,U^{''}_{d}\) are independent. Then   
\begin{multline}\label{eq:controlVariate}
C(i)=E((f(U)-f(U_{1},\ldots ,U_{i},U'_{i+1},\ldots,U'_{d})) \\
 (f(U'_{1},\ldots ,U'_{i},U_{i+1},\ldots,U_{d})-f(U'_{1},\ldots ,U'_{i},U^{''}_{i+1},\ldots,U^{''}_{d})).
\end{multline}
\end{proposition}
\subsection{Numerical algorithm}
\label{sub:numerical}Building upon the previously discussed elements, the algorithm that we have used for our numerical experiments is as follows. It  constructs a vector \((\nu_{0},\dots,\nu_{d})\) and uses it as a proxy for \(\nu^{*}\).   
\begin{enumerate}
\item 
For \(i=0\) to \(d-1\), if  \(i+1\) is a power of \(2\), estimate  \(C(i)\)  by Monte Carlo simulation with \(1000\) samples via Proposition~\ref{pr:CIVarReduc}. \item Set   \(\nu_{0}=C_{}(0)\), and    \(\nu_{d}=0\).  For \(1\leq i\leq d-1\), let \(\nu_{i}=2C(j)\),  where \(j\) is the largest index in \([0,i]\) such that \(j+1\) is a power of \(2\). 
\item For \(i=d-1\) down to \(1\), set \(\nu_{i}\leftarrow\max(\nu_{i},\nu_{i+1})\). Set  \(\nu_{0}\leftarrow\max(\nu_{0},\nu_{1}/2)\).
\item Let \((\nu'_{0},\dots,\nu'_{d})\in\mathbb{R}^{d+1}\) be such that the set  \(\{(t_{i},\nu'_{i}):0\leq i\leq d\}\)
forms the lower hull of the set  \(\{(t_{i},\nu_{i}):0\leq i\leq d\}\). For \(0\leq i\leq d-1\),  set  \(q_{i}=\sqrt{\theta_{i}/\theta_{0}}\), where \(\theta_{i}\) is given by~\eqref{eq:deftheta}. \item  Calculate \(T\) via~\eqref{eq:TimeVarianceUpperBound}.
For \(0\leq i\leq d-1\), set \begin{displaymath}
q_{i}\leftarrow\min(1,\max(q_{i},\frac{T}{t_{i+1}\ln(t_{d}/t_{1})})).
\end{displaymath}\item Run Steps \(1\) through \(3\) of the generic randomized dimension reduction algorithm of \S\ref{sub:generalAlgorithm} using \(q\).
\end{enumerate}
The purpose of Steps~\(3\) and~\(5\) is to reduce the impact on  \(q\) of statistical errors that arise in Step~1.
Because of statistical errors,  Step~1 may underestimate or overestimate the \(C(i)'s\). Step~\(3\) guarantees that the \(\nu_{i}\)'s are non-negative, so that the \(q_{i}\)'s can be calculated in Step~4. Step~5 yields a cap on the RHS of~\eqref{eq:lemmaVariance} by ensuring that the \(q_{i}\)'s are not too small.   Using ~\eqref{eq:TimeVarianceUpperBound} and the proof of Proposition~\ref{pr:explicitDistribution}, and assuming that \(t_{d}\geq2t_{1}\), it can be shown that Step~\(5\)  increases \(T\)  by at most a constant multiplicative factor. 
An alternative way to implement our algorithm
is to skip Steps~1 through~5 and  run the generic algorithm with  \(q_{i}=t_{1}/t_{i+1}\) for \(0\leq i \leq d-1\).   By Proposition~\ref{pr:explicitDistribution}, the resulting
vector \(q\) is optimal up to a logarithmic factor. 
\section{Applications to Markov chains}
\label{se:examples}
In queueing systems, the performance metrics at a specific time instant
are heavily dependent on the last busy cycle, i.e., the events that occurred after the
queue was empty for the last time. Thus, the performance metrics "depend
a lot more on the last random variables driving the system than on the initial ones.
Nevertheless, we can apply our algorithm to queueing systems by  using a time-reversal transformation inspired from~\citep{glynn2014exact}. More generally, using such a time-reversal transformation, this section shows that our algorithm can efficiently estimate the expected value of a function of the  state of a Markov chain at time-step \(d\), for a class of Markov chains driven by independent random variables.

 Let \((X_m)\), \(0\leq m\leq d\), be a Markov chain with state-space
\(F'\) and deterministic initial value \(X_{0}\).
Assume  that  there are independent  random variables \(Y_{i}\),  \(0\leq i\leq d-1\), that take values in  \(F\), and measurable
functions \(g_{i}\) from \(F'\times F\) to \(F'\) such
that \(X_{i+1}=g_{i}(X_{i},Y_{i})\) for \(0\leq i\leq d-1\).     We want to estimate  \(E(g(X_{d}))\)
for a given positive integer \(d\), where \(g\) is a deterministic real-valued
measurable function on \(F'\) such that \(g(X_{d})\) is square-integrable.
  For \(1\leq i\leq d\), set \(U_{i}=Y_{d-i}\). It can be shown by  induction that  \(X_{d}=G_{i}(U_{1},\dots,U_{i},X_{d-i})\), where \(G_{i}\), \(0\leq i\leq d\), is a measurable function from \(F^{i}\times
F'\) to \(F'\), and so there is a real-valued measurable function \(f\) on \(F^d\)
with \(g(X_{d})=f(U_{1},\dots,U_{d})\). We can thus  use our  randomized dimension reduction algorithm to estimate \(E(g(X_{d}))\). Recall that, in iteration \(k+1\) in Step~2 of the generic algorithm  of  \S\ref{sub:generalAlgorithm}, conditioning on \(N_{k}=i\), the first \(i\) arguments of \(f\) are re-drawn, and the remaining arguments are unchanged. This is equivalent to re-drawing the last \(i\) random variables driving the Markov chain, and keeping the first \(d-i\) variables unchanged. In light of above, the generic randomized dimension reduction  algorithm for Markov chains estimation takes as parameter a vector \(q\in A\)  and consists of
the following steps:
\begin{enumerate}
\item 
First iteration.
Generate recursively  \(X_{0},\dots,X_{d}\).
Calculate \(g(X_{d})\).
\item Loop. In iteration \(k+1\), where \(1\leq k\leq n-1\), keep  \(X_{0},\dots,X_{d-N_{k}}\)
unchanged, and calculate recursively \(X_{d-N_{k}+1},\dots,X_{d}\) by re-drawing  \(Y_{d-N_{k}},\dots,Y_{d-1}\), where \(N_{k}\) is a random integer in \([1,d]\) such that \(\mathbb{P}(N_{k}>i)=q_{i}\). Calculate \(g(X_{d})\).
\item Output the average  of \(g\) over the \(n\) copies of \(X_{d}\) generated in the first two steps.\end{enumerate}

We assume that \(g\) and the \(g_{i}\)'s
can be calculated in constant time, and that the expected time needed to simulate each \(Y_{i}\) is upper-bounded by a constant independent of \(d\). Thus, given \(N_{k}\),   the expected time needed to perform iteration \(k+1\)   is \(O(N_{k})\). Hence \(t_{i}\le ci\), for some constant \(c\) independent of \(d\). Proposition~\ref{pr:MarkovRevuz} below shows that, roughly speaking, \(C(i)\) is small if  \(X_{d-i}\) and  \(X_{d}\) are ``almost'' independent. 
\begin{proposition}\label{pr:MarkovRevuz}
For \(0\leq i\leq d\), we have \(C(i)=\var (E(g(X_{d})|X_{d-i})).
\)\end{proposition}
By Proposition~\ref{pr:powerLawCi}, if there are constants \(c'>0\) and \(\gamma<-1\) independent of \(d\) such that \(C(i)\leq c'(i+1)^{\gamma}\) for \(0\leq i\leq d-1\), then  \(R(q;t,\nu^{*})\) is upper-bounded by a constant independent of \(d\), where \(q_{i}=(i+1)^{(\gamma-1)/2}\) for \(0\leq i\leq d-1\).
The analysis in~\citep[Section IV.1a]{asmussenGlynn2007}, combined with Proposition~\ref{pr:MarkovRevuz}, suggests that   \(C(i)\)   decreases exponentially with \(i\)  for a variety of Markov chains.  

For \(x\in\mathbb{F'}\), and \(0\leq i\leq d\), let 
\begin{equation*}
X_{i,x}=G_{i}(U_{1},\dots,U_{i},x).
\end{equation*}In other words,  \(X_{i,x}\) is the state of the chain at time-step \(d\)  if  the chain is at state \(x\) at time-step \(d-i\).  Intuitively, we  expect \(X_{i,x}\) to be close to \(X_{d}\) for large \(i\) if \(X_{d}\) depends mainly on the last \(Y_{j}\)'s. By Proposition~\ref{pr:upperBoundingCi},  if  \(g(X_{i,x})\) is square-integrable,
 \begin{equation}\label{eq:UpperCiMarkovL2}
C(i)\leq||g(X_{d})-g(X_{i,x})||^{2}.
\end{equation}
In the following examples,  we prove   that under certain conditions,   \(R(q;t,\nu^{*})\) is upper-bounded by a constant independent of \(d\)  for an explicit vector \(q\in A\), and so \(T^{\text{tot}}(q,\epsilon)= O(d+ \epsilon^{-2})\). 
\subsection{GARCH volatility model}\label{sub:GARCH}
In the  GARCH(1,1) volatility model (see~\citep[Ch. 23]{Hull14}), the  variance  \(X_{i}\)  of an index return between day \(i\) and day \(i+1\), as estimated at the end of day  \(i\), satisfies the following recursion:
\begin{equation*}
X_{i+1}= w +\alpha X_{i}Y_{i}^{2}+\beta X_{i},
\end{equation*}
 \(i\geq0\), where \( w \), \(\alpha\) and \(\beta\) are positive constants with \(\alpha+\beta<1\), and \(Y_{i}\),  \(i\geq0\), are independent standard Gaussian random variables. The variable \(Y_{i}\) is known at the end of day \(i+1\). At the end of day \(0\), given  \(X_{0}\geq0\), a positive integer \(d\) and a real number \(z\), we want to estimate \(\mathbb{P}(X_{d}> z)\). In this example, \(F=F'=\mathbb{R}\), and \(g_{i}(x,y)= w +\alpha xy^{2}+\beta x\), with   \(g(u)={\bf1}\{u> z\}\) for \(u\in\mathbb{R}\). Proposition~\ref{pr:GARCH} below shows that \(C(i)\) decreases exponentially with \(i\). 

\begin{proposition}\label{pr:GARCH}There is a constant \(\kappa\) independent of \(d\) such that
\(C(i)\leq\kappa(\alpha+\beta)^{i/2}\) for \(0\leq i\leq d-1\).
\end{proposition}
By applying Proposition~\ref{pr:qiChoice} with  \(\vartheta_{i}=ci\) and  \(\nu_i=\kappa(\alpha+\beta)^{i/2}\),  and setting \(q_{i}=(\alpha+\beta)^{i/4}\) for \(0\leq i\leq d-1\), we infer that \(R(q;t,\nu^{*})\) is  upper-bounded by  a constant independent of \(d\).
\subsection{\(G_{t}/D/1 \) queue}\label{sub:singleQueue}
Consider a queue  where   customers arrive at time-step
\(i\), \(1\leq i \leq d\), and are served by a single server in order of arrival. 
Service times are all equal to \(1\).  Assume the system starts empty at time-step
\(0\), and that   \(A_{i}\)    customers arrive
at time-step \(i\), \(0\leq i\leq d\), where  \(A_{0}=0\) and the \(A_{i}\)'s are independent square-integrable random variables. Let \(X_{i}\) be the number of customers waiting in the queue  at time-step \(i\).  Then  \(X_{0}=0\)  and \((X_{i})\) satisfies the Lindley equation \begin{equation*}
X_{i+1}=(X_{i}+Y_{i})^{+},
\end{equation*}for \(0\leq i\leq d-1\), with \(Y_{i}=A_{i+1}-1\).   We want to estimate \(E(X_{d})\). In this example, \(g\) is the identity function, \(F=F'=\mathbb{R}\), and \(g_{i}(x,y)=(x+y)^{+}\). Proposition~\ref{pr:GD1} below shows that \(C(i)\) decreases exponentially with \(i\) under certain conditions on the  service times. 
\begin{proposition}\label{pr:GD1}If  there are constants \(\gamma>0\) and \(\kappa<1\) independent of \(d\) such that
\begin{equation}\label{eq:momentCondGD1}
E(e^{\gamma Y_{i}})\leq\kappa
\end{equation}  for \(0\leq i\leq d-1\), then  \(C(i)\leq \gamma'\kappa^{i}\) for \(0\leq i\leq d-1\), where \(\gamma'\) is a constant independent of \(d\). 
\end{proposition}  By applying Proposition~\ref{pr:qiChoice} with \(\vartheta_{i}=ci\) and \(\nu_i=\gamma'\kappa^{i}\),   and setting  \(q_{i}=\kappa^{i/2}\) for \(0\leq i\leq d-1\), we conclude that, under the assumption of Proposition~\ref{pr:GD1}, \(R(q;t,\nu^{*})\) is  upper-bounded by  a constant independent of \(d\). The assumption in Proposition~\ref{pr:GD1} can be justified as follows. Given \(i\in[0,d-1]\), if   \(E(A_{i})<1\) and the function \(h(\gamma)= E(e^{\gamma Y_{i}})\) is bounded on a neighborhood of \(0\), then \(h'(0)=E(Y_{i})<0\). As   \(h(0)=1\), there is  \(\gamma>0\)   such that  \(h(\gamma)<1\), and~\eqref{eq:momentCondGD1}  holds for  \(\kappa=h(\gamma)\).    The assumption in Proposition~\ref{pr:GD1} says that  \(\gamma\) and \(\kappa\) can be chosen independently of \(i\) and of \(d\). 
\subsection{\(M_t/GI/1\) queue }\label{sub:OtherMC}
Consider  a  \(M_t/GI/1\) queue
 where   customers are served by a single server in order of arrival. 
We assume that   customers arrive
according to a Poisson process with positive and continuous time-varying rate \(\lambda_{s}\le\lambda^{*}\), where \(\lambda^{*}\) is a fixed positive real number.  The service times are assumed to be  i.i.d. and  independent of the arrival times. Assume that the system starts empty at time \(0\). For simplicity, we assume that the number of customers that arrive in any bounded time interval is finite (rather than finite with probability \(1\)). Consider a  customer present in the system at a given time \(s\). If the customer   has been served for a period of length \(\tau\),  its remaining service time is equal to its service time minus \(\tau\), and if the customer   is in the queue,   its remaining service time is equal to its service time. The residual work \(W_{s}\) at time \(s\) is defined as the sum of remaining service times of customers present in the system at   \(s\).   We want to estimate the expectation of \(W_{\theta}\), where  \(\theta\) is a fixed  time.   Let   \(d=\lceil\lambda^{^{*}} \theta\rceil\),  and assume that \(d\geq2\). For \(0\leq i\leq d\), let \(X_{i}=W_{i\theta/d}\) be the  residual work at time \(i\theta/d\). For \(0\leq i\leq d-1\), let \(Y_{i}\) be the vector that consists of  arrival and  service times of customers  that arrive during the interval \((i\theta/d,(i+1)\theta/d]\).   
In this example, \(g\) is the identity function, \(F\) is  equal to the set of real-valued sequences with finite support, and   \(F'=\mathbb{R}\).   Let \(0\leq s<s'\). If no costumers arrive in \((s,s']\)  then \(W_{s'}=(W_{s}-s'+s)^{+}\). On the other hand, if no costumers arrive in \((s,s')\)  and a customer with service time \(S\) arrives at \(s'\), then \(W_{s'}=S+(W_{s}-s'+s)^{+}\).  Thus, given the set of arrival and service times of customers that arrive in \((s,s']\), we can calculate iteratively \(W_{s'}\)   from \(W_{s}\). This implies that  \(X_{i+1}\) is a deterministic measurable function of \(X_{i}\) and \(Y_{i}\), for \(0\leq i\leq d-1\). Proposition~\ref{pr:MG1} below shows that \(C(i)\) decreases exponentially with \(i\) under certain conditions on the arrival and service times.
\begin{proposition}\label{pr:MG1}
For \(0\leq s\leq\theta\), let  \(Z_{\theta}(s)\) be the cumulative service time of costumers that arrive in \([s,\theta]\). Assume there are constants \(\gamma>0\) and \(\kappa<1\) independent of \(d\) such that, for \(0\leq s\leq s'\leq\theta\) and \(s'-s\leq1/\lambda^{*}\),
\begin{equation}\label{eq:prMG1}
E(e^{\gamma (Z_{\theta}(s')-Z_{\theta}(s)-1/\lambda^{*})})\leq\kappa.
\end{equation} Then \(C(i)\leq \gamma'\kappa^{i/2}\) for \(0\leq i\leq d-1\), where \(\gamma'\) is a constant independent of \(d\). \end{proposition}    By applying Proposition~\ref{pr:qiChoice} with \(\vartheta_{i}=ci\) and \(\nu_i=\gamma'\kappa^{i/2}\), and setting \(q_{i}=\kappa^{i/4}\) for \(0\leq i\leq d-1\), we conclude that, under the assumption of Proposition~\ref{pr:MG1}, \(R(q;t,\nu^{*})\) is  upper-bounded by  a constant independent of \(d\). 
The assumption in Proposition~\ref{pr:MG1} can be justified as follows. For \(0\leq s\leq s'\leq\theta\) and \(s'-s\leq1/\lambda^{*}\),
  the cumulative service times of customers that arrive in \([s,s')\) is  \(Z_{\theta}(s')-Z_{\theta}(s)\). If \(E(Z_{\theta}(s')-Z_{\theta}(s))<s'-s\)   and   \(h(\gamma)= E(e^{\gamma (Z_{\theta}(s')-Z_{\theta}(s)-1/\lambda^{*})})\) is bounded on a neighborhood of \(0\), then \(h'(0)<0\). Thus  \(h(\gamma)<1\) for some \(\gamma>0\) and~\eqref{eq:prMG1}  holds for  \(\kappa=h(\gamma)\).    The assumption in Proposition~\ref{pr:MG1} says that  \(\gamma\) and \(\kappa\) can be chosen independently of \(d\), \(s\) and \(s'\). 

 \comment{\subsection{Time-dependent contracting Markov chains}
In~\citep{glynn2014exact}, an algorithm that gives unbiased estimators  for equilibrium expectations associated with real-valued functionals defined on time-homogeneous contracting Markov Chains is described. This subsection defines the related notion of time-dependent contracting Markov Chain and shows that our algorithm ...
 }
\section{Deterministic dimension reduction}
\label{se:deterministic}
This section studies a deterministic dimension reduction algorithm that performs the same steps as the generic randomized dimension reduction algorithm of \S\ref{sub:generalAlgorithm}, but uses a deterministic integral sequence \((N_{k})\),   \(k\geq  1\),   taking values in \([1,d]\), to estimate \(E(f(U))\). As for the randomized algorithm, denote by   \(f_{n}  \) the output of the  deterministic dimension reduction  algorithm, and by \(T_{n}\) its expected running time, where  \(n\) is the number of iterations, including the first one. The sequence  \((N_{k})\),   \(k\geq  1\), may depend on \(n\).  As the algorithm generates \(n\) copies of \(U\), 
the random variable  \(f_{n} \) is an unbiased estimator of \(E(f(U))\). 

 Assume that \(C(d-1)>0\) and let \(\hat q=\arg\min_{q\in A}R(q;t,C)\), where \(C\) denotes the vector \((C(0),\dots,C(d))\). The existence of \(\hat q\) follows from Theorem~\ref{th:OptTimeVariance}. Define the integers \(\mu_{0},\dots,\mu_{d-1}\) recursively as follows. Let \(\mu_{0}=1\) and, for \(1\leq i\leq d-1\),
let  \(\mu_{i}\) be the largest multiple of \(\mu_{i-1}\) in the interval \([0,1/\hat q_{i}] \), i.e.\begin{displaymath}
\mu_{i}=\mu_{i-1}\lfloor\frac{1}{\mu_{i-1}\hat q_{i}}\rfloor.
\end{displaymath} It can be shown by induction that \(\mu_{i}\) is well-defined and positive. Let \(\bar q\)   be the vector in \(A\) defined by \(\bar q_{i}=1/\mu_{i}\),  for \(0\le i\le d-1\). As \(\lfloor x\rfloor\le x<2\lfloor x\rfloor\) for \(x\geq1\), we have \(\mu_{i}\le1/\hat q_{i}<2\mu_{i}\). It follows that, for \(0\le i\le d-1\),  \begin{equation}\label{eq:harBarq}
\hat q_{i}\le \bar q_{i}< 2\hat q_{i}.
\end{equation} Define the sequence  \((\bar N_{k})\),   \(k\geq 1\),    as follows: \begin{equation*}
\bar N_{k}=\max\{i\in[1,d]:k\text{ is a multiple of  }\mu_{i-1}\}.
\end{equation*}  As \(\mu_{0}=1\), such  \(i\) always exists. For \(0\leq i\leq d-1\) and  \(k\geq1\), if \(\bar N_{k}=j\) with \(j>i\),  then  \(k\) is a multiple of \(\mu_{j-1}\), and so    \(k\) is a multiple of \( \mu_{i}\), since  \(\mu_{j-1}/\mu_{i}\) is an integer. Conversely, if     \(k\) is a multiple of \( \mu_{i}\) then, by construction, \(\bar N_{k}>i \). Hence
\begin{equation}\label{eq:modulo}
\bar N_{k}>i \Leftrightarrow k\equiv 0\pmod {\mu_{i}}.
\end{equation}   Given \(i\in[0,d-1]\), the inequality \(\bar N_{k}>i\) occurs once as \(k\) ranges in a set of \(\mu_{i}\)   consecutive positive integers. The sequence   \((\bar N_{k})\) can thus be considered as a deterministic counterpart to the random sequence   \((N_{k})\) generated by the randomized dimension reduction algorithm when \(q=\bar q\). 

Theorem~\ref{th:DDR} 
below gives a lower bound on the performance of the deterministic dimension reduction algorithm for any sequence \((N_{k})\), \(k\geq 1\), and analyses the algorithm when \(N_{k}=\bar N_{k}\) for \( k\geq 1\).   

 \begin{theorem}\label{th:DDR}For \(n\geq1\) and any deterministic sequence \((N_{k})\), \(k\geq 1\), \begin{equation}\label{eq:TvarLowerBoundDet}
T_{n}\var(f_{n})\geq R(\hat q;t,C).
\end{equation}If    \(N_{k}=\bar N_{k}\) for \(k\geq 1\) then, for  \(n\geq1\), \begin{equation}\label{eq:TimeDeterminSpec}
T_{n}=t_{d}+\sum^{d-1}_{i=0}\lfloor(n-1)\bar q_{i}\rfloor(t_{i+1}-t_{i}),
\end{equation}
and 
\begin{equation}\label{eq:VarVander}
n\var(f_{n})\le \sum^{d-1}_{i=0}\frac{C(i)-C(i+1)}{\bar q_{i}}.
\end{equation}
Furthermore, the LHS of~\eqref{eq:VarVander} converges to its RHS as \(n\) goes to infinity.  \end{theorem}
Using again the framework of~\cite{glynn1992asymptotic}, we measure the performance of an estimator via the work-normalized
variance, i.e. the product of the variance
and expected running time. If \(N_{k}=\bar N_{k}\) for \(k\geq1\) then, by Theorem~\ref{th:DDR}, \begin{equation*}
T_{n}\var(f_{n})\rightarrow R(\bar q;t,C)
\end{equation*}
as \(n\) goes to infinity. Furthermore, it follows from~\eqref{eq:RDef} and~\eqref{eq:harBarq} that \(R(\bar q;t,C)\leq2R(\hat q;t,C)\).
Thus, up to a factor of \(2\), the sequence \((\bar N_{k})\) asymptotically  minimizes the work-normalized variance of the deterministic dimension reduction algorithm. Moreover, by definition of \(\hat q\), and since \(C\leq\nu^{*}\), 
\comment{
\begin{eqnarray*}R(\hat q;t,C)
&\leq& R(q^{*};t,C)\\
&\leq& R(q^{*};t,\nu^{*}).
\end{eqnarray*} 
}
\begin{equation*}R(\hat q;t,C)
\leq R(q^{*};t,C)\\
\leq R(q^{*};t,\nu^{*}),
\end{equation*} 
where  \(q^{*}=\arg\min_{q\in A}R(q;t,\nu^{*})\). Hence \(R(\bar q;t,C)\leq2R(q^{*};t,\nu^{*})\).
Similarly, as \(\nu^{*}\leq2C\),\begin{displaymath}
R(q^{*};t,\nu^{*})\leq R(\bar q;t,\nu^{*})\le 2R(\bar q;t,C).
\end{displaymath} Thus, the asymptotic work-normalized variances of the randomized dimension reduction algorithm, with \(q=q^{*}\),  and of the deterministic dimension reduction algorithm, with \(N_{k}=\bar N_{k}\) for  \(k\geq 1\),   are within a factor of \(2\) from each other. 

Proposition~\ref{pr:refereeDeterministic} below shows that if, after generating the first copy of \(U\), we generate  the next \(n(q_{0}-q_{1})\) samples by only changing the first component of the $U$ in the previous
iteration, and the next  \(n(q_{1}-q_{2})\) samples by only changing the first two components of the $U$ in the previous
iteration, and so on, the resulting algorithm is asymptotically less efficient than standard Monte Carlo.    
\begin{proposition}\label{pr:refereeDeterministic} Assume that \(C(d-1)>0\). Let \(q\in A\) with \(q_{d-1}<1\). If  \(N_{k}=i\) for   \(1\leq i\leq d\) and integer \(k\in(n(1-q_{i-1}),n(1-q_{i})]\), then \(n\var(f_{n})\rightarrow\infty \) as \(n\) goes to infinity.
\end{proposition}
\section{Comparison with a class of  multilevel algorithms}\label{se:ML}
We  compare our method to a class of MLMC algorithms, adapted from~\citep{Giles2008},  that   efficiently  estimate \(E(f(U))\) under the assumption that \(f\) is approximated,  in the \(L^{2}\) sense, by functions of its first arguments. Under conditions described in \S\ref{sub:MLMC}, we prove that, up to a constant, the  randomized dimension reduction algorithm is at least as  efficient as this class of   MLMC  algorithms. It should be stressed, however, that there may exist other MLMC algorithms that estimate \(E(f(U))\) more efficiently than  the class of MLMC algorithms described below. 
\subsection{The  MLMC algorithms description and analysis}\label{sub:MLMC}

 Let \(L\) be a positive integer and let \((m_{l})\), \(0\leq l\leq L\), be a strictly  increasing integral sequence, with \(m_{0}=0\) and \(m_{L}=d\). For \(1\le l\leq L\), let  \(\phi_{l}\) be a square-integrable random variable equal to a deterministic measurable function of \(U_{1},\dots,U_{m_{l}}\), with  \(\phi_{L}=f(U)\). The  \(\phi _{l}\)'s are chosen so that,  as \(l\) increases, \(\phi_{l}\) gets closer to \(f(U)\), in the \(L^{2}\) sense.    For instance, \(L\)  can be proportional to \(\ln(d)\),  the \(m_{l}\)'s can increase exponentially with \(l\), and     \(\phi_{l}\) could equal  \(f(U_{1},\dots,U_{m_{l}},\underbrace{x,\dots,x}_{d-m_l})\), for some  \(x\in F\). For   \(1\leq l\leq L\),   let  \(\hat\phi_{l}\) be the average of \(n_{l}\) independent copies of \(\phi_{l}-\phi_{l-1}\) (with \(\phi_{0}\triangleq0)\), where \(n_{l}\) is a positive integer to be specified later. Assume that  the estimators   \(\hat\phi_{1},\dots,\hat\phi_{L}\) are independent. As \begin{equation*}
E(f(U))=\sum^{L}_{l=1}E(\phi_{l}-\phi_{l-1}),
\end{equation*}     \(\hat\phi=\sum^{L}_{l=1}\hat\phi_{l}
\) is an unbiased estimator of \(E(f(U))\). Following the analysis in~\citep{Giles2008}, 
\begin{equation*}
\var(\hat\phi)=\sum^{L}_{l=1}\frac{V_{l}}{n_{l}},
\end{equation*}where \(V_{l}\triangleq\var(\phi_{l}-\phi_{l-1})\) for \(1\leq l\leq L\).   The expected  time needed to simulate \(\hat\phi\) is  \(T_{\text{ML}}
\triangleq\sum^{L}_{l=1}n_{l}\hat t_{l}\), where    \(\hat t_{l}\)  is the expected time needed to simulate \(\phi_{l}-\phi_{l-1}\).   
As the variance of the average of \(n\) i.i.d. square-integrable random variables is proportional to \(1/n\),   for \(\epsilon>0\), we need \(\lceil\var(\hat\phi)\epsilon^{-2}\rceil\)  independent samples   of  \(\hat\phi\)   to achieve an estimator variance at most  \(\epsilon^{2}\). Thus the total expected time \(T^{\text{MLMC}}(\epsilon)\) needed for the MLMC algorithm to estimate \(E(f(U))\) with variance at most \(\epsilon^{2}\)       satisfies the relation
   \begin{equation}\label{eq:MLMCTotalExpected}
T^{\text{MLMC}}(\epsilon)=\Theta(T_{\text{ML}}+T_{\text{ML}}\var(\hat\phi)\epsilon^{-2}).
\end{equation}  
The first term in the RHS of~\eqref{eq:MLMCTotalExpected} accounts for the fact that   \(\hat\phi\)  is simulated at least once. As shown in~\citep{Giles2008}, the work-normalized variance  \(T_{\text{ML}}\var(\hat\phi)\) is minimized when the \(n_{l}\)'s are proportional to \(\sqrt{V_{l}/\hat t_{l}}\) (ignoring the integrality constraints on
the \(n_{l}\)'s), in which case 
\begin{equation}\label{eq:MLMCTimeVariance}
 T_{\text{ML}}\var(\hat\phi)=\bigg(\sum^{L}_{l=1}\sqrt{V_{l}\hat t_{l}}\bigg)^{2}.
\end{equation} In line with \cite[Theorem 3.1]{Giles2008}, if \(\hat t_{l}=O(2^{l})\) and \(||\phi_{l}-\phi_{L}||^{2}=O(2^{\beta l})\), with \(\beta<-1\), where the constants behind the \(O\)-notation do not depend on \(d\), then \(T_{\text{ML}}\var(\hat\phi)\) is upper-bounded by a constant independent of \(d\).
This can be shown by observing that \begin{displaymath}
V_{l}\leq||\phi_{l}-\phi_{l-1}||^{2}\leq(||\phi_{l}-\phi_{L}||+||\phi_{l-1}-\phi_{L}||)^{2}.
\end{displaymath} 
Theorem~\ref{th:multilevel} below  shows that, under certain conditions,  the randomized dimension reduction method is, up to a  multiplicative constant,    at least as  efficient as the class of MLMC methods described above.  Indeed, under the assumptions of Theorem~\ref{th:multilevel}, by~\eqref{eq:corTotq},   \(T^{\text{tot}}(q,\epsilon)=O(d+T_{\text{ML}}\var(\hat\phi)\epsilon^{-2})\). On the other hand,   \(T_{\text{ML}}\ge\hat t_{L}\ge\hat cd\) since \(m_{L}=d\). Thus~\eqref{eq:MLMCTotalExpected} implies that \(T^{\text{MLMC}}(\epsilon)\ge c'(d+T_{\text{ML}}\var(\hat\phi)\epsilon^{-2})\), for some constant \(c'\).    
\begin{theorem}\label{th:multilevel}
Assume that there are constants \(c\) and \(\hat c\) independent of \(d\) such that  \(t_{i}\le ci\) for \(1\leq i\leq d\), and   \(\hat t_{l}\geq \hat cm_{l}\) for \(1\leq l\leq L\), and that \(C(d-1)>0\). Then  \(R(q;t,\nu^{*})\leq (32c/\hat c)T_{\text{ML}}\var(\hat\phi)\)  if, for \(0\leq i\le d-1\),  
\begin{equation*}
q_{i}=\sqrt{\frac{C(i)}{(i+1)C(0)}}.
\end{equation*} 
\end{theorem}
 
\section{Numerical experiments}\label{se:numerExper}
Our simulation experiments, using the examples in \S\ref{se:examples}, 
were implemented  in the C++ programming language. The randomized dimension
reduction algorithm   (RDR)  was
implemented as described in \S\ref{sub:numerical}. The deterministic dimension
reduction algorithm  (DDR) was implemented similarly with  \(N_{k}=\bar N_{k}\).
In both algorithms,     \(n\)  was chosen so that the expected total number of  simulations of the \(U_{i}\)'s  in iterations \(2\) through \(n\)  is approximately   \(10d\). The actual total number of simulations of the \(U_{i}\)'s, denoted by ``Cost'' in   our computer experiments, is about \(11d\) because it includes the \(d\) simulations of the first iteration.    

We have  implemented the multilevel algorithm (MLMC) described in \S\ref{sub:MLMC}, with \(L=\lfloor \log_{2}(d)\rfloor+1 \),  and \(m_{l}=\lfloor2^{l-L}d\rfloor\) for \(1\leq l\leq L\), and   \(\phi_{l}=f(U_{1},\dots,U_{m_{l}},\underbrace{X_{0},\dots,X_{0}}_{d-m_l})\). The \(V_{l}\)'s were estimated by Monte Carlo simulation with \(1000\) samples, and the  \(n_{l}\)'s   were scaled up so that  the actual total number of simulations   of the \(U_{i}\)'s   is about \(11d\). We have also implemented
a  randomized quasi-Monte
Carlo method (QMC) with a random shift~\cite[Section 5.4]{glasserman2004Monte}.   Our implementation  uses  the C++ program available  at 
\commentt{http://web.maths.unsw.edu.au/\texttildelow{}fkuo/sobol}
{\url{http://web.maths.unsw.edu.au/~fkuo/sobol}}
 to generate \(d\)-dimensional Sobol sequences of length \(n=4096\). For
practical reasons linked to computing time and storage cost, the QMC algorithm
was tested for \(d\) up to \(10^{4}\).

In Tables~\ref{tab:garch} through~\ref{tab:MG1VRF} and in the \commentt{appendix}{on-line supplement}, the variable Std  refers to the standard deviation of \(f_{n}\) for the RDR and DDR algorithms,  to the standard deviation of \(\hat \phi\) for the MLMC algorithm, and to the
standard deviation of the Quasi-Monte Carlo estimator for the QMC algorithm.
 The variable Std and a \(90\%\) confidence interval for \(E(f(U))\) were estimated using \(1000\) independent runs of these two algorithms. For the RDR algorithm, a \(90\%\) confidence interval for 
the variable Cost was reported as well. The  variance reduction factor VRF is defined as\begin{displaymath}
\text{VRF}=\frac{d\var(f(U))}{\text{Cost}\times\text{Std}^2}.
\end{displaymath}
We estimated \(\var(f(U))\) by  using \(10000\) independent samples of \(U\).   
\subsection{GARCH volatility model}\label{sub:numerGARCH}
Table~\ref{tab:garch} shows results of our simulations of the GARCH volatility  model for estimating 
\(\mathbb{P}(X_{d}>  z)\),
with \(z=4.4\times10^{-5}\), \(X_{0}=10^{-4}\),  \(\alpha=0.06\), \(\beta=0.9\), and \( w=1.76\times10^{-6}\). As expected, the variable Cost is about \(11d\) for the  RDR, DDR, and MLMC  algorithms. For these algorithms, the variable Cost $\times$ Std\(^2\) is roughly independent of \(d\), and  the variance reduction factors are roughly proportional to \(d\).   In contrast, for the QMC algorithm, the variable Cost $\times$ Std\(^2\) is roughly proportional to \(d\), and the variance reduction factors are roughly constant. The  $90\%$ confidence interval  of
the RDR algorithm running time has a  negligible length in comparison
to the running time.  
The  RDR algorithm outperforms the MLMC algorithm by about a factor of \(10\), and the QMC
algorithm by a factor ranging from \(5\) to \(20\).
In all our numerical experiments, the DDR algorithm outperforms the RDR algorithm by a factor between
\(1\) and \(2\).\begin{table}
\caption{\(\mathbb{P}(X_{d}> z)\)  estimation in GARCH model, with \(z=4.4\times10^{-5}\), using 1000 samples, where \(X_{d}\) is the daily variance at time-step \(d\).}
\begin{scriptsize}
\begin{tabular}{llrrclrr}\hline
    &     & $n$& $90\%$ confidence interval & Std & Cost& Cost $\times$ Std$^2$ &   VRF \\ \hline
$d=1250$ &RDR & $ 277$ & $0.3918 \pm 2.1\times10^{-3}$ & $4.0\times10^{-2}$ & $1.367\times10^{4}\pm 8.9\times10^{1}$ & $21$ & $14$\\
 &DDR & $ 596$ & $0.3935 \pm 1.8\times10^{-3}$ & $3.4\times10^{-2}$ & $1.355\times10^{4}$ & $15$ & $19$\\
 &MLMC & $ 134$ & $0.3946 \pm 6.5\times10^{-3}$ & $1.2\times10^{-1}$ & $1.375\times10^{4}$ & $215$ & $1.4$\\
 &QMC & $ 4096$ & $0.39365 \pm 2.4\times10^{-4}$ & $4.6\times10^{-3}$ & $5.120\times10^{6}$ & $109$ & $2.7$\\

 $d=2500$ &RDR & $ 529$ & $0.3933 \pm 1.4\times10^{-3}$ & $2.8\times10^{-2}$ & $2.745\times10^{4}\pm 1.7\times10^{2}$ & $21$ & $28$\\
 &DDR & $ 1167$ & $0.3939 \pm 1.3\times10^{-3}$ & $2.4\times10^{-2}$ & $2.734\times10^{4}$ & $16$ & $37$\\
 &MLMC & $ 266$ & $0.3919 \pm 4.6\times10^{-3}$ & $8.9\times10^{-2}$ & $2.847\times10^{4}$ & $226$ & $2.6$\\
 &QMC & $ 4096$ & $0.39348 \pm 2.5\times10^{-4}$ & $4.8\times10^{-3}$ & $1.024\times10^{7}$ & $238$ & $2.5$\\

$d=5000$ &RDR & $ 970$ & $0.3923 \pm 1.0\times10^{-3}$ & $2.0\times10^{-2}$ & $5.490\times10^{4}\pm 3.4\times10^{2}$ & $21$ & $56$\\
 &DDR & $ 1899$ & $0.39285 \pm 9.4\times10^{-4}$ & $1.8\times10^{-2}$ & $5.207\times10^{4}$ & $17$ & $71$\\
 &MLMC & $ 524$ & $0.3931 \pm 3.4\times10^{-3}$ & $6.5\times10^{-2}$ & $5.339\times10^{4}$ & $227$ & $5.3$\\
 &QMC & $ 4096$ & $0.39372 \pm 2.4\times10^{-4}$ & $4.6\times10^{-3}$ & $2.048\times10^{7}$ & $435$ & $2.8$\\
\hline\end{tabular}
 \end{scriptsize}
 \label{tab:garch}
\end{table}

 \subsection{\(G_t/D/1\) queue}\label{sub:NumGD1}
Assume that  \(A_{i}\) has a Poisson distribution with time-varying rate \(\lambda_{i}=0.75+0.5\cos(\pi i/50)\),  for \(1\leq i\leq d\), (recall that \(A_{0}=0\)).
  These parameters are taken from~\citep{whitt2016time}. Table~\ref{MD:expectation} estimates \(E(X_{d})\), and   Table~\ref{tab:MDthresh} gives VRFs in the estimation of  \(\mathbb{P}(X_{d}> z)\), for selected values of \(z\). Once again, for  the  RDR, DDR, and MLMC algorithms, the variable Cost $\times$ Std\(^2\) is roughly independent of \(d\), and the variance reduction factors are roughly proportional to \(d\). The VRFs of the  RDR  and DDR  algorithms in Table~\ref{tab:MDthresh}  are greater than or equal to the corresponding VRFs in Table~\ref{MD:expectation}, which confirms the resiliency of these algorithms to discontinuities of \(g\). In contrast, the VRFs of the MLMC algorithm in Table~\ref{tab:MDthresh} are lower than the corresponding VRFs  in Table~\ref{MD:expectation}. The RDR algorithm outperforms the MLMC
algorithm by  a factor  ranging from \(1\) to \(2\)  in Table~\ref{MD:expectation},
and a factor ranging from \(2\) to \(17\)  in Table~\ref{tab:MDthresh}. Table~\ref{MD:expectationShifted} estimates \(E(X_{d})\) for shifted values of \(d\). The results in
Table~\ref{MD:expectationShifted} are similar to those of Table~\ref{MD:expectation},
but the values of \(E(X_{d})\)   in Table~\ref{MD:expectationShifted}
are significantly smaller  than those in Table~\ref{MD:expectation}. This
can be explained by observing that \(\lambda_{d}\) is maximized (resp. minimized)
at the values of \(d\) listed in Table~\ref{MD:expectation} (resp. Table~\ref{MD:expectationShifted}).

\begin{table}
\caption{\(E(X_{d})\) estimation in \(G_{t}/D/1\) queue, 1000 samples,  where \(X_{d}\) is the number of customers in the queue at time-step \(d\).}
\begin{scriptsize}
\begin{tabular}{llcrclrc}\hline
    &     & $n$& $90\%$ confidence interval & Std & Cost& Cost $\times$ Std$^2$ &   VRF \\ \hline
$d=10^4$ &RDR & $ 2.3\times 10^3$ & $5.5243 \pm 4.6\times 10^{-3}$ & $8.8\times 10^{-2}$ & $1.106\times 10^5\pm 6.3\times10^{2}$ & $8.5\times 10^2$ & $1.8\times 10^2$\\
 &DDR & $ 2.8\times 10^3$ & $5.5221 \pm 4.5\times 10^{-3}$ & $8.6\times 10^{-2}$ & $1.032\times 10^5$ & $7.7\times 10^2$ & $2.0\times 10^2$\\
 &MLMC & $ 7.1\times 10^3$ & $5.524 \pm 5.9\times 10^{-3}$ & $1.1\times 10^{-1}$ & $1.076\times 10^5$ & $1.4\times 10^3$ & $1.1\times 10^2$\\

$d=10^{5}$ &RDR & $ 2.3\times 10^4$ & $5.5232 \pm 1.5\times 10^{-3}$ & $2.8\times 10^{-2}$ & $1.101\times 10^6\pm 5.3\times10^{3}$ & $8.6\times 10^2$ & $1.8\times 10^3$\\
 &DDR & $ 3.6\times 10^4$ & $5.5229 \pm 1.5\times 10^{-3}$ & $2.8\times 10^{-2}$ & $1.046\times 10^6$ & $8.3\times 10^2$ & $1.8\times 10^3$\\
 &MLMC & $ 6.3\times 10^4$ & $5.5241 \pm 2.0\times 10^{-3}$ & $3.8\times 10^{-2}$ & $1.096\times 10^6$ & $1.6\times 10^3$ & $9.5\times 10^2$\\

$d=10^{6}$ &RDR & $ 2.3\times 10^5$ & $5.52325 \pm 4.9\times 10^{-4}$ & $9.4\times 10^{-3}$ & $1.103\times 10^7\pm 4.8\times10^{4}$ & $9.8\times 10^2$ & $1.5\times 10^4$\\
 &DDR & $ 2.6\times 10^5$ & $5.52363 \pm 4.5\times 10^{-4}$ & $8.7\times 10^{-3}$ & $1.047\times 10^7$ & $7.9\times 10^2$ & $1.9\times 10^4$\\
 &MLMC & $ 6.6\times 10^5$ & $5.5231 \pm 5.7\times 10^{-4}$ & $1.1\times 10^{-2}$ & $1.119\times 10^7$ & $1.4\times 10^3$ & $1.1\times 10^4$\\
\hline\end{tabular}
 \end{scriptsize}
\label{MD:expectation}
\end{table}

\begin{table}
\caption{VRFs for \(\mathbb{P}(X_{d}> z)\) estimation in  \(G_{t}/D/1\) queue.}
\begin{scriptsize}
\begin{tabular}{llcccccc}\hline
$z$    &     & $0$& $2$ & $4$ & $6$& $8$ &   $10$ \\ \hline
$d=10^4$ &  RDR & $ 2.8\times10^{2}$ & $2.3\times10^{2}$ & $2.1\times10^{2}$ & $2.1\times10^{2}$ & $1.9\times10^{2}$ & $1.8\times10^{2}$\\
&DDR&$4.7\times10^{2}$&$2.8\times10^{2}$&$2.3\times10^{2}$&$2.4\times10^{2}$&$2.2\times10^{2}$&$1.9\times10^{2}$\\
 &  MLMC & $ 2.1\times10^{1}$ & $3.7\times10^{1} $ & $5.1\times10^{1}$ & $6.5\times10^{1}$ & $7.7\times10^{1}$ & $7.2\times10^{1}$\\
$d=10^5$ &  RDR & $ 3.1\times10^{3}$ & $2.2\times10^{3}$ & $2.1\times10^{3}$ & $1.9\times10^{3}$ & $1.7\times10^{3}$ & $1.8\times10^{3}$\\
&DDR&$4.3\times10^{3}$&$ 3.0\times10^{3}$ & $2.4\times10^{3}$ & $2.5\times10^{3}$ & $2.2\times10^{3}$&$2.0\times10^{3}$\\
 &  MLMC & $ 2.1\times10^{2}$ & $3.6\times10^{2}$ & $4.5\times10^{2}$ & $5.9\times10^{2}$ & $5.9\times10^{2}$ & $5.8\times10^{2}$\\
$d=10^6$ &  RDR & $ 3.3\times10^{4}$ & $2.2\times10^{4}$ & $1.9\times10^{4}$ & $1.8\times10^{4}$ & $1.7\times10^{4}$ & $1.9\times10^{4}$\\
&DDR&$ 4.1\times10^{4}$&$2.9\times10^{4}$& $2.2\times10^{4}$& $2.2\times10^{4}$& $1.9\times10^{4}$ & $2.2\times10^{4}$
\\ &  MLMC & $ 2.0\times10^{3}$ & $3.3\times10^{3}$ & $4.4\times10^{3}$ & $5.3\times10^{3}$ & $5.7\times10^{3}$&$5.8\times10^{3}$\\
\hline\end{tabular}
\end{scriptsize}
\label{tab:MDthresh}
\end{table}

\begin{table}
\caption{\(E(X_{d})\) estimation in \(G_{t}/D/1\) queue, 1000 samples,  with
shifted dimensions.}
\begin{scriptsize}
\begin{tabular}{llcrclrc}\hline
    &     & $n$& $90\%$ confidence interval & Std & Cost& Cost $\times$ Std$^2$ &   VRF \\ \hline
$d=10050$ &RDR & $ 8.5\times10^2$ & $0.6599 \pm 4.1\times10^{-3}$ & $7.8\times10^{-2}$ & $1.100\times10^5 \pm 6.1\times10^2$ & $6.7\times10^2$ & $6.2\times10^1$\\
 &DDR & $ 2.4\times10^3$ & $0.6599 \pm 4.1\times10^{-3}$ & $7.9\times10^{-2}$ & $1.065\times10^5$ & $6.6\times10^2$ & $6.3\times10^1$\\
 &MLMC & $ 1.5\times10^3$ & $0.659 \pm 4.5\times10^{-3}$ & $8.6\times10^{-2}$ & $1.106\times10^5$ & $8.2\times10^2$ & $5.0\times10^1$\\

$d=100050$ &RDR & $ 7.8\times10^3$ & $0.6594 \pm 1.3\times10^{-3}$ & $2.5\times10^{-2}$ & $1.101\times10^6 \pm 5.3\times10^3$ & $7.2\times10^2$ & $5.6\times10^2$\\
 &DDR & $ 7.0\times10^3$ & $0.6593 \pm 1.4\times10^{-3}$ & $2.6\times10^{-2}$ & $1.025\times10^6$ & $7.2\times10^2$ & $5.6\times10^2$\\
 &MLMC & $ 1.1\times10^4$ & $0.6587 \pm 1.6\times10^{-3}$ & $3.1\times10^{-2}$ & $1.115\times10^6$ & $1.1\times10^3$ & $3.6\times10^2$\\

$d=1000050$ &RDR & $ 6.7\times10^4$ & $0.66007 \pm 4.5\times10^{-4}$ & $8.6\times10^{-3}$ & $1.101\times10^7 \pm 4.8\times10^4$ & $8.2\times10^2$ & $5.3\times10^3$\\
 &DDR & $ 6.2\times10^4$ & $0.66019 \pm 4.4\times10^{-4}$ & $8.4\times10^{-3}$ & $1.034\times10^7$ & $7.4\times10^2$ & $5.9\times10^3$\\
 &MLMC & $ 1.1\times10^5$ & $0.66033 \pm 5.6\times10^{-4}$ & $1.1\times10^{-2}$ & $1.135\times10^7$ & $1.3\times10^3$ & $3.3\times10^3$\\

\hline\end{tabular}
 \end{scriptsize}
\label{MD:expectationShifted}
\end{table}

\subsection{\(M_t/GI/1\) queue}\label{sub:MtGI1NumerExample}
Assume that  \(\lambda_{s}=0.75+0.5\cos(\pi s/50)\) for \(s\geq0\). These parameters are taken from~\citep{whitt2016time}. Assume further that, for \(j\geq1\), the service time \(S_{j}\) for the \(j\)-th customer has a Pareto distribution with  \(\mathbb{P}(S_{j}\geq z)=(1+z/\alpha)^{-3}  
\)   for \(z\geq0\), for some constant \(\alpha>0\).
  A simple calculation shows that \(E(S_{j})=\alpha/2\).  In our simulations, we have set \(d=\lceil \theta\rceil\). Table~\ref{tab:MG1expectation} gives  our simulation results   for estimating
\(\mathbb{P}(W_{\theta}> 1)\) when \(\alpha=2\), and Table~\ref{tab:MG1VRF} lists VRFs for estimating
\(\mathbb{P}(W_{\theta}> 1)\) for selected values of \(\alpha\).
Here again, for  the  RDR, DDR, and MLMC algorithms, the variable Cost $\times$ Std\(^2\) is roughly independent of \(d\), and the VRFs are roughly proportional to \(d\). The RDR algorithm outperforms the MLMC algorithm by a factor ranging from      \(1\) to  \(10\),
 depending on the value of \(\alpha\).
In Table~\ref{tab:MG1VRF},   the  RDR, DDR, and MLMC algorithms become less efficient as \(\alpha\) increases. This can be explained by noting that, as  \(\alpha\) increases, the length of the last busy cycle 
increases as well, which renders \(W_{\theta}\) more dependent on the first \(Y_{i}\)'s. 
\begin{table}
\caption{\(\mathbb{P}(W_{\theta}> 1)\) estimation in \(M_{t}/GI/1\) queue, \(\alpha=2\), with  1000 samples, where \(W_{\theta}\) is the residual work at time \(\theta\).}
\begin{scriptsize}
\begin{tabular}{llcrclr}\hline
    &     & $n$& $90\%$ confidence interval & Std & Cost& Cost $\times$ Std$^2$  \\ \hline
$\theta=10^4$ &RDR & $ 3.9\times10^3$ & $0.85389 \pm 4.9\times10^{-4}$ & $9.4\times10^{-3}$ & $1.108\times10^5\pm 6.2\times10^{2}$ & $10$ \\
 &DDR & $ 7.6\times10^3$ & $0.85333 \pm 4.0\times10^{-4}$ & $7.7\times10^{-3}$ & $1.037\times10^5$ & $6$ \\
 &MLMC & $ 1.0\times10^4$ & $0.8541 \pm 1.6\times10^{-3}$ & $3.1\times10^{-2}$ & $1.142\times10^5$ & $106$ \\

$\theta=10^5$ &RDR & $ 3.0\times10^4$ & $0.85385 \pm 1.6\times10^{-4}$ & $3.0\times10^{-3}$ & $1.100\times10^6\pm 5.3\times10^{3}$ & $10$ \\
 &DDR & $ 4.8\times10^4$ & $0.85373 \pm 1.3\times10^{-4}$ & $2.5\times10^{-3}$ & $1.031\times10^6$ & $6$ \\
 &MLMC & $ 8.3\times10^4$ & $0.85322 \pm 4.9\times10^{-4}$ & $9.5\times10^{-3}$ & $1.092\times10^6$ & $98$ \\

$\theta=10^6$ &RDR & $ 2.5\times10^5$ & $0.853762 \pm 5.1\times10^{-5}$ & $9.8\times10^{-4}$ & $1.103\times10^7\pm 4.9\times10^{4}$ & $11$ \\
 &DDR & $ 4.5\times10^5$ & $0.853779 \pm 4.4\times10^{-5}$ & $8.4\times10^{-4}$ & $1.040\times10^7$ & $7$ \\
 &MLMC & $ 8.5\times10^5$ & $0.8539 \pm 1.6\times10^{-4}$ & $3.2\times10^{-3}$ & $1.114\times10^7$ & $111$\\

\hline\end{tabular}
 \end{scriptsize}
 \label{tab:MG1expectation}
\end{table}
\begin{table}
\caption{VRF for \(\mathbb{P}(W_{\theta}> 1)\) estimation in  \(M_{t}/GI/1\) queue.}
\begin{scriptsize}
\begin{tabular}{llcccc}\hline
$\alpha$    &     & $0.5$& $1$ & $1.5$ & $2$ \\ \hline
$\theta=10^4$ 
&  RDR & $ 4.3\times10^{2}$ & $2.6\times10^{2}$ & $2.3\times10^{2}$ & $1.3\times10^{2}$\\
&  DDR & $ 5.6\times10^{2}$ & $4.0\times10^{2}$ & $3.0\times10^{2}$ & $2.0\times10^{2}$\\

 &  MLMC & $ 3.2\times10^{2}$ & $1.4\times10^{2} $ & $5.2\times10^{1}$ & $1.2\times10^{1}$\\
$\theta=10^5$ 
&  RDR & $ 4.1\times10^{3}$ & $2.6\times10^{3}$ & $1.8\times10^{3}$ & $1.2\times10^{3}$\\
&  DDR & $ 5.2\times10^{3}$ & $3.8\times10^{3}$ & $2.4\times10^{3}$ & $1.9\times10^{3}$\\
 &  MLMC & $ 2.3\times10^{3}$ & $1.1\times10^{3}$ & $5.3\times10^{2}$ & $1.2\times10^{2}$ \\
$\theta =10^6$
 &  RDR & $ 3.8\times10^{4}$ & $2.7\times10^{4}$ & $1.7\times10^{4}$ & $1.2\times10^{4}$\\
 &  DDR & $ 4.7\times10^{4}$ & $3.5\times10^{4}$ & $2.5\times10^{4}$ & $1.7\times10^{4}$\\

 &  MLMC & $ 2.6\times10^{4}$ & $1.1\times10^{4}$ & $4.4\times10^{3}$ & $1.1\times10^{3}$ \\
\hline\end{tabular}
\end{scriptsize}
\label{tab:MG1VRF}
\end{table}

\section{Conclusion}
\label{se:conclu}
We have described  a randomized dimension reduction algorithm that  estimates $E(f(U))$ via Monte
Carlo  simulation, assuming that \(f\) does not depend equally on all its arguments.  We formally prove
that under some conditions, in order to achieve an estimator variance  \(\epsilon^{2}\), our algorithm requires
\(O(d + \epsilon^{-2})\) computations as opposed to \(O(d \epsilon^{-2})\) under the standard Monte Carlo method. Our algorithm can be used to efficiently estimate the expected value of a function of the state of a  Markov chain at time-step \(d\), for a class of Markov chains driven by random variables.  The
numerical implementation of our algorithm uses a new geometric procedure  of independent interest that solves  in \(O(d)\) time a \(d\)-dimensional optimisation problem that was  previously solved in \(O(d^{3})\) time.    
We have argued intuitively that our method is resilient to discontinuities of \(f\), and have described and analysed a deterministic version of our algorithm. Our numerical experiments confirm  that our approach highly outperforms the  standard Monte Carlo  method for large values of \(d\), and show its high resilience to discontinuities. Whether our approach can be combined with the Quasi-Monte Carlo method to produce a provably efficient estimator is left for future work.  
\section*{Acknowledgments} This research has been presented at the  9th NIPS Workshop on Optimization for Machine Learning, Barcelona, December 2016, the Stochastic  Methods in Finance seminar at \'Ecole des Ponts Paris-Tech, March 2018, and the 23rd International Symposium on Mathematical Programming,  Bordeaux, July 2018.    The author thanks  Bernard Lapeyre, seminar and conference participants for helpful conversations. He is grateful to three anonymous referees,  an anonymous associate editor, and Baris Ata (department editor), for insightful comments and suggestions. This work was achieved through the Laboratory of Excellence on Financial Regulation (Labex ReFi) under the reference ANR-10-LABX-0095. It benefitted from a French government support managed by the National Research Agency (ANR).

\comment{
This can  explained by the fact that the performance of our algorithm (resp. multilevel algorithms) actually relies on the weak dependence of (resp. \(f(U)\)) on \(U_{i+1},\dots,U_{d}\), as \(i\) gets large, as shown formally in    Sections~\ref{se:main} and~\ref{se:ML}. Furthermore, while smoothing techniques are often needed to efficiently implement multilevel~\citep{Giles2008} and QMC algorithms (see~\citep{wangTan2013} and references therein) in the presence of discontinuities, no continuity assumptions are made in the analysis of our algorithm, and we did not  need to smooth \(f\) in our numerical experiments. This can be explained by the fact that  the performance of our algorithm relies on  properties of conditional expectations of \(f\), which can be considered as  smoothed versions of \(f\). Note that it is not always possible to smooth \(f\) explicitly, especially if \(F\)  is multi-dimensional.
}
\commentt{\appendix}
{
\begin{APPENDIX}{}
}

\section{Proof of Proposition~\ref{pr:covf(U)f(U')}}
For \(0\le i\le d\), let \begin{equation*}
f^{(i)}=E(f(U)|U_{i+1},\ldots,U_{d}).
\end{equation*} Thus, \(C(i)=\var(f^{(i)})\). By the tower law, for \(0\le i\le d-1\), \begin{displaymath}
E(f^{(i)}|U_{i+2},\ldots,U_{d})=f^{(i+1)},
\end{displaymath} and so \begin{equation*}
C(i+1)=\var(E(f^{(i)}|U_{i+2},\ldots,U_{d})).
\end{equation*}
As the   variance decreases by taking the conditional expectation,   it follows that \(C(i+1)\leq C(i)\), as desired. 

We now prove~\eqref{eq:ViCov}.  Let \(W=(U'_{1},\ldots ,U'_{i},U_{i+1},\ldots,U_{d})\).
 Since \(U\) and \(W\) are conditionally independent given \(U_{i+1},\ldots,U_{d}\), and \(E(f(W)|U_{i+1},\ldots,U_{d})=f^{(i)}\),
\begin{displaymath}
E(f(U)f(W)|U_{i+1},\ldots,U_{d}))=(f^{(i)})^{2}.
\end{displaymath}
Hence, by the tower law,
\begin{displaymath}E(f(U)f(W))=E((f^{(i)})^{2}).
\end{displaymath}
On the other hand, using the tower law once again, \begin{displaymath}
E(f(U))=E(f(W))=E(f^{(i)}),
\end{displaymath}
and so the RHS of~\eqref{eq:ViCov} is equal to \(\var(f^{(i)})\),
as required.
\commentt{\qed}{\Halmos}
\section{Proof of Theorem~\ref{th:variance}}\label{se:ProofMainLemma}
We first prove Lemma~\ref{le:varianceClassic} below, which follows by classical calculations (see e.g. \citep[Section IV.6a]{asmussenGlynn2007}).
\begin{lemma}\label{le:varianceClassic}Let \((Z_{k})\), \(k\geq1\), be an homogeneous stationary Markov chain in \(\mathbb{R}^{d}\), and let \(g\) be a real-valued  Borel-measurable function on \(\mathbb{R}^{d}\) such that \(g(Z_{1})\) is square-integrable, and \(a_{j}=\cov(g(Z_{1}),g(Z_{1+j}))\) is non-negative for \(j\geq0\). Assume that  \(\sum^{\infty}_{j=1}a_{j}\) is finite. Then 
\begin{equation}\label{eq:lemmaVarianceClassical}
n^{-1}\var(\sum^{n}_{m=1}g(Z_{m}))\le a_{0}+2\sum^{\infty}_{j=1}a_{j}.
\end{equation}Furthermore, the LHS of~\eqref{eq:lemmaVarianceClassical} converges to its RHS as \(n\) goes to infinity. 
\end{lemma}
\commentt{\begin{proof}}{\proof{Proof.}}
Since \((Z_{k})\), \(k\geq1\), is homogeneous and stationary, \(\cov(g(Z_{m})g(Z_{m+j}))=a_{j}\) for \(m\geq1\) and \(j\geq0\). Thus,  \begin{eqnarray*} 
\var(\sum^{n}_{m=1}g(Z_{m}))
&=&\sum^{n}_{m=1}\var(g(Z_{m}))+2\sum_{1\leq m<m+j\leq n} \cov(g(Z_{m})g(Z_{m+j}))\\  
&=&na_{0}+2\sum^{n}_{j=1}(n-j) a_{j}.
 \end{eqnarray*}Hence  
\begin{equation*}
n^{-1}\var(\sum^{n}_{m=1}g(Z_{m}))= a_{0}+2\sum^{n}_{j=1}\frac{n-j}{n}a_{j},
\end{equation*}which implies~\eqref{eq:lemmaVarianceClassical}.  Using Lebesgue's dominated convergence theorem concludes the proof.
\commentt{\end{proof}}{\Halmos\endproof}
We now prove Theorem~\ref{th:variance}.
Since the \(C(i)\)'s and the variance of \(f_{n}\) remain unchanged if we
add a constant to \(f\), we can assume without loss of generality that \(E(f(U))=0\).   For \(0\leq i\leq d-1\), set \(p_{i}=1-q_{i}\). Let  \(m<k\) be two integers in \([1,n]\). For \(0\leq i\leq d-1\), \begin{displaymath}
\mathbb{P}(\max_{m\leq j<k}N_{j}\leq i)= {p_{i}}^{k-m}
\end{displaymath}
and so, for \(1\leq i\leq d-1\), \begin{equation*}
\mathbb{P}(\max_{m\leq j<k}N_{j}= i)= {p_{i}}^{k-m}-{p_{i-1}}^{k-m}.
\end{equation*}
 For \(i\in [1,d]\), conditional on the event  \(\max_{m\leq j<k}N_{j}= i\), the first \(i\)  components of \(V^{(m)}\) and \(V^{(k)}\) are independent, and  the  last \(d-i\)  components of \(V^{(m)}\) and \(V^{(k)}\) are  the same. This is because, if \(N_{j}=i\), with \(m\leq j<k\),  the vector \(V^{(m)}\)  and the first  \(i\)  components of \(V^{(j+1)}\) are independent.   Thus,  by Proposition~\ref{pr:covf(U)f(U')},\begin{equation}\label{eq:covmk}
E(f(V^{(m)})f(V^{(k)})|\max_{m\leq j<k}N_{j}=i)=C(i).
\end{equation} As   \(C(d)=0\),
it follows from Bayes' formula that 
\begin{eqnarray*}
E(f(V^{(m)})f(V^{(k)}))&=&\sum_{i=1}^{d-1}\mathbb{P}(\max_{m\leq j<k}N_{j}= i)C(i)\\&=&\sum^{d-1}_{i=1}({p_{i}}^{k-m}-{p_{i-1}}^{k-m})C(i).
\end{eqnarray*}
Let \(a_{j}=\cov(f(V^{(1)}),f(V^{(1+j)}))\). Thus, for \(j>0\),  \begin{equation*}
a_{j}=\sum^{d-1}_{i=1}({p_{i}}^{j}-{p_{i-1}}^{j})C(i)
\end{equation*}is non-negative, and \begin{eqnarray*}
\sum_{j=1}^\infty a_{j}&=&\sum^{d-1}_{i=1}(\frac{p_{i}}{1-p_{i}}-\frac{p_{i-1}}{1-p_{i-1}})C(i)\\
&=&\sum^{d-1}_{i=1}(\frac{1}{1-p_{i}}-\frac{1}{1-p_{i-1}})C(i)\\
&=&-C(1)+\sum^{d-1}_{i=1}\frac{C(i)-C(i+1)}{q_{i}}\end{eqnarray*} is finite. Since \(a_{0}=C(0)\), it follows that
\begin{equation}\label{eq:infiniteSumaj}
a_{0}+2\sum^{\infty}_{j=1}a_{j}=C(0)-2C(1)+2\sum^{d-1}_{i=1}\frac{C(i)-C(i+1)}{q_{i}}.
\end{equation}
We conclude the proof using Lemma~\ref{le:varianceClassic}.
\commentt{\qed}{\Halmos}
\section{Proof of Proposition~\ref{pr:upperBoundingCi}}
 Let \(U'_{1},\ldots ,U'_{i}\) be random variables  satisfying the conditions of  Proposition~\ref{pr:covf(U)f(U')}, and \(W=(U'_{1},\dots,U'_i,U_{i+1},\dots,U_d)\). Since \((U_1,\dots,U_i)\) and \(W\) are independent, \begin{displaymath}
\cov(f_{i}(U_1,\dots,U_i),f(W))=0.
\end{displaymath} Similarly, \begin{displaymath}
\cov(f_{i}(U_1,\dots,U_i),f_{i}(U'_{1},\dots,U'_i))=\cov(f(U),f_{i}(U'_{1},\dots,U'_i))=0.
\end{displaymath} Thus, by Proposition~\ref{pr:covf(U)f(U')} and  bilinearity of the covariance,
\begin{eqnarray*}
 C(i)&=&\cov(f(U)-f_{i}(U_1,\dots,U_i),f(W)-f_{i}(U'_{1},\dots,U'_i))\\ &\leq&\text{Std}(f(U)-f_{i}(U_1,\dots,U_i))\ \text{Std}(f(W)-f_{i}(U'_{1},\dots,U'_i))\\
 &=&\var(f(U)-f_{i}(U_1,\dots,U_i)).
\end{eqnarray*} 
The last equation follows by observing that \(f(W)-f_{i}(U'_{1},\dots,U'_i)\overset d=f(U)-f_{i}(U_1,\dots,U_i)\).
\commentt{\qed}{\Halmos}
\section{Proofs of Propositions~\ref{pr:powerLawCi} and~\ref{pr:varfnSqrCi}} 
We first prove the following.
\begin{proposition}\label{pr:sqrtqiGen}
Let \(\nu=(\nu_{0},\dots,\nu_{d})\) be an element of \(\mathbb{R}^{d}\times \{0\}\). Assume that \(\nu_{0},\dots,\nu_{d-1}\) are positive, and that the sequence \((\nu_{i}/t_{i+1})\),  \(0\leq i\leq d-1\), is decreasing. For \(0\leq i\leq d-1\), set \(q_{i}=\sqrt{{(t_{1}\nu_{i})}/{(\nu_{0} t_{i+1})}}.\)
Then \begin{displaymath}
R(q;t,\nu)\leq4\left(\sum^{d-1}_{i=0}\sqrt{\nu_{i}}(\sqrt{t_{i+1}}-\sqrt{t_i})\right)^2.
\end{displaymath}
 \end{proposition}
 \commentt{\begin{proof}}{\proof{Proof.}}
Applying the inequality \(x-y\leq2\sqrt{x}(\sqrt{x}-\sqrt{y})\), which holds for   \(x\geq0\) and \( y\geq0\), to \(x=\nu_{i}\) and \(y=\nu_{i+1}\) yields
\begin{eqnarray*}
\sum^{d-1}_{i=0}\frac{\nu_{i}-\nu_{i+1}}{q_{i}}
&\le& 2\sqrt{\frac{\nu_{0}}{t_{1}}}\sum_{i=0}^{d-1}(\sqrt{\nu_{i}}-\sqrt{\nu_{i+1}})\sqrt{t_{i+1}}
\\&=&2\sqrt{\frac{\nu_{0}}{t_{1}}}\sum_{i=0}^{d-1}\sqrt{\nu_{i}}(\sqrt{t_{i+1}}-\sqrt{t_i}).
\end{eqnarray*} 
Similarly, since \( t_{i+1}-t_{i}\leq2\sqrt{t_{i+1}}(\sqrt{t_{i+1}}-\sqrt{t_i})\), \begin{displaymath}
\sum^{d-1}_{i=0}q_{i}(t_{i+1}-t_{i})\leq2\sqrt{\frac{t_{1}}{\nu_{0} }}\sum^{d-1}_{i=0}\sqrt{\nu_{i}}(\sqrt{t_{i+1}}-\sqrt{t_i}).
\end{displaymath}Taking the product completes the proof.\commentt{\end{proof}}{\Halmos\endproof}
We now prove Proposition~\ref{pr:powerLawCi}. Set  \(t'=(t'_{0},\dots,t'_{d})\),
with \(t'_{i}=ci\), \(0\leq i\leq d\), and let \(\nu=(\nu_{0},\dots,\nu_{d-1},0)\), with   \(\nu_{i}=c'\,(i+1)^{\gamma}\), \(0\leq i \leq d-1\).  Thus   \(R(q;t,\nu^{*})\leq R(q;t',2\nu)\) since  \(t\leq t'\) and \(\nu^{*}\leq 2\nu\).  
By Proposition~\ref{pr:sqrtqiGen}, \begin{displaymath}
R(q;t',2\nu)\leq8cc'\left(\sum^{d-1}_{i=0}(i+1)^{\gamma/2}(\sqrt{{i+1}}-\sqrt{i})\right)^2.
\end{displaymath} As \(\sqrt{{i+1}}-\sqrt{i}\leq(i+1)^{-1/2}\) for \(i\geq0\),  it follows
that   
\begin{equation*}
R(q;t,\nu^{*})\leq8 cc' (\sum_{i=1}^{d}i^{(\gamma-1)/2})^{2}.
\end{equation*}    
The inequality \begin{displaymath}
\sum_{i=1}^{d}i^{(\gamma-1)/2}\leq1+\int^{d}_{1}x^{(\gamma-1)/2}\,dx
\end{displaymath}implies that \begin{equation*}
\sum^{d-1}_{i=0}i^{(\gamma-1)/2} \le \begin{cases} 1+2/(\gamma+1), &\mbox{ } \gamma<-1, 
\\ 1+\ln(d), & \mbox{ } \gamma=-1, 
\\ 2\, d^{(\gamma+1)/2}/(\gamma+1), & \mbox{ } -1<\gamma<0. 
\end{cases} 
\end{equation*}
We conclude that there is a constant \(c_{1}\) such that~\eqref{eq:UpperRPower} holds. This concludes the proof of Proposition~\ref{pr:powerLawCi}.   
\comment{Corollary~\ref{cor:Ttotal} then   implies the existence of a constant \(c_{2}\) such that~\eqref{eq:UpperTPower} holds. }
\commentt{\qed}{\Halmos}

We now prove Proposition~\ref{pr:varfnSqrCi}.
 By applying Proposition~\ref{pr:sqrtqiGen} to  \(\nu=(2C(0),\dots,2C(d-1),0)\), it follows that  \(R(q;t,\nu)\) is upper-bounded by the RHS of~\eqref{eq:qiforExplicitBoundGen}. Since \(R(q;t,\nu^{*})\le R(q;t,\nu)\), this implies~\eqref{eq:qiforExplicitBoundGen}.
\commentt{\qed}{\Halmos}
\section{Proof of Proposition~\ref{pr:explicitDistribution}}
Since \(R(q;t,\nu)\) is increasing with respect to \(\nu\), \begin{eqnarray*}
R(q;t,\nu^{*})&\geq& R(q;t,C(0),\dots,C(d))\\
&=&(\sum^{d-1}_{i=0}\frac{C(i)-C(i+1)}{q_{i}})(\sum^{d-1}_{i=0}q_{i}(t_{i+1}-t_{i})) .\end{eqnarray*}
However, as  \(\sum^{d-1}_{j=0}q_{j}
(t_{j+1}-t_{j})\ge q_{i}t_{i+1}\) for \(0\leq i\leq d-1\), and since \((C(i))\) is a decreasing sequence,
\begin{equation*}
(\sum^{d-1}_{i=0}\frac{C(i)-C(i+1)}{q_{i}})(\sum^{d-1}_{j=0}q_{j}(t_{j+1}-t_{j}))
\ge\sum^{d-1}_{i=0}(C(i)-C(i+1))t_{i+1}.
\end{equation*}
This implies~\eqref{eq:prWLowerB} since 
\begin{equation}\label{eq:integrationParParties}
\sum^{d-1}_{i=0}(C(i)-C(i+1))t_{i+1}
=\sum^{d-1}_{i=0}C(i)(t_{i+1}-t_{i}).
\end{equation}

Assume now that \(q_{i}=t_{1}/t_{i+1}\) for \(0\leq i\leq d-1\). Since \(R(q;t,\nu)\) is increasing with respect to \(\nu\), 
\begin{eqnarray*}
R(q;t,\nu^{*})&\leq& R(q;t,2C(0),\dots,2C(d))\\
&=&2(\sum^{d-1}_{i=0}\frac{C(i)-C(i+1)}{q_{i}})(\sum^{d-1}_{i=0}q_{i}(t_{i+1}-t_{i}))\\ &=&2(\sum^{d-1}_{i=0}(C(i)-C(i+1))t_{i+1})(\sum^{d-1}_{i=0}\frac{t_{i+1}-t_{i}}{t_{i+1}}) .\end{eqnarray*}
Furthermore, as \((t_{i+1}-t_{i})/t_{i+1}\leq \ln (t_{i+1}/t_{i})\) for \(1\leq i\leq d-1\),\begin{displaymath}
\sum^{d-1}_{i=0}\frac{t_{i+1}-t_{i}}{t_{i+1}}\leq1+\ln (\frac{t_{d}}{t_{1}}).
\end{displaymath}  
Using~\eqref{eq:integrationParParties} once again yields~\eqref{eq:prWUpperB}. \commentt{\qed}{\Halmos}
\section{Relation with the ANOVA decomposition and the truncation dimension}
\label{se:anova}This section assumes that \(f\) is a square-integrable function on \([0,1]^{d}\), and that each   \(U_{i}\) is  uniformly distributed on \([0,1]\), with \(\var(f(U))>0\). Consider a decomposition of   \(f\) in the following form:\begin{equation}\label{eq:ANOVA}
f=\sum_{Y\subseteq\{1,\dots,d\}}f_{Y},
\end{equation} where \(f_Y\) is a measurable function on \([0,1]^{d}\) and  \(f_{Y}(u)\) depends on \(u\) only through  \((u_{j})_{j\in Y}\),  for \(u=(u_{1},\dots,u_{d})\in[0,1]^{d}\). For instance, \(f_{\emptyset}\) is a constant, \(f_{\{j\}}(u)\) is a function of \(u_{j}\), and \(f_{\{j,k\}}(u)\) is a function of \((u_{j},u_{k})\). The relation \eqref{eq:ANOVA} is called ANOVA representation of \(f\) if, for \(Y\subseteq\{1,\dots,d\}\),  any vector \(u\in[0,1]^{d}\), and any  \(j\in Y\),
\begin{equation*}
\int^{1}_{0} f_{Y}(u_{1},\dots,u_{j-1},x,u_{j+1},\dots,u_{d})\,dx=0.
\end{equation*}
It can be shown~\citep[p. 272]{sobol2001global} that there is a unique ANOVA representation of \(f\), that  the \(f_{Y}\)'s are square-integrable, and that   \begin{displaymath}
\var(f(U))=\sum_{Y\subseteq\{1,\dots,d\}}\sigma_{Y}^{2},
\end{displaymath}
where \(\sigma_{Y}\) is the standard deviation of \(f_{Y}(U)\). Furthermore,  for \(0\leq i\leq d-1\),    
\begin{equation*} 
E(f(U)|U_{i+1},\ldots,U_{d})=\sum_{Y\subseteq\{i+1,\dots,d\}}f_{Y}(U),
\end{equation*} and the covariance between \(f_{Y}(U)\) and \(f_{Y'}(U)\) is null if \(Y\neq Y'\). Hence,   for \(0\leq i\leq d-1\), 
\begin{equation}\label{eq:CiGlobalSensitivity}
C(i)=\sum_{Y\subseteq\{i+1,\dots,d\}}\sigma_{Y}^{2}.
\end{equation}It follows that \(C(i)\) is equal to the variance corresponding to the subset \({\{i+1,\dots,d\}}\), as defined in~\cite[Eq. 4]{sobol2001global}.
 Thus, \(C(i)/C(0)\) is equal to the global sensitivity index \(S_{\{i+1,\dots,d\}}\) for the subset \(\{i+1,\dots,d\}\)  (see~\cite[Definition 3]{sobol2001global}).
  
Proposition~\ref{pr:effectiveDim} below  relates the performance of our algorithm to the truncation dimension \(d_{t}\) of \(f\),
  defined in~\citep{owen2003} as\begin{equation*}
d_{t}\triangleq \frac{\sum_{Y\subseteq\{1,\dots,d\},Y\neq\varnothing}\max(Y)\sigma_{Y}^{2}}{\var(f(U))}.
\end{equation*}
Under the conditions in Proposition~\ref{pr:effectiveDim}, if  \(t_{d}=\Theta(d)\) and \(d_{t}\) is upper bounded by a constant independent of \(d\), the asymptotic (as \(n\) goes to infinity) work-normalized variance  of our algorithm
is \(O(\ln(d)\var(f(U)))\), whereas the  work-normalized variance  of the standard
Monte Carlo algorithm is  \(\Theta(d\var(f(U)))\).    
\begin{proposition}\label{pr:effectiveDim}
Assume that there is a real number \(c\) such that
\(t_{i}\le ci\) for \(1\leq i\leq d\), and that \(q_{i}=1/(i+1)\) for \(0\leq i\leq d-1\). Then  \begin{equation*}
R(q;t,\nu^{*})\leq2c(1+\ln(d))d_{t}\var(f(U)).
\end{equation*}\end{proposition} 
\commentt{\begin{proof}}{\proof{Proof.}}
For  \(0\leq i\leq d-1\), we can rewrite~\eqref{eq:CiGlobalSensitivity} as 
\begin{equation*}
C(i)=\sum_{Y\subseteq\{1,\dots,d\},Y\neq\varnothing}{\bf1}\{i<\min(Y)\}\sigma_{Y}^{2}.
\end{equation*}Hence, \begin{eqnarray*}
\sum ^{d-1}_{i=0}C(i)
&=&\sum_{Y\subseteq\{1,\dots,d\},Y\neq\varnothing}\sum^{d-1}_{i=0}{\bf1}\{i<\min(Y)\}\sigma_{Y}^{2}\\
&=&\sum_{Y\subseteq\{1,\dots,d\},Y\neq\varnothing}\min(Y)\sigma_{Y}^{2}
\\&\le& d_{t}\var(f(U)).
\end{eqnarray*}
Let \(t'_{i}=ci\) for \(0\leq i\leq d\), and \(t'=(t'_{0},\dots,t'_{d})\). The proof of~\eqref{eq:prWUpperB} shows that this relation still holds if
\(t\) is replaced by any increasing sequence of length \(d+1\) starting at
\(0\). Replacing \(t\) by \(t'\) implies that  \begin{eqnarray*}
R(q;t',\nu^{*})
&\le&2c(1+\ln(d))\sum ^{d-1}_{i=0}C(i)\\
&\le& 2c(1+\ln(d))d_{t}\var(f(U)).
\end{eqnarray*}
Moreover, \(R(q;t,\nu^{*})\le R(q;t',\nu^{*})\)  since \(t\leq t'\). This completes the proof.
\commentt{\end{proof}}{\Halmos\endproof}     
\section{Proof of Theorem~\ref{th:OptTimeVariance}}
We use the  following proposition, whose proof follows immediately from~\eqref{eq:numonotone}.\begin{proposition}
\label{pr:lowerW}If   \(\nu\in \mathbb{R}^{d}\times\{0\}\) and  \(\nu'\in \mathbb{R}^{d}\times\{0\}\)
are such that \(\nu'\leq\nu\),  and  \(q\in A\), then \(R(q;t,\nu')\leq R(q;t,\nu)\),
 with equality if  \(\nu_{0}=\nu'_{0}\) and, for \(1\leq i\leq d-1\),  \((\nu_{i}-\nu'_{i})(q_{i-1}- q_{i})=0\). \end{proposition}
 By definition of
the lower hull,  \((\theta_{i})\), \(0\leq i\leq d-1\), is an increasing sequence. Furthermore, \(\theta_{d-1}<0\) since it is equal to the slope of a segment joining \((t_{i},\nu_{i})\) to  \((t_{d},0)\), for some \(i\in[0,d-1]\). Hence \(\theta_{i}<0\) for   \(0\leq i\leq d-1\), and so \(q^{*}\) is well defined and belongs to \(A\). Furthermore,   \((\nu'_{i})\), \(0\leq i\leq d\), is a decreasing sequence, and \(\nu'_{d}=\nu_{d}=0\). On the other hand, by~\eqref{eq:RDef},
\begin{equation*}
 R(q^{*};t,\nu')=\bigg(\sum_{i=0}^{d-1}\sqrt{(\nu'_{i}-\nu'_{i+1})(t_{i+1}-t_{i})}\bigg
)^2.\end{equation*} Since, by the Cauchy-Schwartz inequality, for all non-negative sequences \((x_{i})\) and \((y_{i})\),\begin{displaymath}
(\sum^{d-1}_{i=0}\sqrt{x_{i}y_{i}})^{2}\le(\sum^{d-1}_{i=0}x_{i})(\sum^{d-1}_{i=0}y_{i}),
\end{displaymath}it follows that \(R(q^{*};t,\nu')\leq R(q;t,\nu')\) for \(q\in A\). Furthermore, by Proposition~\ref{pr:lowerW},   \(R(q;t,\nu')\le\ R(q;t,\nu)\), and so  \(R(q^{*};t,\nu')\leq R(q;t,\nu)\). On the other hand,  \((\nu_{i}-\nu'_{i})(q^{*}_{i-1}- q^{*}_{i})=0\) for \(1\leq i\leq d-1\). This is because, if  \(\nu_{i}\neq\nu'_{i}\), then the point \((t_{i},\nu_{i})\) does not belong to the lower hull of the set  \(\{(t_{j},\nu_{j}):0\leq i\leq d\}\). Hence  \((t_{i},\nu'_{i})\) belongs to the segment \(( (t_{i-1},\nu'_{i-1}),(t_{i+1},\nu'_{i+1}) \), which implies that \(\theta_{i-1}= \theta_{i}\) and \(q^{*}_{i-1}= q^{*}_{i}\).   Thus, as   \(\nu_{0}=\nu'_{0}\),   Proposition~\ref{pr:lowerW} shows that    \(R(q^{*};t,\nu)=\ R(q^{*};t,\nu')\).  This implies~\eqref{eq:optTimeVariance}
and that  \(R(q^{*};t,\nu)\leq R(q;t,\nu)\) for \(q\in A\), as desired.
\commentt{\qed}{\Halmos}
\section{Proof of Proposition~\ref{pr:CIVarReduc}}
The random vectors \(W^{(1)}=(U_{1},\ldots ,U_{i},U'_{i+1},\ldots,U'_{d})\), \(W^{(2)}=(U'_{1},\ldots ,U'_{i},U_{i+1},\ldots,U_{d})\), and \(W^{(3)}=(U'_{1},\ldots ,U'_{i},U^{''}_{i+1},\ldots,U^{''}_{d})\) have the same distribution as \(U\).
Hence \begin{displaymath}
E((f(U)-f(W^{(1)}))(f(W^{(2)})-f(W^{(3)}))=\cov(f(U)-f(W^{(1)}),f(W^{(2)})-f(W^{(3)})).
\end{displaymath}As \(W^{(1)}\) and \(W^{(2)}\) are independent,   \begin{displaymath}
\cov(f(W^{(1)}),f(W^{(2)})=0.
\end{displaymath} Similarly,    \begin{displaymath}
\cov(f(W^{(1)}),f(W^{(3)})=\cov(f(U),f(W^{(3)})=0.
\end{displaymath} By bilinearity of the covariance, it follows that \begin{displaymath}
E((f(U)-f(W^{(1)}))(f(W^{(2)})-f(W^{(3)}))=\cov(f(U),f(W^{(2)}).
\end{displaymath}
We conclude the proof using Proposition~\ref{pr:covf(U)f(U')}.
\commentt{\qed}{\Halmos}
\section{Proof of Proposition~\ref{pr:MarkovRevuz}}
We first prove the following Markov property.
\begin{proposition}\label{pr:MonotoneClass}
Let   \(i\in[0,d]\). If \(H\) is a bounded random variable which is measurable with respect to the \(\sigma\)-algebra generated by \(X_{i},Y_{i},\dots,Y_{d-1}\),  then  
\begin{equation}\label{eq:conditionalExpectationMarkov}
E(H|Y_{0},\dots,Y_{i-1})=E(H|X_{i}).
\end{equation}
 \end{proposition}
 \commentt{\begin{proof}}{\proof{Proof.}}
Let \(\mathcal{H}\) be the vector space of bounded real-valued random variables \(H\) satisfying~\eqref{eq:conditionalExpectationMarkov}. Clearly, the constant random variables belong to  \(\mathcal{H}\). Let \((H_{m})\), \(m\geq0\), be an increasing sequence of  positive elements of    \(\mathcal{H}\) such that \(H=\sup_{m\geq0}H_{m}\) is bounded. For \(m\geq0\), \begin{equation}
\label{eq:Hn}E(H_{m}|Y_{0},\dots,Y_{i-1})=E(H_{m}|X_{i}).
\end{equation} By the conditional  Lebesgue dominated convergence theorem~\cite[Theorem 2, p. 218]{shiryaev1996probability}, the LHS (resp. RHS) of~\eqref{eq:Hn}   converges to \(E(H|Y_{0},\dots,Y_{i-1})\)  (resp. \(E(H|X_{i})\)) as \(m\) goes to infinity, and so \(H\in\mathcal{H}\).

 Let \(\mathcal{G}\)  (resp.  \(\mathcal{G}')\)  be the set of  bounded real-valued  random variables which are measurable with respect to the \(\sigma\)-algebra generated by \(X_{i}\) (resp. \((Y_{i},\dots,Y_{d-1})\)),  and let \(\mathcal{C}\) be the set of random variables of the form \(GG'\), with     \(G\in\mathcal{G}\) and \(G'\in\mathcal{G}'\).
For      \(G\in\mathcal{G}\) and \(G'\in\mathcal{G}'\), \begin{eqnarray*}
E(GG'|Y_{0},\dots,Y_{i-1})
&=&GE(G'|Y_{0},\dots,Y_{i-1})
\\
&=&GE(G').
\end{eqnarray*} 
The first equation holds since \(X_{i}\) is a measurable function of \(Y_{0},\dots,Y_{i-1}\), which implies that \(G\)  is measurable with respect to the \(\sigma\)-algebra generated by \(Y_{0},\dots,Y_{i-1}\). The second equation follows from the independence of  \((Y_{i},\dots,Y_{d-1})\) and \((Y_{0},\dots,Y_{i-1})\). Similarly, since  \((Y_{i},\dots,Y_{d-1})\) and  \(X_{i}\) are independent,
\begin{equation*}
E(GG'|X_{i})=GE(G'|X_{i})=GE(G').
\end{equation*}
Thus, \(GG'\in\mathcal{H}\),  and so \(\mathcal{C}\subseteq\mathcal{H}\). As \(\mathcal{C}\) is closed under pointwise multiplication, by the monotone class theorem~\cite[Theorem 2.2, p. 3]{Yor99}, \(\mathcal{H}\) contains all bounded random variables which are measurable with respect to the \(\sigma\)-algebra generated by the elements of \(\mathcal{C}\). Since \(Y_{i},\dots,Y_{d-1}\) and \(X_{i}\) belong to \(\mathcal{C}\), it follows that  \(\mathcal{H}\) contains all bounded random variables which are measurable with respect to the \(\sigma\)-algebra generated by \(X_{i},Y_{i},\dots,Y_{d-1}\). This completes the proof.
\commentt{\end{proof}}{\Halmos\endproof}
  
  As \(g(X_{d})\) is a measurable function of \((X_{i},Y_{i},\dots,Y_{d-1})\), for any integer \(m\), the random variable \(\min(m,g^{+}(X_{d}))\) belongs to \(\mathcal{H}\). By the conditional  Lebesgue dominated convergence theorem, taking the limit as \(m\) goes to infinity implies that
\begin{equation*}
E(g^{+}(X_{d})|Y_{0},\dots,Y_{i-1})=E(g^{+}(X_{d})|X_{i}).
\end{equation*} A similarly equation holds for \(g^{-}(X_{d})\).  Thus, \begin{equation*}
E(g(X_{d})|Y_{0},\dots,Y_{i-1})=E(g(X_{d})|X_{i}).
\end{equation*}
Replacing \(i\) with \(d-i\) implies that  
\begin{equation*}
E(g(X_{d})|U_{i+1},\dots,U_{d})=E(g(X_{d})|X_{d-i}).
\end{equation*}Taking the variance of both sides concludes the proof.
\commentt{\qed}{\Halmos}
\section{Proof of Proposition~\ref{pr:GARCH}}
For \(0\leq i\leq d-1\), set \(Z_{i}=\alpha Y_{i}^{2}+\beta\), so that \(X_{i+1}= w+Z_{i}X_{i}\).  It can be shown by induction on \(i\) that  \(X_{d}= w+ w Z_{d-1}Z'+Z'' \) for \(2\leq i\leq d\),  where 
 \begin{equation*}
Z'=\sum^{i-1}_{j=1}\prod^{j}_{k=2}Z_{d-k}
\end{equation*}
  and
\begin{equation*}
Z''=Z_{d-i}\cdots Z_{d-1}X_{d-i}.
\end{equation*}
   By convention,  the product over an empty set is equal to \(1\). Since \(X_{i,0}\) is equal to the state of the Markov chain at step \(d\) if \(X_{d-i}=0\), it follows that \( X_{i,0}= w+ w Z_{d-1}Z'\). Note that \(E(Z'')=(\alpha+\beta)^{i}E(X_{d-i})\) as \(E(Y_{i}^{2})=1\). By~\citep[Eq. 23.13]{Hull14}, for \(0\leq m\leq d\), \begin{equation*}
E(X_{m})\leq\max(X_{0},\frac{ w }{1-\alpha-\beta}),
\end{equation*} and so \begin{displaymath}
E(Z'')\leq\max(X_{0},\frac{ w }{1-\alpha-\beta})(\alpha+\beta)^{i}.
\end{displaymath} Since the density of \(Y_{d-1}\) is upper-bounded by \(1/2\),
for \(\gamma\leq\gamma'\),
\begin{eqnarray}\label{eq:zdmdensity}
\mathbb{P}(\gamma\leq Z_{d-1}\leq \gamma')
\nonumber&=&2 \mathbb{P}(\sqrt{\frac{(\gamma-\beta)^{+}}{\alpha}}\leq Y_{d-1}\leq \sqrt{\frac{(\gamma'-\beta)^{+}}{\alpha}})\\
\nonumber&\le& \sqrt{\frac{(\gamma'-\beta)^{+}}{\alpha}} - \sqrt{\frac{(\gamma-\beta)^{+}}{\alpha}}\\
&\le& \sqrt{\frac{\gamma'-\gamma}{\alpha}}.
 \end{eqnarray} 
The last equation follows from the inequality \(\sqrt{y'}-\sqrt{y}\leq\sqrt{y'-y}\), which holds for \(0\leq y\leq y' \). On the other hand, as \(X_{i,0}\leq X_{d}\), \begin{displaymath}
g(X_{d})-g(X_{i,0}) = \begin{cases} 1 &\mbox{if } X_{i,0}\leq z<X_{d}, \\ 
0 & \mbox{otherwise.} \end{cases}
\end{displaymath}Thus, by~\eqref{eq:UpperCiMarkovL2},\begin{eqnarray*}
C(i)&\le&||g(X_{d})-g(X_{i,0})||^{2}\\
&=& 
\mathbb{P}(X_{i,0}\leq z< X_{d}).
 \end{eqnarray*}But 
\begin{eqnarray*}
\mathbb{P}(X_{i,0}\leq z< X_{d})
&=& \mathbb{P} (\frac{z-Z''- w}{ w Z'}<Z_{d-1}\leq \frac{z- w}{ w Z'})\\
&=& E(\mathbb{P} (\frac{z-Z''- w}{ w Z'}<Z_{d-1}\leq \frac{z- w}{ w Z'}|Z',Z''))\\&\le&E(\sqrt{\frac{Z''}{\alpha w Z'}})\\
&\le&\kappa(\alpha+\beta)^{i/2} ,\end{eqnarray*} 
 where \(\kappa\) is a constant  that depends only on \( w\), \(X_{0}\), \(\alpha\) and \(\beta\). The third equation follows from~\eqref{eq:zdmdensity} and the independence of \(Z_{d-1}\) and \((Z',Z'')\). The last equation follows from Jensen's inequality and the inequality \(Z'\geq1\).
Hence \(C(i)\le\kappa(\alpha+\beta)^{i/2}\) for \(2\leq i\leq d\). This inequality also holds for \(0\leq i\leq1\) by replacing \(\kappa\) with \(\max(\kappa,1/(\alpha+\beta))\).
\commentt{\qed}{\Halmos}
\comment{
\begin{proposition}
If \(Y\) and \(Y'\) are independent and \(X\) is a measurable function of  \(Y\), and \(X'\) is a measurable function of \((X,Y')\), then \(E(X'|Y)=E(X'|X)\).
\end{proposition}
}
\section{Proof of Proposition~\ref{pr:GD1}}
By classical calculations (see, e.g., \citep[\S I, Eq.~(1.4)]{asmussenGlynn2007}), it can be shown by induction on \(d\) that \(X_{d}=\max _{0\leq j\leq d}S_{j}\), where  \(S_{j}=\sum^{d-1}_{k=d-j}Y_{k}\) for \(1\leq  j\le d\), with \(S_{0}=0\). For  \(0\leq i\leq d-1\), since \(X_{i,0}\)  is the number of customers in the queue at time-step \(d\) if there are no costumers in the queue at time-step \(d-i\), it can be shown by induction on \(d\) that \(X_{i,0}=\max _{0\leq j\leq i}S_{j}\).  Hence \begin{displaymath}
X_{d}=\max(X_{i,0},\max _{i+1\leq j\leq d}{S_{j}}).
\end{displaymath} Thus 
\begin{equation*}
X_{d}-X_{i,0}\leq \max _{i+1\leq j\leq d}{S_{j}}^{+},
\end{equation*}and so \begin{eqnarray*}
C(i)&\leq&||X_{d}-X_{i,0}||^{2}\\
&\leq& \sum _{j=i+1}^d||{S_{j}}^{+}||^{2}.
\end{eqnarray*}
On the other hand, since the \(Y_{i}\)'s are independent,   \(E(e^{\gamma S_{j}})\leq\kappa^{j}\) for \(0\leq j\leq d\). Furthermore, as \((x^{+})^{2}\leq2e^{x}\) for \(x\in\mathbb{R}\), \(\gamma^{2}{(S_{j}^{+})}^{2}\leq 2e^{\gamma S_{j}}\). Taking expectations implies that \(\gamma^{2}||{S_{j}}^{+}||^{2}\leq 2\kappa^{j}  
\), and so \(C(i)\leq \gamma'\kappa^{i}\), where \(\gamma'=2\gamma^{-2}/(1-\kappa)\). \commentt{\qed}{\Halmos}
\section{Proof of Proposition~\ref{pr:MG1}}
It can be shown by induction on the number of arrivals in \((\theta',\theta]\)  (see also~\citep{whitt2017RareEvent})  that, if the system is empty at time \(\theta'\in[0,\theta]\), 
\begin{equation*}
W_{\theta}=\sup_{s\in[\theta',\theta]}\{Z_{\theta}(s)-(\theta -s)\}.
\end{equation*}
 Hence 
\begin{equation*}
X_{d}=\sup_{s\in[0,\theta]}\{Z_{\theta}(s)-(\theta -s)\}.
\end{equation*} Also, for \(0\leq i\leq d\), since \(X_{i,0}\) is the residual work at time  \(\theta\) if the residual work at time \((d-i)\theta/d\) is \(0\), \begin{equation*}
X_{i,0}=\sup_{s\in[(d-i)\theta/d,\theta]}\{Z_{\theta}(s)-(\theta -s)\}.
\end{equation*}   Thus,
by a calculation similar to the proof of Proposition~\ref{pr:GD1},\begin{equation*}
X_{d}-X_{i,0}\leq \sup_{0\leq s\leq(d-i) \theta/d}\{Z_{\theta}(s)-(\theta -s)\}^{+}.
\end{equation*}For non-negative integer \(j\), set \begin{displaymath}
S_{j}=Z_{\theta}((\theta-\frac{j}{\lambda ^{*}})^{+})-\frac{j-1}{\lambda ^{*}}.
\end{displaymath}Fix  \(s\in[0,(d-i) \theta/d]\), and let \(j=\lceil(\theta-s)\lambda^{*}\rceil\). As \(\theta-{j}/{\lambda ^{*}}\leq s\) and \(Z_{\theta}\) is a decreasing function, 
\begin{displaymath}
Z_{\theta}(s)\leq Z_{\theta}((\theta-\frac{j}{\lambda ^{*}})^{+}).
\end{displaymath} Since \(\theta-s\geq(j-1)/\lambda ^{*}\), this implies that
\(Z_{\theta}(s)-(\theta -s)\leq S_{j}.
\) But  \(2j\geq i\) as  \(j\geq i\lambda^{*} \theta/d\) and \(\lambda^{*}\theta\geq d/2\). Hence \begin{equation*}
X_{d}-X_{i,0}\leq \sup_{j\geq i/2}S_{j}^{+}.
\end{equation*}
By calculations similar to the proof of Proposition~\ref{pr:GD1}, it  follows that\begin{displaymath}
C(i)\leq \sum _{j\geq i/2}||{S_{j}}^{+}||^{2}.
\end{displaymath}On the other hand, \eqref{eq:prMG1} implies that   \(E(e^{\gamma (S_{j+1}-S_{j})})\leq\kappa\) for \(j\geq0\), and so  \(E(e^{\gamma S_{j}})\leq e^{\gamma /\lambda^{*}}\kappa^{j}\). We conclude the proof in a way similar to the proof of Proposition~\ref{pr:GD1}.
\commentt{\qed}{\Halmos}

\commentt{}
{
\end{APPENDIX}
\bibliography{poly}
\ECSwitch
\ECHead{Supplementary Material}
}

\section{Proof of Theorem~\ref{th:CLT}}
 If \(\mathcal{C}\) is a collection of random variables, denote by \(\sigma[\mathcal{C}]\) the  \(\sigma\)-algebra generated by the elements of \(\mathcal{C}\). Let \((\xi_{k})\), \(k\geq1\), be a stationary sequence of real-valued random variables. For \(k\geq1\), define the \(\sigma\)-algebras   \(\mathcal{F}_{k}=\sigma[\xi_{n}:1\leq n\leq k]\) and   \(\mathcal{F}'_{k}=\sigma[\xi_{n}: n\geq k]\).  Following~\cite[p.~203]{billingsley1999convergence},
for \(n\geq1\), let \begin{displaymath}
\varphi_{n}=\sup_{k\geq1}\sup\{\mathbb{P} (B'|B)-\mathbb{P} (B'):B\in\mathcal{F}_{k}, \mathbb{P}(B)>0, B'\in\mathcal{F}'_{k+n}\}.
\end{displaymath}
The sequence \((\varphi_{n})\) is used to study the mixing proprieties of the sequence  \((\xi_{k})\). Theorem~\ref{th:Billingsley} below establishes a functional central limit theorem on the partial sums of   \((\xi_{k})\)
under certain conditions on \((\varphi_{n})\).
Theorem~\ref{th:Billingsley}  follows immediately from    Theorem~19.2 and the discussion on p. 203 of~\citep{billingsley1999convergence}. Let \(D_{\infty}\) denote the set of real-valued functions on the interval \([0,\infty)\)  that are right-continuous and have left-hand limits, endowed with the Skorokhod topology (see~\cite[Section 16]{billingsley1999convergence}).   
\begin{theorem}[\citet{billingsley1999convergence}]\label{th:Billingsley} Assume that  \((\xi_{k})\), \(k\geq1\), is stationary, that    \(\xi_{1}\) is square-integrable, and that 
\begin{equation}
\label{eq:mixingphin}\sum_{n=0}^{\infty}\sqrt{\varphi_{n}}<\infty.
\end{equation} 
Then the series \begin{equation}\label{eq:defSigma}
\bar\sigma^{2}=\var(\xi_{1})+2\sum^{\infty}_{k=2}\cov(\xi_{1},\xi_{k})
\end{equation}
converges absolutely. If \(\bar\sigma>0\)  and \(S_{n}=\sum^{n}_{k=1}(\xi_{k}-E(\xi_{k}))\) then, as  \(n\rightarrow\infty\), \begin{equation}\label{eq:functionnalCLT}
\frac{S_{\lfloor ns\rfloor}}{\sqrt{n}}\Rightarrow \bar\sigma B_{s}
\end{equation}in the sense of \(D_{\infty}\), where \(B_{s}\) is a standard Brownian motion.\end{theorem} 
We show that the conditions of Theorem~\ref{th:Billingsley} hold when \(\xi_{k}=f(V^{(k)})\) for \(k\geq1\).
As \((V^{(k)})\), \(k\geq1\), is a stationary Markov chain, the sequence \((f(V^{(k)}))\), \(k\geq1\), is stationary. We first prove the following lemma, which follows intuitively from the fact that if all components of  the copy of \(U\)  are redrawn at a certain iteration, future copies of \(U\) are independent of past ones.
\begin{lemma}\label{le:indep} For positive integers \(n\) and \(k\), if \(B\in\mathcal{F}_{k} \) and \( B'\in\mathcal{F}'_{k+n}\), then \(B\) and \(B'\cap I\) are independent, where \begin{displaymath}
I=I_{k,n}=\{\omega\in\Omega:\exists j\in[k,k+n-1],N_{j}(\omega)=d\}.
\end{displaymath}\end{lemma}
\commentt{\begin{proof}}{\proof{Proof.}}
   Let \(\mathcal{G}_{k}=\sigma[N_{j},U^{(j+1)}:j\geq k]\).  
  Fix   \(l\geq k+n\). By construction, if \(\omega\in I\), there is \(j\in[k,k+n-1]\) is such that  \(V^{(j+1)}(\omega)=U^{(j+1)}(\omega)\).  Hence,  if   \(\omega\in I\), the vector  \(V^{(l)}\) does not depend \(V^{(k)}\). More precisely, it can be shown by induction that there is a measurable function \(H:F^{d(l-k)}\times\mathbb{R}^{(l-k)}\rightarrow\mathbb{F}^{d}\) such that, for  \(\omega\in I\),  
\begin{equation*}
V^{(l)}(\omega)=H(U^{(k+1)}(\omega),\dots,U^{(l)}(\omega),N_k(\omega),\dots,N_{l-1}(\omega)).
\end{equation*}
For \(\omega\in\Omega\), let
 \begin{equation*}
G(\omega)=f(H(U^{(k+1)}(\omega),\dots,U^{(l)}(\omega),N_k(\omega),\dots,N_{l-1}(\omega))).
\end{equation*}   Thus,     \(f(V^{(l)}(\omega))=G(\omega)\) for  \( \omega \in I\). As  the maps \(f\) and \(H\) are measurable, and since the random variables \(U^{(k+1)},\dots,U^{(l)}\) and \(N_k,\dots,N_{l-1}\) are measurable with respect  to \(\mathcal{G}_{k}\),  the random variable \(G\) is measurable with respect  to \(\mathcal{G}_{k}\) as well. 
  For  \(z\in\mathbb{R}\),  let \(B_{z}=\{\omega\in\Omega:f(V^{(l)}(\omega))<z\}\). Since  \(I\in\mathcal{G}_{k}\) and\begin{displaymath}
B_{z}\cap I=\{\omega\in \Omega:G(\omega)<z\}\cap I,
\end{displaymath}  we have \(B_{z}\cap I\in\mathcal{G}_{k}\). Thus \(B_{z}\in\mathcal{G}'_{k}\) for any real number \(z\), where\begin{displaymath}
\mathcal{G}'_{k}=\{B\in\mathcal{F}:B\cap I\in\mathcal{G}_{k}\}.
\end{displaymath}It is easy to check that  \(\mathcal{G}'_{k}\) is a \(\sigma\)-algebra.  We conclude that  \(f(V^{(l)})\) is measurable with respect to \(\mathcal{G}'_{k}\), for any   \(l\geq k+n\), and so   \(\mathcal{F}'_{k+n}\subseteq \mathcal{G}'_{k}\).
Consequently, if \(B\in\mathcal{F}_{k} \) and \( B'\in\mathcal{F}'_{k+n}\), then  \(B'\in\mathcal{G}'_{k}\), and so \(B'\cap I\in\mathcal{G}_{k}\). Since  \(\mathcal{F}_{k}\) and \(\mathcal{G}_{k}\) are independent, so are   \(B\) and \(B'\cap I\). 
\commentt{\end{proof}}{\Halmos\endproof}
We now prove the theorem. For positive integers \(n\) and \(k\), let \(B\in\mathcal{F}_{k} \) with \(\mathbb{P}(B)>0\) and \( B'\in\mathcal{F}'_{k+n}\), and define \(I\) as in Lemma~\ref{le:indep}. Thus,\begin{displaymath}
\mathbb{P}(B'\cap I|B)=\mathbb{P}(B'\cap I).
\end{displaymath}
 Since   \begin{displaymath}
\mathbb{P}( B')\le\mathbb{P}(B'\cap I)+\mathbb{P}( I^{c}),
\end{displaymath} 
it follows that
\begin{displaymath}
\mathbb{P}(B')\le\mathbb{P}( B'|B)+\mathbb{P}( I^{c}).
\end{displaymath} 
Replacing \(B'\) with its complement and noting that   \(\mathbb{P}( I^{c})=(1-q_{d-1})^{n}\), we conclude that
\begin{displaymath}
|\mathbb{P}( B'|B)-\mathbb{P}(B')|\leq (1-q_{d-1})^{n}.
\end{displaymath}  Hence  \(\varphi_{n}\leq  (1-q_{d-1})^{n}\), and so \eqref{eq:mixingphin} holds. Thus the conclusions of Theorem~\ref{th:Billingsley} hold for the sequence \((f(V^{(k)}))\), \(k\ge1\). It follows from~\eqref{eq:infiniteSumaj} and~\eqref{eq:defSigma} that \(\bar \sigma=\sigma\).  
By~\eqref{eq:functionnalCLT},  as  \(n\rightarrow\infty\),

 \begin{equation}\label{eq:functionnalCLTfn}
\frac{{\lfloor ns\rfloor}(f_{\lfloor ns\rfloor}-E(f(U)))}{\sqrt{n}}\Rightarrow \bar\sigma B_{s}
\end{equation} 
 in \(D_{\infty}\). Setting \(s=1\), which can be justified by applying the  continuous mapping theorem~\cite[Theorem 2.7]{billingsley1999convergence}  with the projection map~\cite[Theorem 16.6]{billingsley1999convergence},   implies~\eqref{eq:simpleCLT}. 
Let \(\tau_{k}\) denote the
random amount time to generate \(V^{(k)}\) and calculate \(f(V^{(k)})\). Thus \(E(\tau_{k})=T\). By the strong law of large numbers, with probability \(1\), \begin{equation}\label{eq:SLLN}
\frac{1}{n}\sum^{n}_{k=1}\tau_{k}\rightarrow T
\end{equation} as \(n\) goes to infinity.  By~\eqref{eq:functionnalCLTfn},~\eqref{eq:SLLN} and~\cite[Theorem 1]{glynn1992asymptotic}, it follows that \begin{equation*}
\sqrt{c}(f_{\tilde N(c)}-E(f(U))) \Rightarrow N(0,T\sigma^{2} ),
\end{equation*} as \(c\rightarrow\infty\). Since \(T\sigma^2 = R(q;t, \nu^{*})\), this implies~\eqref{eq:CLTComputingTime}.

\section{Proof of Theorem~\ref{th:DDR} and of Proposition~\ref{pr:refereeDeterministic}}
\label{se:ProofDeterministic}For convenience, set \(N_{0}=\bar N_{0}=d\). For \(0\leq i\leq d-1\), let\begin{equation}\label{eq:defSi}
\mathcal{S}_{i}=\mathcal{S}_{i}(n)=\{k\in[0,n-1]:N_{k}>i\}.
\end{equation} Let  \(u_{i}(0)=0,u_{i}(1),\dots,u_{i}(|\mathcal{S}_{i}|-1)\) be the elements of \(\mathcal{S}_{i}\)  sorted in increasing order.  Set  \(u_{i}(|\mathcal{S}_{i}|)=n\) and let \(Q(i)=Q(i,n)=\sum^{|\mathcal{S}_{i}|}_{l=1}(u_{i}(l)-u_{i}(l-1))^{2}\). Note that \(\mathcal{S}_{0}=\{0,\dots,n-1\}\) and \(Q(0)=n\). The following lemma gives an alternative characterization of \(Q(i)\).
\begin{lemma}\label{le:balance}
For \(0\leq i\leq d-1\),
\begin{displaymath}
\sum_{1\leq m<k\leq n}{\bf1} \{\max_{m\leq j<k}N_{j}\le i\}=\frac{1}{2}\big(Q(i)-n).
\end{displaymath}\end{lemma}
\commentt{\begin{proof}}{\proof{Proof.}}
Given integers \(m\) and \(k\) with \(1\leq m<k\leq n\), the condition \(\max_{m\leq j<k}N_{j}\le i\) holds if and only if \([m,k)\cap\mathcal{S}_{i}=\emptyset\).
It is thus equivalent to the existence of an integer \(l\in[1,|\mathcal{S}_{i}|]\) such that \(u_{i}(l-1)<m<k\leq u_{i}(l)\).
Hence \begin{displaymath}
\sum_{1\leq m<k\leq n}{\bf1} \{\max_{m\leq j<k}N_{j}\le i\}=\frac{1}{2}\sum^{|\mathcal{S}_{i}|}_{l=1}(u_{i}(l)-u_{i}(l-1))(u_{i}(l)-u_{i}(l-1)-1).
\end{displaymath} 
We conclude the proof by noting that \begin{displaymath}
\sum^{|\mathcal{S}_{i}|}_{l=1}(u_{i}(l)-u_{i}(l-1))=n.
\end{displaymath} \commentt{\end{proof}}{\Halmos\endproof}      
Lemma~\ref{le:varfndUpperBound} below relates the variance of \(f_{n}\) to the \(Q(i)\)'s and \(C(i)\)'s.
\begin{lemma}\label{le:varfndUpperBound}
For \(n\geq1\),
\begin{equation*}
\var(f_{n})=n^{-2}\sum_{i=0}^{d-1}Q(i)(C(i)-C(i+1)).
\end{equation*}
\end{lemma}
\commentt{\begin{proof}}{\proof{Proof.}}
As in the  proof of Theorem~\ref{th:variance}, we  assume without loss of generality that \(E(f(U))=0\). Let  \(m<k\) be two integers in \([1,n]\). By arguments similar to those that lead to~\eqref{eq:covmk}, \begin{displaymath}
E(f(V^{(m)})f(V^{(k)}))=C(\max_{m\leq j<k}N_{j}).
\end{displaymath}
Since, for \(0\leq l\leq d\), \begin{eqnarray*}
C(l)&=&\sum_{i=l}^{d-1}(C(i)-C(i+1))\\
&=&\sum_{i=0}^{d-1}{\bf1}\{l\leq i\}(C(i)-C(i+1)),
\end{eqnarray*}it follows that \begin{equation*}
E(f(V^{(m)})f(V^{(k)}))=\sum_{i=0}^{d-1}{\bf1} \{\max_{m\leq j<k}N_{j}\le i\}(C(i)-C(i+1)).
\end{equation*}
As
 \begin{equation*}
\var(f_{n})=n^{-2}(\sum_{m=1}^nE((f(V^{(m)}))^{2})+2\sum_{1\leq m<k\leq n}E(f(V^{(m)})f(V^{(k)}))),
\end{equation*}
and \(E((f(V^{(m)}))^{2})=C(0)\) for \(1\leq m\leq n\), we conclude that \begin{equation*}
\var(f_{n})=n^{-2}(nC(0)+2\sum_{i=0}^{d-1}\sum_{1\leq m<k\leq n}{\bf1} \{\max_{m\leq j<k}N_{j}\le i\}(C(i)-C(i+1))).
\end{equation*}
Thus, by Lemma~\ref{le:balance},

\begin{equation*}
\var(f_{n})=n^{-2}(nC(0)+\sum_{i=0}^{d-1}(Q(i)-n)(C(i)-C(i+1))),
\end{equation*}which, after some simplifications, completes the proof.
\commentt{\end{proof}}{\Halmos\endproof}     
We now prove Theorem~\ref{th:DDR}. By the Cauchy-Schwartz inequality, for \(0\leq i\leq d-1\) and  any sequence \((x_{l})\), \(1\leq l\leq \mathcal{S}_{i}\), \begin{displaymath}
(\sum^{|\mathcal{S}_{i}|}_{l=1}x_{l})^{2}\le|\mathcal{S}_{i}|(\sum^{|\mathcal{S}_{i}|}_{l=1}x_{l}^{2}).
\end{displaymath}
Replacing \(x_{l}\) with \(u_{i}(l)-u_{i}(l-1)\) shows that \(n^{2}\leq|\mathcal{S}_{i}|Q(i)\). By Lemma~\ref{le:varfndUpperBound}, it follows that \begin{equation*}
\var(f_{n})\geq \sum_{i=0}^{d-1}\frac{C(i)-C(i+1)}{|\mathcal{S}_{i}|}.
\end{equation*}
 For \(0\leq k\leq n-1\), the expected running time of iteration \(k+1\) is \(t_{i}\) if \(N_{k}=i\). Furthermore, for \(i\in[1,d]\), there are \(|\mathcal{S}_{i-1}|-|\mathcal{S}_{i}|\)  integers \(k\) in \([0,n-1]\) with \(N_{k}=i\), where \(\mathcal{S}_{d}\triangleq\emptyset\), and so
\begin{equation}\label{eq:TimeDetermin}
T_{n}=\sum^{d}_{i=1}(|\mathcal{S}_{i-1}|-|\mathcal{S}_{i}|)t_{i}=\sum^{d-1}_{i=0}|\mathcal{S}_{i}|(t_{i+1}-t_{i}).
\end{equation}Thus, \begin{equation*}
T_{n}\var(f_{n})\geq R(q;t,C),
\end{equation*}where \(q_{i}=|\mathcal{S}_{i}|/n\) for \(0\leq i\leq d-1\). This implies~\eqref{eq:TvarLowerBoundDet}. 

Assume now that \(N_{k}=\bar N_{k}\) for \(k\geq1\).
Then, by~\eqref{eq:modulo}, for \(0\leq i\leq d-1\),\begin{equation}\label{eq:SiVander}
\mathcal{S}_{i}=\{k\in[0,n-1]:  k \text{ is a multiple of }\mu_{i}\}.
\end{equation}  Thus  \(|\mathcal{S}_{i}|=1+\lfloor(n-1)\bar q_{i}\rfloor\) which, by~\eqref{eq:TimeDetermin}, implies~\eqref{eq:TimeDeterminSpec}.   Since  \(u_{i}(l)=l\mu_{i}\) for \(0\leq l\leq |\mathcal{S}_{i}|-1\),  
\begin{eqnarray*} 
Q(i)
&=&\mu_{i}\sum^{|\mathcal{S}_{i}|-1}_{l=1}(u_{i}(l)-u_{i}(l-1))
+(n-\max(\mathcal{S}_{i}))^{2} \\&=&\mu_{i}\max\mathcal{S}_{i}+(n-\max(\mathcal{S}_{i}))^{2}.
\end{eqnarray*} 
As \(0\leq n-\max(\mathcal{S}_{i})\leq \mu_{i}\), it follows that \(\mu_{i}( n-\mu_{i})\leq Q(i)\leq \mu_{i}n.
\)
By Lemma~\ref{le:varfndUpperBound}, this implies~\eqref{eq:VarVander} and shows that, as \(n\) goes to infinity, the LHS of~\eqref{eq:VarVander} converges to its RHS. 
\commentt{\qed}{\Halmos}

We now prove Proposition~\ref{pr:refereeDeterministic}. Assume that \((N_{k})\) satisfies the conditions in the proposition, and that \(nq_{d-1}>1\).  Then \(\mathcal{S}_{d-1}\) consists of the  integers belonging to \(\{0\}\cup(n-nq_{d-1},n)\), and so \(Q(d-1)\ge n^{2}(1-q_{d-1})^{2}\). By Lemma~\ref{le:varfndUpperBound}, it follows that    \(\var(f_{n})\ge (1-q_{d-1})^{2}C(d-1)\), which completes the proof.
\commentt{\qed}{\Halmos}
\section{Proof of Theorem~\ref{th:multilevel}}
We first prove the following lemma.   
 \begin{lemma}\label{le:LowerBoundMLMC}
 Let \((\nu_{i})\), \(0\leq i\leq d\), be a decreasing sequence such that \(\nu_{m_l}\leq\var(\phi_{L}-\phi_{l})\) for \(0\leq l\leq L\), with \(\nu_{d}=0\). Then \begin{displaymath}
\sum_{i=0}^{d-1}\sqrt{\frac{\nu_{i}}{i+1}}\leq2\sum^{L}_{l=1}\sqrt{m_{l}V_{l}}.
\end{displaymath}  
\end{lemma}
\commentt{\begin{proof}}{\proof{Proof.}}
Since the sequence \((\nu_{i})\) is decreasing and \((i+1)^{-1/2}\leq2(\sqrt{i+1}-\sqrt{i})\) for   \(i\geq0\),  \begin{displaymath}
\sum_{i=j}^{k-1}\sqrt{\frac{\nu_{i}}{i+1}}
\leq2(\sqrt{k}-\sqrt{j})\sqrt{\nu_{j}},
\end{displaymath} for \(0\leq j\leq k\leq d\). Hence,
\begin{eqnarray*}
\sum_{i=0}^{d-1}\sqrt{\frac{\nu_{i}}{i+1}}
&=&\sum^{L-1}_{l=0}\sum_{i=m_{l}}^{m_{l+1}-1}\sqrt{\frac{\nu_{i}}{i+1}}
\\&\leq&2\sum^{L-1}_{l=0}(\sqrt{m_{l+1}}-\sqrt{m_{l}})\sqrt{\nu_{m_{l}}}\\
&\leq&2\sum^{L-1}_{l=0}(\sqrt{m_{l+1}}-\sqrt{m_{l}})\text{Std}(\phi_{L}-\phi_{l})\\
&=&2\sum^{L}_{l=1}\sqrt{m_{l}}(\text{Std}(\phi_{L}-\phi_{l-1})-\text{Std}(\phi_{L}-\phi_{l}))
\\&\le&2\sum^{L}_{l=1}\sqrt{m_{l}V_{l}},
\end{eqnarray*} where the last equation follows by sub-linearity of the standard deviation. \commentt{\end{proof}}{\Halmos\endproof}
We now prove Theorem~\ref{th:multilevel}.
Since \(\phi_{l}\) is square-integrable and is a measurable function of  \(U_{1},\dots,U_{m_{l}}\),
by  Proposition~\ref{pr:upperBoundingCi}, \(C(m_{l})\leq\var(\phi_{L}-\phi_{l})\) for \(0\leq l\leq L\). By Proposition~\ref{pr:covf(U)f(U')}, the sequence \((C(i))\), \(0\leq i\leq d\), is decreasing, and so it satisfies the conditions of Lemma~\ref{le:LowerBoundMLMC}. Thus,  \begin{displaymath}
\sum_{i=0}^{d-1}\sqrt{\frac{C(i)}{i+1}}\leq2\sum^{L}_{l=1}\sqrt{m_{l}V_{l}}.
\end{displaymath}Furthermore, by  Proposition~\ref{pr:varfnSqrCi}, 
\begin{equation*}
R(q;t,\nu^{*})
\leq8c\left(\sum^{d-1}_{i=0}(\sqrt{i+1}-\sqrt{i})\sqrt{C(i)}\right)^2.
\end{equation*} Since \(\sqrt{i+1}-\sqrt{i}\leq(i+1)^{-1/2}\), it follows that  
\begin{equation*}
R(q;t,\nu^{*})
\leq32 c\left(\sum^{L}_{l=1}\sqrt{m_{l}V_{l}}\right)^2.
\end{equation*} Using~\eqref{eq:MLMCTimeVariance} concludes
the proof.
\commentt{\qed}{\Halmos}
\section{Relation with  splitting and conditional Monte Carlo } \label{se:splitting}
Like the splitting algorithm described in~\cite[Section V.5]{asmussenGlynn2007}
when \(d=2\),
our method samples more often important random variables.    This section explores further the relation between our method and the splitting and conditional Monte Carlo methods.  Fix \(n\geq1\) and assume for simplicity that the sequence \((N_{k})\) is deterministic. We first analyse the relation between our method and the conditional Monte Carlo method in the general case, then show that
 the generic dimension reduction algorithm for Markov chains estimation can  be efficiently cast as a  splitting algorithm.
\subsection{Relation with conditional Monte Carlo}
Define  \(\mathcal{S}_{i}\)  via~\eqref{eq:defSi} for \(0\leq i\leq d-1\), and set  \(\mathcal{S}_{d}=\{0\}\).  For \(0\leq i\leq d\) and \(m\in\mathcal{S}_{i}\),   let \begin{equation}\label{eq:SmiDef}
\mathcal{S}_{i,m}=\{k\in[m,n-1]:(m,k]\cap\mathcal{S}_{i}=\emptyset\},
\end{equation}and\begin{equation*}
f_{i,m}=\frac{1}{|\mathcal{S}_{i,m}|}\sum^{}_{k\in \mathcal{S}_{i,m}}f(V^{(k+1)}).
\end{equation*}
Thus, if \((m,m')\) are consecutive elements of \(\mathcal{S}_{i}\cup\{n\}\), then   \(\mathcal{S}_{i,m}=\{m,\dots, m'-1\}\).  In particular, for \(0\leq m\leq n-1\), we have \(\mathcal{S}_{0,m}=\{m\}\),   and so  \(f_{m,0}=f(V^{(m+1)})\). Similarly, \(\mathcal{S}_{d,0}=\{0,\dots,n-1\}\) and \(f_{d,0}=f_{n}\). For \(0\leq i\leq d\),    \(m\in\mathcal{S}_{i}\),   and \(k \in\mathcal{S}_{i,m}\), we have  \(N_{l}\leq i\) for \(m< l\le k\), and so  the  last \(d-i\)  components of \(V^{(k+1)}\) and \(V^{(m+1)}\) are  the same. In other words, \begin{equation}\label{eq:Vjkp1Split}
V^{(k+1)}_{j}=V_{j}^{(m+1)},   \text{ for } 0\leq i<j\leq  d \text{ and } k \in\mathcal{S}_{i,m}. \end{equation} 
 We can thus view \(f_{i,m}\)  as a discrete analog to   \(E(f(V^{(m+1)})|V_{i+1}^{(m+1)},\dots,V_{d}^{(m+1)})\).
The following proposition shows that the random variables \(f_{i,m}\), for  \(0\leq i\leq d\) and \(m\in\mathcal{S}_{i}\),  can be calculated inductively via~\eqref{eq:CMCSplitt}. This is reminiscent of the conditional Monte Carlo method, where an expectation  is estimated via an average of conditional expectations.  
\begin{proposition}
For \(1\leq i\leq d\) and \(m\in\mathcal{S}_{i}\),  
\begin{equation}\label{eq:CMCSplitt}
f_{i,m}=\sum_{k\in \mathcal{S}_{i-1}\cap\mathcal{S}_{i,m}}\frac{|\mathcal{S}_{i-1,k}|}{|\mathcal{S}_{i,m}|}f_{i-1,k},
\end{equation} and
\begin{equation}\label{eq:SmiSizes}
|\mathcal{S}_{i,m}|=\sum_{k\in\mathcal{S}_{i-1}\cap\mathcal{S}_{i,m} }|\mathcal{S}_{i-1,k}|.
\end{equation}     
\end{proposition}
\commentt{\begin{proof}}{\proof{Proof.}}
We  first show that \begin{equation}\label{eq:simPartition}
\mathcal{S}_{i,m}=\bigcup_{k\in\mathcal{S}_{i-1}\cap\mathcal{S}_{i,m} }\mathcal{S}_{i-1,k}.
\end{equation} If  \(k\in\mathcal{S}_{i-1}\cap\mathcal{S}_{i,m}\) and \(k'\in\mathcal{S}_{i-1,k}\),
we have \((m,k]\cap\mathcal{S}_{i}=\emptyset\) and  \((k,k']\cap\mathcal{S}_{i-1}=\emptyset\).
As \(\mathcal{S}_{i}\subseteq \mathcal{S}_{i-1}\) and \(m\leq k\leq k'\), this implies that  \((m,k']\cap\mathcal{S}_{i}=\emptyset\), and so   \(k'\in\mathcal{S}_{i,m}\).    Thus  \begin{equation*}
\bigcup_{k\in\mathcal{S}_{i-1}\cap\mathcal{S}_{i,m} }\mathcal{S}_{i-1,k}\subseteq\mathcal{S}_{i,m}.
\end{equation*}    Conversely, given \(k'\in\mathcal{S}_{i,m}\), let\begin{equation}\label{eq:kmax0}
k=\max([0,k']\cap\mathcal{S}_{i-1}).
\end{equation} Since \(m\in[0,k']\cap\mathcal{S}_{i-1}\), the integer \(k\) is well-defined and \(m\leq k\).     Since \(k\leq k'\) and \((m,k']\cap\mathcal{S}_{i}=\emptyset\), we have \((m,k]\cap\mathcal{S}_{i}=\emptyset\). Hence \(k\in\mathcal{S}_{i,m}\), and so  \(k\in\mathcal{S}_{i-1}\cap\mathcal{S}_{i,m} \).  Furthermore,      \((k,k']\cap\mathcal{S}_{i-1}=\emptyset\) by~\eqref{eq:kmax0}.  Thus  \(k'\in\mathcal{S}_{i-1,k}\), and so  
   \begin{equation*}
\mathcal{S}_{i,m}\subseteq\bigcup_{k\in\mathcal{S}_{i-1}\cap\mathcal{S}_{i,m} }\mathcal{S}_{i-1,k}.
\end{equation*}This implies~\eqref{eq:simPartition}.  Moreover,  if \(k\in \mathcal{S}_{i-1}\) and \(j\in \mathcal{S}_{i-1,k}\), then \((k,j]\cap\mathcal{S}_{i-1}=\emptyset\), and so \(k=\max([0,j]\cap\mathcal{S}_{i-1}).\) Thus,  if \(k\) and \(k'\) are distinct elements of \(\mathcal{S}_{i-1}\),
the sets \(\mathcal{S}_{i-1,k}\) and \(\mathcal{S}_{i-1,k'}\) are disjoint.
Together with~\eqref{eq:simPartition}, this immediately implies~\eqref{eq:CMCSplitt} and~\eqref{eq:SmiSizes}.            
\commentt{\end{proof}}{\Halmos\endproof}     
  
\subsection{The Markov chains case}
Using the same notation and assumptions as in \S\ref{se:examples}, we  show
how to cast the generic dimension reduction algorithm for Markov chains estimation as a  splitting algorithm. For each integer \(k\) in  \([1,n]\), define the Markov chain  \((X_{i}^{(k)})\), \(0\leq i\leq d\), by induction on \(i\) as follows: \(X_{0}^{(k)}=X_{0}\) and \(X^{(k)}_{i+1}=g_{i}(X^{(k)}_{i},V^{(k)}_{d-i})\) for \(0\leq i\leq d-1\). Then it can be shown by induction that \(X_{d}^{(k)}=G_{d}(V^{(k)},X_{0})\), and that \(g(X_{d}^{(k)})=f(V^{(k)})\). The generic dimension reduction algorithm for Markov chains estimation
described in \S\ref{se:examples}  thus outputs the average of \(g(X_{d}^{(1)}),\dots,g(X_{d}^{(n)})\).
Furthermore, it can be shown by induction that, for \(0\leq i \leq d\) and \(1\leq k\leq n\), the random variable \(X_{i}^{(k)}\) is a deterministic function of \((V^{(k)}_{j})\), \(d-i<j\leq d\). On the other hand, by~\eqref{eq:SmiDef}, for \(0\leq i\leq d\) and  \(0\le k\le n-1\),   if \(m=\max([0,k]\cap\mathcal{S}_{d-i})\) then \(k\in\mathcal{S}_{d-i,m}\). The integer \(m\) is well-defined since \(0\in\mathcal{S}_{d-i}\). By~\eqref{eq:Vjkp1Split}, the last \(i\) components of   \(V^{(k+1)}\) and \(V^{(m+1)}\) are  the same, and so \(X^{(k+1)}_{i}=X_{i}^{(m+1)}\). However, for \(0\leq i\leq d-1\) and \(k\in\mathcal{S}_{d-i-1}\), we have \(V^{(k+1)}_{d-i}=U^{(k+1)}_{d-i}\) since \(N_{k}\geq d-i\), and so \begin{equation}\label{eq:splitInductionMarkov}
 X^{(k+1)}_{i+1}=g_{i}(X^{(m+1)}_{i},U^{(k+1)}_{d-i}), \text{ with }m=\max([0,k]\cap\mathcal{S}_{d-i}).  \end{equation}We can  calculate  \(X_{i}^{(k+1)}\) by induction on \(i\) via~\eqref{eq:splitInductionMarkov}  for  \(0\leq i\leq d\) and  \(k\in\mathcal{S}_{d-i}\). Recalling that \(\mathcal{S}_{0}=\{0,\dots,n-1\}\), this allows us to simulate \(X_{d}^{(1)},\dots,X_{d}^{(n)}\). The generic dimension reduction algorithm for Markov chains estimation described in \S\ref{se:examples} can thus be rewritten as follows: \begin{enumerate}
\item Generate   \(N_{k}\) for \(1\leq k\leq n-1\) and calculate
the sets \(\mathcal{S}_{i}\), \(0\leq i\leq d\). 
\item Set \(X_{0}^{(1)}=X_{0}\).
\item For \(i=0,\dots,d-1\) and \(k\in\mathcal{S}_{d-i-1}\), sample a copy \(U^{(k+1)}_{d-i}\) of \(Y_{i}\) and calculate  \(X_{i+1}^{(k+1)}\) via~\eqref{eq:splitInductionMarkov}. \item Output the average  of \(g(X_{d}^{(1)}),\dots,g(X_{d}^{(n)})\).
\end{enumerate}
Step~3 generates one or several copies of \(X_{i+1}\) for each copy of \(X_{i}\), and so the above algorithm may be viewed as a splitting algorithm. Assume now that  \(N_{k}=\bar N_{k}\) for \(k\geq1\).
 Step~1 can then be implemented via~\eqref{eq:SiVander}.   In~\eqref{eq:splitInductionMarkov},  \(m=0\) if \(i=0\) and, by~\eqref{eq:SiVander},  \(m=\lfloor k/\mu_{d-i}\rfloor \mu_{d-i}\) if \(1\leq i\leq d-1\).  The expected running times of this algorithm and of the algorithm in~\S\ref{se:examples}  are  within a constant from each other as they are  both 
 proportional to the number of sampled copies of the \(U_{i}\)'s. \section{Further numerical experiments}\label{se:furtherNumerExper}
\subsection{Comparison with Quasi-Monte Carlo}
Consider the \(G_t/D/1\) queue with the same parameters as in~\S\ref{sub:NumGD1}. Table~\ref{tab:MD1ExpectationQuasi} estimates \(E(X_{d})\), and   Table~\ref{tab:MDthreshQuasi} gives VRFs in the estimation of  \(\mathbb{P}(X_{d}> z)\), for selected values of \(z\). Our numerical results for the RDR and DDR algorithms are similar to those of~\S\ref{sub:NumGD1}.
Once again, for  the RDR and DDR algorithms, the variable Cost $\times$ Std\(^2\) is roughly independent of \(d\), and the variance reduction factors are roughly proportional to \(d\). In contrast, for the QMC algorithm, the variable Cost $\times$ Std\(^2\) is roughly proportional to \(d\), and the variance reduction factors are roughly constant. The VRFs of the RDR and DDR algorithms in Table~\ref{tab:MDthreshQuasi} are, in general, greater than or equal to the corresponding VRFs in Table~\ref{tab:MD1ExpectationQuasi}, which confirms the resiliency of these algorithms to discontinuities of \(g\). In contrast, the VRFs of the QMC algorithm  in Table~\ref{tab:MDthreshQuasi} are lower than the corresponding VRFs  in Table~\ref{tab:MD1ExpectationQuasi}. The RDR and DDR algorithms  outperform the QMC
algorithm. The VRFs of the QMC algorithm in Tables~\ref{tab:MD1ExpectationQuasi} and~\ref{tab:MDthreshQuasi} are of the same order of magnitude as those obtained
by \citet{EcuyerLemieux2000}, who
have reduced the variance in the simulation of a \(M/M/1 \) queue by a  factor
ranging between \(5\) and \(10\) via lattice rules. 
   
 \begin{table}
\caption{Comparison with QMC in \(E(X_{d})\) estimation in \(G_{t}/D/1\) queue, 1000 samples,  where \(X_{d}\) is the number of customers in the queue at time-step \(d\).}
\begin{scriptsize}
\begin{tabular}{llcrclrr}\hline
    &     & $n$& $90\%$ confidence interval & Std & Cost& Cost $\times$ Std$^2$ &   VRF \\ \hline
$d=2500$ &RDR & $ 649$ & $5.525 \pm 8.9\times 10^{-3}$ & $1.7\times 10^{-1}$ & $2.736\times 10^4 \pm 1.5\times 10^2$ & $8.0\times 10^2$ & $48$\\
 &DDR & $ 782$ & $5.526 \pm 8.6\times 10^{-3}$ & $1.7\times 10^{-1}$ & $2.608\times 10^4$ & $7.2\times 10^2$ & $53$\\
 &QMC & $ 4096$ & $5.5235 \pm 1.0\times 10^{-3}$ & $2.0\times 10^{-2}$ & $1.024\times 10^7$ & $3.9\times 10^3$ & $10$\\

$d=5000$ &RDR & $ 1225$ & $5.521 \pm 6.5\times 10^{-3}$ & $1.2\times 10^{-1}$ & $5.462\times 10^4 \pm 3.2\times 10^2$ & $8.4\times 10^2$ & $92$\\
 &DDR & $ 1400$ & $5.518 \pm 6.2\times 10^{-3}$ & $1.2\times 10^{-1}$ & $5.174\times 10^4$ & $7.3\times 10^2$ & $105$\\
 &QMC & $ 4096$ & $5.5225 \pm 8.2\times 10^{-4}$ & $1.6\times 10^{-2}$ & $2.048\times 10^7$ & $5.1\times 10^3$ & $15$\\

$d=10000$ &RDR & $ 2295$ & $5.5243 \pm 4.6\times 10^{-3}$ & $8.8\times 10^{-2}$ & $1.106\times 10^5 \pm 6.3\times 10^2$ & $8.5\times 10^2$ & $182$\\
 &DDR & $ 2822$ & $5.5221 \pm 4.5\times 10^{-3}$ & $8.6\times 10^{-2}$ & $1.032\times 10^5$ & $7.7\times 10^2$ & $202$\\
 &QMC & $ 4096$ & $5.5234 \pm 1.1\times 10^{-3}$ & $2.0\times 10^{-2}$ & $4.096\times 10^7$ & $1.7\times 10^4$ & $9$\\

\hline\end{tabular}
 \end{scriptsize}
\label{tab:MD1ExpectationQuasi}
\end{table}

\begin{table}
\caption{VRFs for \(\mathbb{P}(X_{d}> z)\) estimation in  \(G_{t}/D/1\) queue.}
\begin{scriptsize}
\begin{tabular}{llrrrrrr}\hline
$z$    &     & $0$& $2$ & $4$ & $6$& $8$ &   $10$ \\ \hline
$d=2500$ &  RDR & $ 83$ & $60$ & $59$ & $49$ & $50$ & $51$\\
&DDR&$125$&$76$&$67$&$71$&$63$&$61$\\
 &  QMC & $ 1.8$ & $2.2 $ & $2.5$ & $2.7$ & $2.1$ & $1.7$\\
$d=5000$ &  RDR & $ 149$ & $120$ & $109$ & $101$ & $101$ & $107$\\
&DDR&$215$&$ 165$ & $132$ & $122$ & $116$&$110$\\
 &  QMC & $ 1.1$ & $1.8$ & $2.8$ & $2.4$ & $2.2$ & $1.8$\\
$d=10000$ &  RDR & $ 280$ & $227$ & $206$ & $208$ & $188$ & $179$\\
&DDR&$ 470$&$280$& $228$& $237$& $220$ & $189$
\\ &  QMC & $ 1.2$ & $2$ & $2.5$ & $2.2$ & $2.3$&$1.9$\\
\hline\end{tabular}
\end{scriptsize}
\label{tab:MDthreshQuasi}
\end{table}
\subsection{\(G_t/D/1\) queue with time-varying
amplitude}\label{sub:timeVaryingAmplitude}
Consider a \(G_t/D/1\) queue where  \(A_{i}\) has a Poisson distribution with time-varying rate
\begin{equation*}
\lambda_{i}=(1-\frac{1}{\ln (i+2)})(0.75+0.5\cos(\frac{\pi i}{50})),
\end{equation*}   for \(1\leq i\leq d\) (recall that \(A_{0}=0\)).
Thus, up to a time-varying multiplicative factor,   \(\lambda_{i}\)  has
the same expression as in~\S\ref{sub:NumGD1}. Table~\ref{MDTrend:expectation} estimates \(E(X_{d})\), and   Table~\ref{tab:MDthreshTrend} gives VRFs in the estimation of  \(\mathbb{P}(X_{d}> z)\), for selected values of \(z\). Our numerical results are similar to those of~\S\ref{sub:NumGD1}, except
that there is a big difference in the means at different times.
 Once again, for  the RDR, DDR, and MLMC algorithms, the variable Cost $\times$ Std\(^2\) is roughly independent of \(d\), and the variance reduction factors are roughly proportional to \(d\). The VRFs of the RDR and DDR algorithms in Table~\ref{tab:MDthreshTrend}  are, in most cases, greater than or equal to the corresponding VRFs in Table~\ref{MDTrend:expectation}, which confirms the resiliency of these algorithms to discontinuities of \(g\). In contrast, the VRFs of the MLMC algorithm in Table~\ref{tab:MDthreshTrend} are lower than the corresponding VRFs  in Table~\ref{MDTrend:expectation}. The RDR algorithm outperforms the MLMC
algorithm by  a factor  ranging from \(1\) to \(2\)  in Table~\ref{MDTrend:expectation},
and a factor ranging from \(2\) to \(14\)  in Table~\ref{tab:MDthreshTrend}.

\begin{table}
\caption{\(E(X_{d})\) estimation in \(G_{t}/D/1\) queue with time-varying
amplitude, 1000 samples,  where \(X_{d}\) is the number of customers in the queue at time-step \(d\).}
\begin{scriptsize}
\begin{tabular}{llcrclcc}\hline
    &     & $n$& $90\%$ confidence interval & Std & Cost& Cost $\times$ Std$^2$ &   VRF \\ \hline
$d=10^4$ &RDR & $ 2.7\times 10^3$ & $3.693 \pm 3.5\times 10^{-3}$ & $6.7\times 10^{-2}$ & $1.103\times 10^5 \pm 6.1\times 10^2$ & $4.9\times 10^2$ & $2.1\times 10^2$\\
 &DDR & $ 3.8\times 10^3$ & $3.6955 \pm 3.5\times 10^{-3}$ & $6.6\times 10^{-2}$ & $1.029\times 10^5$ & $4.5\times 10^2$ & $2.3\times 10^2$\\
 &MLMC & $ 8.7\times 10^3$ & $3.696 \pm 4.1\times 10^{-3}$ & $7.9\times 10^{-2}$ & $1.084\times 10^5$ & $6.8\times 10^2$ & $1.5\times 10^2$\\

$d=10^5$ &RDR & $ 2.6\times 10^4$ & $4.0264 \pm 1.2\times 10^{-3}$ & $2.3\times 10^{-2}$ & $1.099\times 10^6 \pm 5.4\times 10^3$ & $5.8\times 10^2$ & $1.9\times 10^3$\\
 &DDR & $ 2.9\times 10^4$ & $4.0255 \pm 1.2\times 10^{-3}$ & $2.3\times 10^{-2}$ & $1.040\times 10^6$ & $5.7\times 10^2$ & $1.9\times 10^3$\\
 &MLMC & $ 7.3\times 10^4$ & $4.0271 \pm 1.4\times 10^{-3}$ & $2.8\times 10^{-2}$ & $1.103\times 10^6$ & $8.5\times 10^2$ & $1.3\times 10^3$\\

$d=10^6$ &RDR & $ 2.5\times 10^5$ & $4.2573 \pm 3.7\times 10^{-4}$ & $7.2\times 10^{-3}$ & $1.098\times 10^7 \pm 4.9\times 10^4$ & $5.6\times 10^2$ & $2.1\times 10^4$\\
 &DDR & $ 2.7\times 10^5$ & $4.25687 \pm 3.8\times 10^{-4}$ & $7.3\times 10^{-3}$ & $1.051\times 10^7$ & $5.6\times 10^2$ & $2.1\times 10^4$\\
 &MLMC & $ 7.3\times 10^5$ & $4.2566 \pm 4.6\times 10^{-4}$ & $8.8\times 10^{-3}$ & $1.127\times 10^7$ & $8.7\times 10^2$ & $1.3\times 10^4$\\
\hline\end{tabular}
 \end{scriptsize}
\label{MDTrend:expectation}
\end{table}

\begin{table}
\caption{VRFs for \(\mathbb{P}(X_{d}> z)\) estimation in  \(G_{t}/D/1\) queue
with time-varying amplitude.}
\begin{scriptsize}
\begin{tabular}{llcccccc}\hline
$z$    &     & $0$& $2$ & $4$ & $6$& $8$ &   $10$ \\ \hline
$d=10^4$ &  RDR & $ 3.1\times10^{2}$ & $2.4\times10^{2}$ & $2.2\times10^{2}$ & $2.1\times10^{2}$ & $2.3\times10^{2}$ & $2.2\times10^{2}$\\
&DDR&$4.4\times10^{2}$&$2.7\times10^{2}$&$2.4\times10^{2}$&$2.6\times10^{2}$&$3.0\times10^{2}$&$2.8\times10^{2}$\\
 &  MLMC & $ 3.8\times10^{1}$ & $6.1\times10^{1} $ & $7.4\times10^{1}$ & $8.7\times10^{1}$ & $7.8\times10^{1}$ & $7.9\times10^{1}$\\
$d=10^5$ &  RDR & $ 3.3\times10^{3}$ & $2.3\times10^{3}$ & $2.0\times10^{3}$ & $2.0\times10^{3}$ & $1.9\times10^{3}$ & $2.1\times10^{3}$\\
&DDR&$3.9\times10^{3}$&$ 2.8\times10^{3}$ & $2.6\times10^{3}$ & $2.4\times10^{3}$ & $2.0\times10^{3}$&$2.4\times10^{3}$\\
 &  MLMC & $ 3.3\times10^{2}$ & $5.3\times10^{2}$ & $6.3\times10^{2}$ & $7.5\times10^{2}$ & $6.8\times10^{2}$ & $7.6\times10^{2}$\\
$d=10^6$ &  RDR & $ 3.5\times10^{4}$ & $2.3\times10^{4}$ & $2.0\times10^{4}$ & $1.9\times10^{4}$ & $1.8\times10^{4}$ & $1.8\times10^{4}$\\
&DDR&$ 4.5\times10^{4}$&$3.0\times10^{4}$& $2.6\times10^{4}$& $2.3\times10^{4}$& $2.3\times10^{4}$ & $2.4\times10^{4}$
\\ &  MLMC & $ 3.2\times10^{3}$ & $4.7\times10^{3}$ & $6.2\times10^{3}$ & $7.8\times10^{3}$ & $7.3\times10^{3}$&$6.5\times10^{3}$\\
\hline\end{tabular}
\end{scriptsize}
\label{tab:MDthreshTrend}
\end{table}
\subsection{A multi-frequency \(G_t/D/1\) queue}\label{sub:multiperiod}
Consider a \(G_t/D/1\) queue where  \(A_{i}\) has a Poisson distribution with time-varying rate
\begin{equation*}
\lambda_{i}=0.75+0.2\cos(\frac{\pi i}{50})+0.1\cos(\frac{\pi i}{5000})+0.05\cos(\frac{\pi i}{500000}),
\end{equation*}   for \(1\leq i\leq d\), with \(A_{0}=0\).
Table~\ref{tab:multiperiodic} estimates \(E(X_{d})\). 
 Once again, for  the RDR, DDR, and MLMC algorithms, the variable Cost $\times$ Std\(^2\) is roughly independent of \(d\), and the variance reduction factors are roughly proportional to \(d\).    The RDR algorithm outperforms the MLMC
algorithm by  about a factor  of \(1.5\).

\begin{table}
\caption{\(E(X_{d})\) estimation in a multi-frequency \(G_{t}/D/1\) queue, 1000 samples,  where \(X_{d}\) is the number of customers in the queue at time-step \(d\).}
\begin{scriptsize}
\begin{tabular}{llcrclrr}\hline
    &     & $n$& $90\%$ confidence interval & Std & Cost& Cost $\times$ Std$^2$ &   VRF \\ \hline
$d=10^{4}$ &RDR & $ 4.2\times10^2$ & $5.6 \pm 1.6\times10^{-2}$ & $3.0\times10^{-1}$ & $1.104\times10^5 \pm 6.6\times10^2$ & $1.0\times10^4$ & $2.5\times10^1$\\
 &DDR & $ 9.8\times10^2$ & $5.616 \pm 1.4\times10^{-2}$ & $2.6\times10^{-1}$ & $1.028\times10^5$ & $7.2\times10^3$ & $3.4\times10^1$\\
 &MLMC & $ 2.1\times10^3$ & $5.607 \pm 2.0\times10^{-2}$ & $3.8\times10^{-1}$ & $1.065\times10^5$ & $1.6\times10^4$ & $1.6\times10^1$\\

$d=10^{5}$ &RDR & $ 6.0\times10^3$ & $5.1944 \pm 4.5\times10^{-3}$ & $8.7\times10^{-2}$ & $1.101\times10^6 \pm 5.6\times10^3$ & $8.4\times10^3$ & $2.7\times10^2$\\
 &DDR & $ 7.8\times10^3$ & $5.1961 \pm 4.0\times10^{-3}$ & $7.7\times10^{-2}$ & $1.030\times10^6$ & $6.1\times10^3$ & $3.6\times10^2$\\
 &MLMC & $ 2.0\times10^4$ & $5.187 \pm 5.9\times10^{-3}$ & $1.1\times10^{-1}$ & $1.108\times10^6$ & $1.4\times10^4$ & $1.6\times10^2$\\

$d=10^{6}$ &RDR & $ 4.8\times10^4$ & $5.6118 \pm 1.7\times10^{-3}$ & $3.3\times10^{-2}$ & $1.102\times10^7 \pm 5.0\times10^4$ & $1.2\times10^4$ & $2.1\times10^3$\\
 &DDR & $ 1.0\times10^5$ & $5.6111 \pm 1.5\times10^{-3}$ & $3.0\times10^{-2}$ & $1.042\times10^7$ & $9.1\times10^3$ & $2.8\times10^3$\\
 &MLMC & $ 1.8\times10^5$ & $5.6113 \pm 2.2\times10^{-3}$ & $4.2\times10^{-2}$ & $1.124\times10^7$ & $2.0\times10^4$ & $1.3\times10^3$\\

\hline\end{tabular}
 \end{scriptsize}
 \label{tab:multiperiodic}
\end{table}

\subsection{Comparison with a long-run  average estimator}
In several examples, when \(d\) is large, the Markov chain \((X_{m})\), \(0\leq
m\leq d\), has some notion
of stationarity, and so \(E(g(X_d))\)  can be estimated  via a suitable
long-run average. A drawback of such an estimator is that it is   biased.
We compare below a long-run average  estimator with the RDR and DDR algorithms. In Tables~\ref{tab:garchBias} through~\ref{tab:MG1expectationLongRun}, the RDR and DDR algorithms  use $10^9/d$  samples, while the long-run algorithm uses \(11\times10^9/d \)  samples. Thus, for
each    of the three algorithms and each \(d\),  the  total number of simulations of the $U_i$'s throughout the independent samples is roughly \(11\times10^9\). The bias of the long-run average  estimator is calculated by taking the  difference
with the DDR estimator.

For the GARCH volatility example of~\S\ref{sub:numerGARCH}, a natural long-run average  estimator for $\Pr(X_d > z)$ is 
\begin{equation*}
\frac{1}{d}\sum^{d}_{i=1}{\bf1}\{X_{i}>z\}.
\end{equation*}
Table~\ref{tab:garchBias}  compares this estimator with the RDR and DDR estimators.
 The work-normalized variance of each of the three algorithms is roughly independent
of \(d\). The work-normalized variance of the long-run average  estimator  is smaller than those of the RDR and DDR algorithms by about
a factor of \(3\) and \(2\), respectively.  The bias of the long-run average  estimator decreases
as  \(d\) increases, but   is
much larger than the standard deviations of the long-run average  and DDR estimators.
\begin{table}
\caption{\(\mathbb{P}(X_{d}> z)\)  estimation in GARCH model, with \(z=4.4\times10^{-5}\), where \(X_{d}\) is the daily variance at time-step \(d\). }
\begin{scriptsize}
\begin{tabular}{llrrclrr}\hline
    &     & $n$& $90\%$ confidence interval & Std & Cost& Cost $\times$ Std$^2$ &   Bias \\ \hline
$d=1250$ &RDR & $ 277$ & $0.393471 \pm 7.3\times10^{-5}$ & $3.9\times10^{-2}$ & $1.368\times10^4 \pm 3.2$ & $21$ & $0$\\
 &DDR & $ 596$ & $0.393483 \pm 6.2\times10^{-5}$ & $3.4\times10^{-2}$ & $1.355\times10^4$ & $16$ & $0$\\
 &Long-run & $-$ & $0.415353 \pm 4.4\times10^{-5}$ & $7.9\times10^{-2}$ & $1.250\times10^3$ & $7.7$ & $0.0219$\\
$d=2500$ &RDR & $ 529$ & $0.39339 \pm 7.2\times10^{-5}$ & $2.8\times10^{-2}$ & $2.742\times10^4 \pm 8.5$ & $21$ & $0$\\
 &DDR & $ 1167$ & $0.393489 \pm 6.3\times10^{-5}$ & $2.4\times10^{-2}$ & $2.734\times10^4$ & $16$ & $0$\\
 &Long-run & $-$ & $0.404431 \pm 4.4\times10^{-5}$ & $5.6\times10^{-2}$ & $2.500\times10^3$ & $7.9$ & $0.0109$\\
$d=5000$ &RDR & $ 970$ & $0.39346 \pm 7.4\times10^{-5}$ & $2.0\times10^{-2}$ & $5.494\times10^4 \pm 23$ & $22$ & $0$\\
 &DDR & $ 1899$ & $0.393506 \pm 6.5\times10^{-5}$ & $1.8\times10^{-2}$ & $5.207\times10^4$ & $16$ & $0$\\
 &Long-run &$-$  & $0.398968 \pm 4.4\times10^{-5}$ & $4.0\times10^{-2}$ & $5.000\times10^3$ & $7.9$ & $0.0055$\\

\hline

\end{tabular}

 \end{scriptsize}
 \label{tab:garchBias}
\end{table}

For the \(G_{t}/D/1\) queue example of~\S\ref{sub:NumGD1} (resp.~\S\ref{sub:timeVaryingAmplitude}), the arrival rate is periodic (resp. almost periodic) with  period  \(100\), and so a long-run average   estimator for \(E(X_{d})\)
is 
\begin{equation*}\label{eq:LongRunGTD1}
\frac{1}{\lceil d/100\rceil}\sum^{\lceil d/100\rceil-1}_{i=0}X_{d-i*100}.
\end{equation*}
Tables~\ref{tab:MDexpectationLong-Run} and~\ref{tab:MDexpectationLong-RunVaryingAmplitude} compares this estimator with the RDR and DDR estimators. The work-normalized variance of the long-run average  estimator is larger than those of the RDR and DDR estimators by 
a factor ranging from \(1.5\) and \(2.2\). The bias of the long-run algorithm
is not reported in Table~\ref{tab:MDexpectationLong-Run} because it is not statistically significant. In Table~\ref{tab:MDexpectationLong-RunVaryingAmplitude}, however, the bias of the long-run average  estimator is
much larger than the standard deviations of the long-run average  and DDR estimators.

\begin{table}
\caption{\(E(X_{d})\) estimation in \(G_{t}/D/1\) queue,  where \(X_{d}\) is the number of customers in the queue at time-step \(d\), with \(A_{i}\sim\text{Poisson}(0.75+0.5\cos(\pi i/50))\)  for \(1\leq i\leq d\).}
\begin{scriptsize}
\begin{tabular}{llcrclr}\hline
    &     & $n$& $90\%$ confidence interval & Std & Cost& Cost $\times$ Std$^2$  \\ \hline
$d=10^4$ &RDR & $ 2.3\times10^3$ & $5.52351 \pm 4.8\times10^{-4}$ & $9.2\times10^{-2}$ & $1.100\times10^5 \pm 6.2\times10^1$ & $9.2\times10^2$ \\
 &DDR & $ 2.8\times10^3$ & $5.52333 \pm 4.4\times10^{-4}$ & $8.5\times10^{-2}$ & $1.032\times10^5$ & $7.4\times10^2$ \\
 &Long-run & $ -$ & $5.5238 \pm 6.2\times10^{-4}$ & $3.9\times10^{-1}$ & $1.000\times10^4$ & $1.5\times10^3$ \\

$d=10^5$ &RDR & $ 2.3\times10^4$ & $5.52324 \pm 4.8\times10^{-4}$ & $2.9\times10^{-2}$ & $1.100\times10^6 \pm 1.7\times10^3$ & $9.4\times10^2$\\
 &DDR & $ 3.6\times10^4$ & $5.52339 \pm 4.7\times10^{-4}$ & $2.8\times10^{-2}$ & $1.046\times10^6$ & $8.4\times10^2$ \\
 &Long-run & $ -$ & $5.5238 \pm 6.2\times10^{-4}$ & $1.2\times10^{-1}$ & $1.000\times10^5$ & $1.5\times10^3$ \\

$d=10^6$ &RDR & $ 2.3\times10^5$ & $5.52325 \pm 4.9\times10^{-4}$ & $9.4\times10^{-3}$ & $1.103\times10^7 \pm 4.8\times10^4$ & $9.8\times10^2$\\
 &DDR & $ 2.6\times10^5$ & $5.52363 \pm 4.5\times10^{-4}$ & $8.7\times10^{-3}$ & $1.047\times10^7$ & $7.9\times10^2$ \\
 &Long-run & $ -$ & $5.5238 \pm 6.1\times10^{-4}$ & $3.9\times10^{-2}$ & $1.000\times10^6$ & $1.5\times10^3$ \\

\hline\end{tabular}
 \end{scriptsize}
\label{tab:MDexpectationLong-Run}
\end{table}

\begin{table}
\caption{\(E(X_{d})\) estimation in \(G_{t}/D/1\) queue,  where \(X_{d}\) is the number of customers in the queue at time-step \(d\), with \(A_{i}\sim\text{Poisson}((1-{1}/{\ln (i+2)})(0.75+0.5\cos({\pi i}/{50})))\)}
\begin{scriptsize}
\begin{tabular}{llcrclrr}\hline
    &     & $n$& $90\%$ confidence interval & Std & Cost& Cost $\times$ Std$^2$ &   Bias\\ \hline
$d=10^4$ &RDR & $ 2.7\times10^3$ & $3.69517 \pm 3.5\times10^{-4}$ & $6.7\times10^{-2}$ & $1.100\times10^5 \pm 6.0\times10^1$ & $5.0\times10^2$ & $0$\\
 &DDR & $ 3.8\times10^3$ & $3.69535 \pm 3.4\times10^{-4}$ & $6.5\times10^{-2}$ & $1.029\times10^5$ & $4.4\times10^2$ & $0$\\
 &Long-run & $ -$ & $3.4862 \pm 4.8\times10^{-4}$ & $3.1\times10^{-1}$ & $1.000\times10^4$ & $9.5\times10^2$ & $-0à.209$\\

$d=10^{5}$ &RDR & $ 2.6\times10^4$ & $4.02661 \pm 3.7\times10^{-4}$ & $2.3\times10^{-2}$ & $1.100\times10^6 \pm 1.7\times10^3$ & $5.7\times10^2$ & $0$\\
 &DDR & $ 2.9\times10^4$ & $4.02676 \pm 3.7\times10^{-4}$ & $2.2\times10^{-2}$ & $1.040\times10^6$ & $5.3\times10^2$ & $0$\\
 &Long-run & $ -$ & $3.8854 \pm 5.1\times10^{-4}$ & $1.0\times10^{-1}$ & $1.000\times10^5$ & $1.1\times10^3$ & $-0.141$\\

$d=10^{6}$ &RDR & $ 2.5\times10^5$ & $4.2573 \pm 3.7\times10^{-4}$ & $7.2\times10^{-3}$ & $1.098\times10^7 \pm 4.9\times10^4$ & $5.6\times10^2$ & $0$\\
 &DDR & $ 2.7\times10^5$ & $4.25687 \pm 3.8\times10^{-4}$ & $7.3\times10^{-3}$ & $1.051\times10^7$ & $5.6\times10^2$ & $0$\\
 &Long-run & $ -$ & $4.1582 \pm 5.3\times10^{-4}$ & $3.4\times10^{-2}$ & $1.000\times10^6$ & $1.1\times10^3$ & $-0.099$\\

\hline\end{tabular}
 \end{scriptsize}
\label{tab:MDexpectationLong-RunVaryingAmplitude}
\end{table}

Consider now the  \(M_{t}/GI/1\) queue example of~\S\ref{sub:MtGI1NumerExample}, and assume that \(\theta\) is an integer. In our experiments, we have set \(d=\theta\) and \(X_{i}=W_{i}\) for \(0\leq i\leq d\),  and so a long-run average  estimator for 
 \(\mathbb{P}(W_{\theta}> 1)\) is 
\begin{equation*}
\frac{1}{\lceil d/100\rceil}\sum^{\lceil d/100\rceil-1}_{i=0}{\bf1}\{X_{d-i*100}>1\}.
\end{equation*}
Table~\ref{tab:MG1expectationLongRun} compares this estimator with the RDR and DDR estimators. The work-normalized variance of the long-run average  estimator is larger than those of the RDR and DDR estimators by about
a factor of \(1.3\) and \(2\), respectively. Here again, the bias of the long-run algorithm
is not reported because it is not statistically significant.
 \begin{table}
\caption{\(\mathbb{P}(W_{\theta}> 1)\) estimation in \(M_{t}/GI/1\) queue, \(\alpha=2\), where \(W_{\theta}\) is the residual work at time \(\theta\).
}
\begin{scriptsize}
\begin{tabular}{llcrclr}\hline
    &     & $n$& $90\%$ confidence interval & Std & Cost& Cost $\times$ Std$^2$    \\ \hline
$d=10^4$ &RDR & $ 3.9\times 10^3$ & $0.853775 \pm 4.9\times10^{-5}$ & $9.3\times10^{-3}$ & $1.100\times10^5 \pm 6.1\times10^1$ & $9.6$ \\
 &DDR & $ 7.6\times 10^3$ & $0.853788 \pm 3.9\times10^{-5}$ & $7.6\times10^{-3}$ & $1.037\times10^5$ & $5.9$ \\
 &Long-run & $ -$ & $0.853734 \pm 5.6\times10^{-5}$ & $3.6\times10^{-2}$ & $1.000\times10^4$ & $13$ \\

$d=10^5$ &RDR & $ 3.0\times 10^4$ & $0.85385 \pm 5.0\times10^{-5}$ & $3.0\times10^{-3}$ & $1.100\times10^6 \pm 1.7\times10^3$ & $10$ \\
 &DDR & $ 4.8\times 10^4$ & $0.853799 \pm 4.1\times10^{-5}$ & $2.5\times10^{-3}$ & $1.031\times10^6$ & $6.5$\\
 &Long-run & $ -$ & $0.853837 \pm 5.6\times10^{-5}$ & $1.1\times10^{-2}$ & $1.000\times10^5$ & $13$ \\

$d=10^6$ &RDR & $ 2.5\times 10^5$ & $0.853762 \pm 5.1\times10^{-5}$ & $9.8\times10^{-4}$ & $1.103\times10^7 \pm 4.9\times10^4$ & $11$ \\
 &DDR & $ 4.5\times 10^5$ & $0.853779 \pm 4.4\times10^{-5}$ & $8.4\times10^{-4}$ & $1.040\times10^7$ & $7.3$ \\
 &Long-run & $ -$ & $0.853847 \pm 5.6\times10^{-5}$ & $3.6\times10^{-3}$ & $1.000\times10^6$ & $13$ \\

\hline\end{tabular}
 \end{scriptsize}
 \label{tab:MG1expectationLongRun}
\end{table}

In summary, for large \(d\) and Markov chains with periodic features, \(E(g(X_{d}))\) can be  estimated via a suitable biased long-run average  estimator. The order of magnitude of the bias  depends on
the application and on the value of  \(d\). The bias of the long-run average  estimator is  difficult to evaluate without using an alternative estimator, though. In the examples in this subsection, the work-normalized variances of the long-run average, RDR and DDR estimators have the same order of magnitude. In the  \(G_t/D/1\) queue example of~\S\ref{sub:multiperiod}, however, the arrival rate is periodic with  period  \(10^{6}\). This example does not seem to admit a suitable long-run  average estimator for the values of \(d\) listed in Table~\ref{tab:multiperiodic}.

\commentt{\bibliography{poly}}
{}
\end{document}